\documentclass{aa} 
\usepackage{txfonts}
\usepackage[authoryear]{natbib}
\usepackage{lscape}                                
\usepackage{color}
\usepackage{graphicx}
\usepackage{longtable}
\newcommand{\elecd}{$n_{\rm e}$}
\newcommand{\te}{$T_{\rm e}$}
\newcommand{\hb}{H$\beta$}

\newcommand{\foiii}{[O~{\sc iii}]}

\newcommand{\foii}{[O~{\sc ii}]}

\newcommand{\fsii}{[S~{\sc ii}]}
\newcommand{\fsiii}{[S~{\sc iii}]}

\newcommand{\fnii}{[N~{\sc ii}]}

\newcommand{\mgii}{Mg~{\sc ii}}
\newcommand{\fariii}{[Ar~{\sc iii}]}
\newcommand{\fariv}{[Ar~{\sc iv}]}
\newcommand{\farv}{[Ar~{\sc v}]}
\newcommand{\fcaii}{[Ca~{\sc ii}]}
\newcommand{\fclii}{[Cl~{\sc ii}]}
\newcommand{\fcliii}{[Cl~{\sc iii}]}
\newcommand{\fcliv}{[Cl~{\sc iv}]}

\newcommand{\fneiii}{[Ne~{\sc iii}]}
\newcommand{\fneiv}{[Ne~{\sc iv}]}
\newcommand{\fnev}{[Ne~{\sc v}]}

\newcommand{\fniqii}{[Ni~{\sc ii}]}
\newcommand{\fniqiii}{[Ni~{\sc iii}]}

\newcommand{\ffeii}{[Fe~{\sc ii}]}
\newcommand{\ffeiii}{[Fe~{\sc iii}]}
\newcommand{\ffeiv}{[Fe~{\sc iv}]}

\newcommand{\oiii}{O~{\sc iii}}
\newcommand{\nitroi}{N~{\sc i}}
\newcommand{\nii}{N~{\sc ii}}
\newcommand{\niii}{N~{\sc iii}}
\newcommand{\sili}{Si~{\sc i}}
\newcommand{\silii}{Si~{\sc ii}}
\newcommand{\oi}{O~{\sc i}}
\newcommand{\oii}{O~{\sc ii}}
\newcommand{\ci}{C~{\sc i}}
\newcommand{\cii}{C~{\sc ii}}
\newcommand{\ciii}{C~{\sc iii}}

\newcommand{\nei}{Ne~{\sc i}}
\newcommand{\neii}{Ne~{\sc ii}}

\newcommand{\sii}{S~{\sc ii}}
\newcommand{\siii}{S~{\sc iii}}

\newcommand{\cli}{Cl~{\sc i}}

\newcommand{\feiii}{Fe~{\sc iii}}

\newcommand{\hi}{H\,{\sc i}}
\newcommand{\hii}{H~{\sc ii}}

\newcommand{\hei}{He~{\sc i}}
\newcommand{\heii}{He~{\sc ii}}

\newcommand{\tsq}{\emph{$t^2$}}
\newcommand{\mc}{\multicolumn}
\newcommand{\nodata}{---}

\def\msun{\mbox{{\rm M}$_\odot$}}

\begin{document}

\title{Analysis of chemical abundances in planetary nebulae with  [WC] central stars. II. Chemical abundances and the abundance discrepancy factor.
\thanks{Based on data obtained at Las Campanas Observatory, Carnegie Institution.} 
} 
\author{
Jorge Garc\'ia-Rojas\inst{1,2}, 
 Miriam Pe\~na\inst{3}, Christophe Morisset\inst{1,3}, Gloria Delgado-Inglada\inst{3}, Adal Mesa-Delgado\inst{4},  and Mar\'ia Teresa Ruiz\inst{5}} 
\offprints{J. Garc\'ia-Rojas} 
\institute{
1 Instituto de Astrof\'isica de Canarias (IAC), E-38200 La Laguna, Tenerife, Spain\\ 
2 Universidad de La Laguna, Dept. Astrof\'{\i}sica. E-38206 La Laguna, Tenerife, Spain \\
3 Instituto de Astronom\'ia, Universidad Nacional Aut\'onoma de M\'exico,
Apdo. Postal 70264, M\'ex. D. F., 04510 M\'exico\\
4 Instituto de Astrof\'{\i}sica, Facultad de F\'{\i}sica, Pontificia Universidad Cat\'olica de Chile, Av. Vicu\~na Mackenna 4860,782-0436 Macul, Santiago, Chile.\\
5 Departamento de Astronom\'ia, Universidad de Chile, Casilla 36 D, Las Condes, Santiago, Chile.\\
\email{jogarcia@iac.es; miriam@astro.unam.mx; morisset@astro.unam.mx; gloria@astro.unam.mx; amesad@astro.puc.cl; mtruiz@das.uchile.cl}  }
\date{Received xxxxxxxxxxx; accepted xxxxxxxxxxxx} 

\titlerunning{PNe with a [WC] central stars. II.}

\authorrunning{Garc{\'\i}a-Rojas et al.} 


\abstract 
{} 
{ We present the abundance analysis of 12 planetary nebulae ionized by [WC]-type stars and weak-emission-line stars ($wels$) obtained from high-resolution spectrophotometric data. Our main aims are to determine the chemical composition of the nebulae and to study the behaviour of the abundance discrepancy problem (ADF) in this type of planetary nebulae.} 
{The detection of a large number of optical recombination lines (ORLs) and collisionally excited lines (CELs) from different ions (O$^+$, O$^{++}$, C$^{++}$, C$^{+3}$ and Ne$^{++}$) were presented previously. Most of the ORLs were reported for the first time in these PNe, which increased the sample of PNe with detected faint ORLs. 
Ionic abundances were determined from the available CELs and ORLs, using previously determined physical conditions. Based on these two sets of ionic abundances, we derived the total chemical abundances in the nebulae using suitable ionization correction factors (when available).} 
{In spite of the [WC] nature of the central stars, moderate ADF(O$^{++}$) in the range from 1.2 to 4 were found for all the objects. We found that when the quality of the spectra is high enough, the ORLs O$^{++}$/H$^+$ abundance ratios obtained from different multiplets excited mainly by recombination are very similar. Possible dependence of ADFs on some nebular characteristics such as surface brightness and nebular diameter were analysed, but we found no correlation. Abundances derived from CELs were corrected by determining the $t^2$ parameter. O abundances for PNe, derived from ORLs, are in general higher than the solar abundance. 
We derived the C/O ratio from ORLs and N/O and $\alpha$-element/O ratios from CELs and found that these PNe are, in average, richer in N and C than the average of large PN samples.  About half of our sample is C-rich (C/O$>$1). The growth of $\alpha$-elements is corelated with the O abundance. Comparing the N/O and C /O ratios with those derived from stellar evolution models, we estimate that about half of our PNe have progenitors with initial masses similar to or larger than 4 M$_\odot$. No correlation was found between the stellar [WC]-type and the nebular chemical abundances. }
{}

\maketitle 


\section{Introduction
\label{intro}}

Planetary nebulae (PNe) are produced at the final stages of low to intermediate-mass stars (LIMS, initial masses between~1 and 8 {\msun} and ages from 0.1 to 9 Gyr)  by ejection of the outer stellar atmosphere shells.  Thus, the chemical composition of PNe contains elements present in the interstellar medium (ISM) at the time of the star formation, and elements processed in the stellar nucleus that are dredged-up to the stellar surface at different moments of the stellar evolution. Of the central stars of PNe, about 10\% belong to the [WC] class. These stars are H-deficient, their atmospheres are  He,  C, and O-rich and they present intense winds, with terminal velocities ranging from several hundred to several thousand km s$^{-1}$ and mass loss rates of about 10$^{-5}$$-$10$^{-6}$ {\msun} yr$^{-1}$  \citep[see e.g.,][and references therein]{ koesterke01}.  The surrounding PNe, on the other hand, show a normal level of H (H-rich gas).

Owing to their H-deficient composition, [WC] central stars are helium-burning objects. It has been found that [WC] central stars are mainly distributed in [WC]-late types ([WC8] to [WC11]) with stellar temperatures in the range from 30 kK to about 70 kK, that are surrounded by young and dense nebulae, and [WC]-early types ([WC4] to [WO2), with stellar temperatures higher than 100 kK that are surrounded by evolved nebulae. Thus there is no defined evolutionary status for these stars, and they evolve from the asymptotic giant branch (AGB) phase as do the other PN nuclei. Three different scenarios have been proposed for the formation of a [WC] central star: a late thermal pulse (LTP), a very late thermal pulse (VLTP) and an asymptotic giant branch final thermal pulse (AFTP) \citep[see][for a description of these scenarios]{herwig01,blocker01}.
Regarding their distribution in the Galaxy, a number of Wolf-Rayet planetary nebulae (hereinafter WRPNe) have been discovered in the Galactic bulge \citep{gornyetal04}, and recently it has been found that WRPNe in the disk  belong to a thinner disk than the average of PNe, probably indicating that they are young objects \citep{penaetal13}.

A few years ago we started an analysis of the chemical composition of PNe around [WC] central stars (WRPNe), aiming to determine these abundances via collisionally excited lines (CELs) and optical recombination lines (ORLs) and to compare the two results. Three WRPNe were studied by \citet{garciarojasetal09}.  In the present work, we analyze the chemical abundances for a sample of 12 Galactic PNe with [WC] or weak-emission-lines ({\it wels}) central stars.  In a previous paper \citep[][hereafter paper I]{garciarojasetal12}  we have presented the high-resolution spectrophotometric data for these objects, obtained at Las Campanas Observatory (LCO, Carnegie Institution) with the 6.5m Magellan telescope Clay and the double echelle Magellan Inamori Kyocera Spectrograph (MIKE). An analysis of the physical conditions (electron temperatures, {\te}, and electron densities, {\elecd}) for each object, along with a discussion of the diagnostics that better represent {\te} and {\elecd} for each nebulae, was presented there\footnote{In Paper I, the diagnostic diagram for M\,1-61 in figure 7 shows an incorrect {\te}({\fnii}) due to a typo in the data file. The correct value is higher and agree with the value presented in their table 4.}. 
In the present paper we use the results of Paper I to determine  ionic and total chemical abundances for He and several heavy\footnote{Although it is not strictly correct, it is a widely used convention in nebular astrophysics to refer to all elements heavier than He as heavy, and we followed this convention.} elements. We have added  two additional objects from \citet{garciarojasetal09} (PB\,8 and NGC\,2867). As in \citet{garciarojasetal09}, the main focus of this work is 
on the abundance discrepancy problem, that is, the difference between the chemical composition of the photoionized plasma as derived from CELs and from ORLs. These differences are commonly found  in photoionized nebulae and reach to a factor of 2-3 in {\hii} regions \citep{garciarojasesteban07} and up to 70 in PNe, with an averaged value of about 3 \citep{mcnabbetal13}.
These discrepancies are parametrized through the abundance discrepancy factor (ADF) which is defined as

\centerline{ADF(X$^{+i}$) = (X$^{+i}$/H$^+)_{\rm ORLs}$ / (X$^{+i}$/H$^+)_{\rm CELs}$,} 

\noindent where X$^{i+}$ is the $i+$ ionic abundance of an element X and H$^+$ is the abundance of ionized hydrogen.
Since one of the possible mechanisms proposed to cause high ADFs are tiny metal-rich inclusions embedded in the nebular plasma \citep[][and references therein]{liuetal06}, we consider that WRPNe are nebulae with high probability of having such inclusions, because the central stars are H-deficient and present in their stellar winds He, C, and O-rich material.

In the following we present the ionic abundances (Sect.~\ref{ionic}) for He and for heavy elements; for some of these elements, we have derived abundances from both CELs and ORLs. Total abundances and a discussion of the obtained results are presented in Sects.~\ref{totalabundances} and \ref{discuss_ab}. In Sect.~\ref{adfs}, the ADFs are presented and discussed. A final discussion is presented in Sect.~\ref{discuss}. Our conclusions can be found in Sect.~\ref{conclu}.

\section{Ionic chemical abundances
\label{ionic}}

\subsection{He$^+$ and He$^{++}$ abundances
\label{he_abund}}

\setcounter{table}{0}
\begin{table}
\begin{tiny}
\caption{He$^+$ and He$^{++}$ abundances$^{\rm a}$.}
\label{helium}
\begin{tabular}{l@{\hspace{2pt}}c@{\hspace{2pt}}c@{\hspace{2pt}}c@{\hspace{2pt}}c@{\hspace{2pt}}c@{\hspace{2pt}}c@{\hspace{2pt}}}
\noalign{\hrule} \noalign{\vskip3pt}
  			&          Cn\,1-5 &        Hb\,4 	        &     He\,2-86 	& M\,1-25       &    M\,1-30   &      M\,1-32     \\
\noalign{\smallskip} \noalign{\hrule} \noalign{\smallskip}
Lines			& 12		    	& 11	  		& 12 		& 16  		& 16 	 	& 13		\\ 
$\tau_{3889}$		& 6.23$\pm$1.09		& 13.51$\pm$2.78     	& 17.67$\pm$1.62& 13.22$\pm$2.04& 12.93:	& 5.51$\pm$2.35	\\ 
$\chi^2$ 		& 6.9		  	& 10.2		 	& 33.4		& 9.3	 	& 15.3		& 10.5		\\ 
He$^+$/H$^+$$^{\rm b}$	& 1380$\pm$28    	& 934$\pm$43	  	& 1226$\pm$24 	& 1236$\pm$35  	& 1306$\pm$35  	& 1257$\pm$47	\\ 
			& 			&			&		&		&		&		\\ 
He$^{++}$/H$^+$		& \nodata    		&203$\pm$6	  	& \nodata 	& 0.39$\pm$0.16  & \nodata  	& 3.75$\pm$0.44	\\ 
\noalign{\smallskip} \noalign{\hrule} \noalign{\smallskip}
  			&  M\,1-61   		& M\,3-15      		&  NGC\,5189   &    NGC\,6369 &    PC\,14   	&       Pe\,1-1    \\
\noalign{\smallskip} \noalign{\hrule} \noalign{\smallskip}
Lines			& 16		    	& 12	  		&  13 		& 14  		& 13 	 	& 14		\\ 
$\tau_{3889}$		& 16.50$\pm$1.87	& 14.21$\pm$2.28     	& 3.20$\pm$0.72 & 8.48$\pm$1.83& 5.88$\pm$1.41	& 10.50$\pm$1.67\\ 
$\chi^2$ 		& 28.3		  	& 17.0		 	&  7.9		& 9.3	 	& 6.0		& 7.3		\\ 
He$^+$/H$^+$$^{\rm b}$	& 1086$\pm$32    	& 1055$\pm$35	  	&  847$\pm$21 	& 1033$\pm$32  	& 1036$\pm$31  	& 1054$\pm$34	\\ 
			& 			&			&		&		&		&		\\ 
He$^{++}$/H$^+$		& \nodata    		& \nodata	  	&  376$\pm$6	& 3.96$\pm$0.36 & 41.5$\pm$1.2  & 0.47$\pm$0.19	\\ 
\noalign{\smallskip} \noalign{\hrule} \noalign{\smallskip}
\end{tabular}
\begin{description}
\item[$^{\rm a}$] In units of 10$^{-4}$. Errors correspond to the uncertainties in line flux measurements.
\item[$^{\rm b}$] It includes all uncertainties related to line intensities, {\elecd}, $\tau_{3889}$, and {\tsq}.
\end{description}
\end{tiny}
\end{table}

We have detected several {\hei} emission lines in the spectra of each nebulae. 
These lines arise mainly from recombination, but some of them can be affected by collisional excitation and 
self-absorption effects. 

We used the effective recombination coefficients of \citet{storeyhummer95} 
for {\hi} and those computed by \citet{porteretal05}, with the interpolation formulae provided by 
\citet{porteretal07} for {\hei}. 
The collisional contribution
was estimated from \citet{saweyberrington93} and \citet{kingdonferland95}, and the optical depth in 
the triplet lines was derived from the computations by \citet{benjaminetal02}. 
We determined the He$^+$/H$^+$ ratio from a maximum-likelihood method \citep[MLM, ][]{peimbertetal00, 
apeimbertetal02}. 
To self-consistently determine {\elecd}({\hei}), {\te}({\hei}), He$^+$/H$^+$ ratio and the optical depth in the {\hei} 
$\lambda$3889 line, ($\tau_{3889}$), we used the adopted density obtained from 
the CEL ratios for each object (see Table~\ref{helium}) and a set of $I$({\hei})/$I$({\hi}) line ratios. 
For all objects, we employed at least 12 observational constraints 
({\hei} lines + {\elecd}); each of which depends upon the four unknown quantities, and each dependence is 
unique. Finally, we obtained the best value for the four unknowns and the temperature fluctuations parameter, $t^2$ (see Sect.~\ref{tsquared}),  by minimizing $\chi^2$.
The obtained $\chi^2$ parameters are showed in Table~\ref{helium}; these parameters indicate a reasonable goodness of the fits, 
taking into account the degrees of freedom in each case.

Several {\heii} emission lines were measured in the spectra of Hb\,4, M\,1-25, M\,1-32, NGC\,5189, NGC\,6369, PC\,14, and Pe\,1-1. 
We used the brightest lines to compute the He$^{++}$/H$^+$ ratio using the recombination coefficients computed 
by \citet{storeyhummer95}. There is a very good agreement between the results obtained from the different lines, therefore we finally adopted the He$^{++}$/H$^+$ average, weighted by the uncertainties of each individual line. Final results 
are presented in Table~\ref{helium}. For the remaining objects, which have a lower ionization degree, we detected no {\heii} lines.

\subsection{Heavy-element ionic abundances from CELs
\label{abund_cels}}

We derived the ionic abundance ratios, X$^{+i}$/H$^+$, for several heavy-element ions from CELs. Computations were made
using {\sc pyneb}\footnote{http://www.iac.es/proyecto/PyNeb} \citep{luridianaetal12}, a python based package that generalizes the tasks of the {\sc iraf} package {\sc nebular}, except for Fe, Ni, and Ca ions, for which we used our own scripts. As pointed out in paper I, we assumed for each object a three-zone ionization scheme, where temperatures are represented by {\te}({\fnii}), {\te}({\foiii}) and {\te}({\fariv}), when available, for the low-, medium- and high-ionization zones, respectively. In Table~\ref{zones} we summarize the physical conditions used for each ion. In general, we used {\te}({\fnii})-{\elecd}(low to medium) for ions with ionization potential, IP $<$ 17 eV; and {\te}({\foiii})-{\elecd}(low to medium) for ions with 17 eV $<$ IP $<$ 39 eV. For high-ionization species, those with IP $>$ 39 eV (see Table~\ref{zones}), we have computed the abundances taking into account two sets of physical conditions: set A, which adopts {\te}({\foiii}) and {\elecd}(low to medium) and set B, which uses {\te}({\fariv}) and {\elecd}(high). Note that this only affects the  Cl, Ne, and Ar ionic abundances. We discuss the results in Sect.~\ref{totalabundances}.
The lines selected to compute ionic abundances are showed in Table~\ref{lines}. The ionic abundances obtained are presented in Table~\ref{ionic_ab}. Errors in the ionic abundances were computed from the uncertainties in the line fluxes through Monte Carlo simulations.

\setcounter{table}{1}
\begin{table}
\begin{tiny}
\caption{Physical conditions for abundance determinations.}
\label{zones}
\begin{tabular}{ll}
\noalign{\hrule} \noalign{\vskip3pt}
Phys. cond & Ions \\
\noalign{\smallskip} \noalign{\hrule} \noalign{\smallskip}
{\te}({\fnii}), {\elecd}(low to medium)$^{\rm a}$ & N$^+$, O$^+$, S$^+$, Cl$^+$, Fe$^+$ and Fe$^{++}$ \\
{\te}({\foiii}), {\elecd}(low to medium)$^{\rm a}$ & O$^{++}$,  S$^{++}$, Cl$^{++}$, Ar$^{++}$ and Fe$^3+$\\
{\te}({\fariv})$^{\rm b}$, {\elecd}(high)$^{\rm a}$ & Cl$^{+3}$, Ar$^{+3}$, Ar$^{+4}$, Ne$^{++}$, Ne$^{+3}$ and Ne$^{+4}$\\
\noalign{\smallskip} \noalign{\hrule} \noalign{\smallskip}
\end{tabular}
\begin{description}
\item[$^{\rm a}$] See table 6 of paper I.
\item[$^{\rm b}$] When available. If not available we used {\te}({\foiii})
\end{description}
\end{tiny}
\end{table}

The atomic data used for our computations are the same as in paper I and are presented in table 6 of \citet{garciarojasetal09} except for the transition probabilities of {\fneiii}, {\fneiv}, and {\fnev}, which were updated to the values given by \citet{galavisetal97}, \citet{beckeretal89}, and \citet{galavisetal97}, respectively, and the transition probabilities of {\fsii} and {\fsiii} that were updated to the values provided by \citet{podobedovaetal09}. For the Fe, Ni, and Ca ions, the atomic data are described below. 

\setcounter{table}{2}
\begin{table}
\begin{tiny}
\caption{Lines used for abundance determinations.}
\label{lines}
\begin{tabular}{ll}
\noalign{\hrule} \noalign{\vskip3pt}
Ion & Line \\
\noalign{\smallskip} \noalign{\hrule} \noalign{\smallskip}
N$^+$  &  {\fnii} $\lambda$$\lambda$6548, 6584 \\
O$^+$  &  {\foii} $\lambda$$\lambda$3726+29, 7320+30 \\
O$^{++}$  &  {\foiii} $\lambda$$\lambda$4959, 5007 \\
Ne$^{++}$  & {\fneiii} $\lambda$$\lambda$3868, 3967 \\
Ne$^{+3}$  &   {\fneiv} $\lambda$$\lambda$4714+15, 4724+25  \\
Ne$^{+4}$  &  {\fnev} $\lambda$3425   \\
S$^+$  &	{\fsii} $\lambda$$\lambda$6717+31, 4068+76 \\
S$^{++}$  & {\fsiii} $\lambda$9069	\\
Cl$^+$  &	{\fclii} $\lambda$$\lambda$8578, 9123 \\
Cl$^{++}$  & {\fcliii} $\lambda$5517+37	\\
Cl$^{+3}$  & {\fcliv} $\lambda$8045	\\
Ar$^{++}$  & {\fariii} $\lambda$$\lambda$7136, 7751	\\
Ar$^{+3}$  & {\fariv} $\lambda$$\lambda$4711+40	\\
Ar$^{+4}$ &  {\farv} $\lambda$7005 	    \\
Fe$^{+}$   & {\ffeii} $\lambda$8616 (7155)	 \\
Fe$^{++}$  & {\ffeiii} $\lambda$$\lambda$4658, 4701, 4733, 4754, 4769, 4881	 \\
Fe$^{+3}$  & {\ffeiv} $\lambda$6740 	   \\
Ni$^{++}$  & {\fniqiii} $\lambda$$\lambda$6000, 6401, 6533   \\
Ca$^{+}$   &  {\fcaii} $\lambda$$\lambda$7291, 7324    \\
\noalign{\smallskip} \noalign{\hrule} \noalign{\smallskip}
\end{tabular}
\end{tiny}
\end{table}

Several {\ffeii} lines have been detected in our spectra. Most of them are affected by continuum fluorescence effects 
\citep[][]{rodriguez99, verneretal00}. Fortunately, in the objects Cn\,1-5, He\,2-86, M\,1-32, and Pe\,1-1 we detected the {\ffeii} $\lambda$8616 line, 
which is almost insensitive to the effects of UV pumping. In addition, for  M\,1-30 and PC\,14, we estimated the Fe$^{+}$ abundance using the {\ffeii} $\lambda$7155 line,  which is least sensitive to fluorescence effects apart from the {\ffeii} $\lambda$8616 line \citep{verneretal00}. For these two objects, we assumed 
I($\lambda$7155)/I($\lambda$8616) $\sim$ 1 \citep{rodriguez96}. To compute the final Fe$^{+}$ abundances we used the atomic data presented in \cite{bautistapradhan98} and we resolved a 159-level atom. 

At least one, and in most of the cases, several {\ffeiii} lines unaffected by fluorescence, were detected in our PNe. Only the spectrum of NGC\,6369 did not present any {\ffeiii} line. 
To calculate the Fe$^{++}$/H$^+$ ratio, we 
implemented a 34-level model atom that uses collision strengths from \citet{zhang96} and the 
transition probabilities of \citet{quinet96} as well as the new transitions found by 
\citet{johanssonetal00}. The average value of the Fe$^{++}$ abundance was obtained from several individual emission lines for 
the different objects, except for Hb\,4, M\,3-15, and NGC\,5189, where only one (Hb\,4) or two lines (M\,3-15 and NGC\,5189) were measured. 

One {\ffeiv} line was detected in He\,2-86 at $\lambda$6739.79. The Fe$^{+3}$/H$^+$ ratio was derived 
using a 33-level model atom where all collision strengths are those calculated by \citet{zhangpradhan97} 
and the transition probabilities are those recommended by 
\citet{froesefischerrubin98}, and those from \citet{garstang58} for the transitions not considered 
by \citeauthor{froesefischerrubin98}.

Several {\fniqii} lines have been measured in several objects, but they are 
strongly affected by continuum fluorescence \citep{lucy95}. 
To compute Ni$^+$ abundances it is necessary to use a 76-level model that includes continuum fluorescence excitation. Therefore, a proper description of 
the ionizing continuum in the zone covered by the slit is necessary for each PNe, that is, we would need to compute the distance between the central star and the area covered by our slit, in addition to the stellar radius and effective temperature. All these requirements are very difficult to be carried out for most of our objects, given the uncertainties in the distances to the PNe and that many of them are almost completely covered by our slit. Because of the relatively low importance of this ion in the total Ni abundance, even for low-ionization objects such as {\hii} regions \citep[see][]{mesadelgadoetal09}, we finally did not compute Ni$^+$ abundances for our PNe, because in general they are more highly ionized objects than average {\hii} regions. 

We measured several {\fniqiii} lines in our PNe. These lines are not expected to be affected by fluorescence given the structure of the ion \citep{bautista01}, in the same way as the {\feiii} lines. The Ni$^{++}$/H$^+$ ratio was derived using a 126-level model atom and the atomic data of \citet{bautista01}.

Two {\fcaii} lines at $\lambda\lambda$7291 and 7324  were detected in the spectrum of He\,2-86 and only one at $\lambda$7291  in  M\,1-25, M\,1-32 and Pe\,1-1. The {\fcaii} line at $\lambda$7291 in He\,2-86 was not reported in paper I; we obtained an intensity of I({\fcaii})/I({\hb})=0.0095$\pm$0.0013 (I({\hb})=100) for this object. To derive the Ca$^+$ abundance, we solved a five-level model atom using the single atomic data set available for this ion \citep{melendezetal07}. Note that this is the first determination of the Ca$^+$ abundance in a PNe ever, and, because we did not detect additional ionization stages of this ion, this places a lower limit to the gas-phase Ca/H ratio in these objects.

\setcounter{table}{3}
\begin{table*}
\begin{tiny}
\begin{center}
\caption{Ionic abundances from CELs.}
\label{ionic_ab}
\begin{tabular}{lccccccc}
\noalign{\smallskip} \noalign{\smallskip} \noalign{\hrule} \noalign{\smallskip}
 & \multicolumn{7}{c}{12 + log(X$^{+i}$/H$^+$)} \\
\noalign{\smallskip} \noalign{\hrule} \noalign{\smallskip}
                 Ion &        Cn1-5 &          Hb4 &       He2-86 &        M1-25 &        M1-30 &        M1-32 &        M1-61\\
\noalign{\smallskip} \noalign{\hrule} \noalign{\smallskip}
              N$^+$  &8.03$\pm$0.04 &7.41$\pm$0.08 &7.48$\pm$0.07 &8.01$\pm$0.06 &8.34$\pm$0.05 &8.25$\pm$0.08 &6.95$\pm$0.07\\
              O$^+$  &8.09$\pm$0.07 &7.56$\pm$0.12 &7.64$\pm$0.13 &8.48$\pm$0.12 &8.65$\pm$0.11 &8.54$\pm$0.15 &7.35$\pm$0.14\\
           O$^{++}$  &8.69$\pm$0.04 &8.61$\pm$0.04 &8.75$\pm$0.04 &8.63$\pm$0.04 &8.53$\pm$0.05 &8.27$\pm$0.04 &8.65$\pm$0.04\\
          Ne$^{++}$$^{\rm a}$  &8.27$\pm$0.04 &8.07$\pm$0.06/7.63$\pm$0.12 &8.23$\pm$0.05/8.43$\pm$0.12 &7.47$\pm$0.10 &7.67$\pm$0.08 &7.23$\pm$0.06 &8.05$\pm$0.05/8.32$\pm$0.19\\
          Ne$^{+3}$$^{\rm a}$  &   \nodata    &   \nodata    &   \nodata    &   \nodata    &   \nodata    &   \nodata    &  \nodata    \\
          Ne$^{+4}$$^{\rm a}$  &   \nodata    &   \nodata    &   \nodata    &   \nodata    &   \nodata    &   \nodata    &  \nodata    \\
              S$^+$  &6.37$\pm$0.08 &5.65$\pm$0.43 &5.90$\pm$0.14 &6.31$\pm$0.18 &6.33$\pm$0.12 &6.56$\pm$0.18 &5.55$\pm$0.10\\
           S$^{++}$  &7.00$\pm$0.05 &6.73$\pm$0.08 &6.95$\pm$0.08 &7.12$\pm$0.06 &7.17$\pm$0.07 &7.02$\pm$0.10 &6.80$\pm$0.08\\
             Cl$^+$  &4.89$\pm$0.05 & 4.33$\pm$0.11  &4.40$\pm$0.08 &4.74$\pm$0.08 &4.93$\pm$0.08 &5.06$\pm$0.11 &4.09$\pm$0.08\\
          Cl$^{++}$  &5.40$\pm$0.04 &5.04$\pm$0.06 &5.32$\pm$0.05 &5.42$\pm$0.05 &5.54$\pm$0.05 &5.25$\pm$0.06 &5.05$\pm$0.06\\
          Cl$^{+3}$  &3.94$\pm$0.06 &4.79$\pm$0.08/4.55$\pm$0,10 &4.50$\pm$0.06/4.60$\pm$0.08 &2.88$\pm$0.10 &   \nodata    &   \nodata    &4.28$\pm$0.06/4.42$\pm$0.11\\
          Ar$^{++}$  &6.61$\pm$0.04 &6.33$\pm$0.07 &6.52$\pm$0.13 &6.64$\pm$0.05 &6.74$\pm$0.06 &6.53$\pm$0.06 &6.39$\pm$0.05\\
          Ar$^{+3}$$^{\rm a}$  &5.11$\pm$0.11 &6.04$\pm$0.05/5.66$\pm$0.11 &5.57$\pm$0.07/5.78$\pm$0.11 &   \nodata    &4.42$\pm$0.25 &4.35$\pm$0.23 &5.39$\pm$0.08/5.65$\pm$0.17\\
          Ar$^{+4}$$^{\rm a}$&   \nodata    &4.32$\pm$0.07/4.02$\pm$0.10 &   \nodata    &   \nodata    &   \nodata    &   \nodata    &  \nodata    \\
          Fe$^{+}$   &5.36$\pm$0.10 &   \nodata    &4.72$\pm$0.13 &5.10$\pm$0.12 &4.65$\pm$0.10 &6.05$\pm$0.16 &4.13$\pm$0.14 \\
          Fe$^{++}$  &5.95$\pm$0.06 &4.78$\pm$0.25 &5.52$\pm$0.07 &5.87$\pm$0.07 &5.49$\pm$0.07 &6.56$\pm$0.08 &4.80$\pm$0.11/4.85$\pm$0.07 \\
          Fe$^{+3}$  &   \nodata    &   \nodata    &5.73$\pm$0.13 &   \nodata    &   \nodata    &   \nodata    &  \nodata    \\
          Ni$^{++}$  &4.68$\pm$0.11 &   \nodata    &4.38$\pm$0.32 &   \nodata    &4.28$\pm$0.11 &5.54$\pm$0.11 &  \nodata    \\
          Ca$^{+}$   &   \nodata    &   \nodata    &2.19$\pm$0.23 &2.66$\pm$0.12 &   \nodata    &3.85$\pm$0.12 &  \nodata    \\
	  
\noalign{\smallskip} \noalign{\hrule} \noalign{\smallskip}
                 Ion &        M3-15 &      NGC5189 &      NGC6369 &         PC14 &        Pe1-1 &          PB8 &      NGC2867\\
\noalign{\smallskip} \noalign{\hrule} \noalign{\smallskip}
              N$^+$  &6.75$\pm$0.08 &8.12$\pm$0.05 &6.51$\pm$0.05 &6.81$\pm$0.05 &7.39$\pm$0.07 &6.80$\pm$0.08 &6.92$\pm$0.04\\
              O$^+$  &7.23$\pm$0.11 &8.28$\pm$0.13 &6.88$\pm$0.08 &7.43$\pm$0.09 &8.00$\pm$0.13 &7.36$\pm$0.13 &7.40$\pm$0.07\\
           O$^{++}$  &8.80$\pm$0.05 &8.42$\pm$0.04 &8.52$\pm$0.04 &8.74$\pm$0.04 &8.56$\pm$0.04 &8.75$\pm$0.05 &8.44$\pm$0.04\\
          Ne$^{++}$$^{\rm a}$ &8.01$\pm$0.14 &7.96$\pm$0.06/7.46$\pm$0.10 &7.89$\pm$0.04/7.53$\pm$0.16 &8.17$\pm$0.05/7.67$\pm$0.21 &7.92$\pm$0.05 &8.11$\pm$0.06 &7.80$\pm$0.05\\
          Ne$^{+3}$$^{\rm a}$&   \nodata    &7.91$\pm$0.30/6.80$\pm$0.36 &   \nodata    &   \nodata    &   \nodata    &   \nodata    &7.57$\pm$0.28\\
          Ne$^{+4}$$^{\rm a}$&   \nodata    &6.46$\pm$0.06/5.88$\pm$0.11 &   \nodata    &   \nodata    &   \nodata    &   \nodata    & 5.48$\pm$0.06\\
            S$^{+}$  &5.35$\pm$0.25 &6.57$\pm$0.07 &5.22$\pm$0.06 &5.56$\pm$0.10 &6.02$\pm$0.12 &5.08$\pm$0.08 &5.52$\pm$0.05\\
           S$^{++}$  &6.82$\pm$0.09 &6.88$\pm$0.06 &6.50$\pm$0.05 &6.77$\pm$0.05 &6.61$\pm$0.06 &6.98$\pm$0.07 &6.37$\pm$0.05\\
           Cl$^{+}$  &3.91$\pm$0.11 &4.99$\pm$0.06 &3.83$\pm$0.08 &4.08$\pm$0.09 &4.45$\pm$0.07 &   \nodata    &  \nodata    \\
          Cl$^{++}$  &5.20$\pm$0.06 &5.10$\pm$0.04 &4.90$\pm$0.04 &5.10$\pm$0.04 &5.01$\pm$0.06 &5.29$\pm$0.08 &4.80$\pm$0.05\\
          Cl$^{+3}$  &4.45$\pm$0.08 &4.47$\pm$0.05/4.17$\pm$0.07 &4.40$\pm$0.04/4.20$\pm$0.10 &4.63$\pm$0.04/4.34$\pm$0.12 &3.77$\pm$0.05 &   \nodata    &4.40$\pm$0.04\\
          Ar$^{++}$  &6.40$\pm$0.07 &6.43$\pm$0.05 &6.17$\pm$0.04 &6.27$\pm$0.04 &6.31$\pm$0.05 &6.58$\pm$0.05 &5.96$\pm$0.04\\
          Ar$^{+3}$$^{\rm a}$&5.68$\pm$0.09 &5.82$\pm$0.04/5.38$\pm$0.09 &5.44$\pm$0.05/5.14$\pm$0.14 &5.70$\pm$0.05/5.27$\pm$0.18 &4.55$\pm$0.18 &4.98$\pm$0.15 &5.63$\pm$0.06\\
          Ar$^{+4}$$^{\rm a}$ &   \nodata    &4.82$\pm$0.05/4.46$\pm$0.08 &   \nodata    &   \nodata    &   \nodata    &   \nodata    &4.40$\pm$0.07\\
          Fe$^{+}$   &   \nodata    &   \nodata    &   \nodata    &3.87$\pm$0.23 &4.71$\pm$0.16 &   \nodata    &  \nodata    \\
          Fe$^{++}$  &5.04$\pm$0.13 &4.40$\pm$0.17 &   \nodata    &4.62$\pm$0.11 &4.98$\pm$0.13 &   \nodata    &  \nodata    \\
          Fe$^{+3}$  &   \nodata    &   \nodata    &   \nodata    &   \nodata    &   \nodata    &   \nodata    &  \nodata    \\
          Ni$^{++}$  &   \nodata    &   \nodata    &   \nodata    &   \nodata    &   \nodata    &   \nodata    &  \nodata    \\
          Ca$^{+}$   &   \nodata    &   \nodata    &   \nodata    &   \nodata    &2.54$\pm$0.17 &   \nodata    &  \nodata    \\
	  
\noalign{\smallskip} \noalign{\hrule} \noalign{\smallskip}
\end{tabular}
\end{center}
\begin{description}
\item[$^{\rm a}$] For these ions we computed abundances considering low to medium ionization and high-ionization physical conditions when different (see text).
\end{description}
\end{tiny}
\end{table*}

\subsection{Heavy-element ionic abundances from ORLs}

Multiple permitted lines of heavy-element ions, such as {\ci}, {\cii}, {\ciii}, {\nitroi}, 
{\nii}, {\niii}, {\oi}, {\oii}, {\oiii}, {\nei}, {\neii}, {\sili}, {\silii}, {\sii}, {\siii}, {\cli}, and {\mgii} were detected in our spectra.   Many of these lines are affected by 
fluorescence effects, as has been pointed out in \citet[][and references therein]{garciarojasetal09} or are blended with strong telluric emissions, making their intensities unreliable.

Several of these detected permitted lines are heavy-element ORLs, mainly of 
{\oi}, {\oii}, {\cii}, {\ciii}, and {\nii}, but also of {\neii} and {\mgii}, therefore we were able to measure ionic abundances of the corresponding ions. 
Most of these lines are detected for the first time in these PNe. To derive abundances we used the atomic data compiled in Table 6 of \citet{garciarojasetal09} for {\oii}, {\cii}, 
{\ciii}, {\nii}, and {\neii} ORLs. For {\oi} and {\mgii} the atomic data are described below. 

We followed a similar strategy as was used in previous papers \citep[see][and references therein]{garciarojasetal09, garciarojasesteban07}, selecting the most adequate multiplets of each ion to compute the ionic abundace from ORLs.

\subsubsection{Permitted lines of oxygen}

\begin{figure}[!htb] 
\begin{center}
\includegraphics[width=\columnwidth]{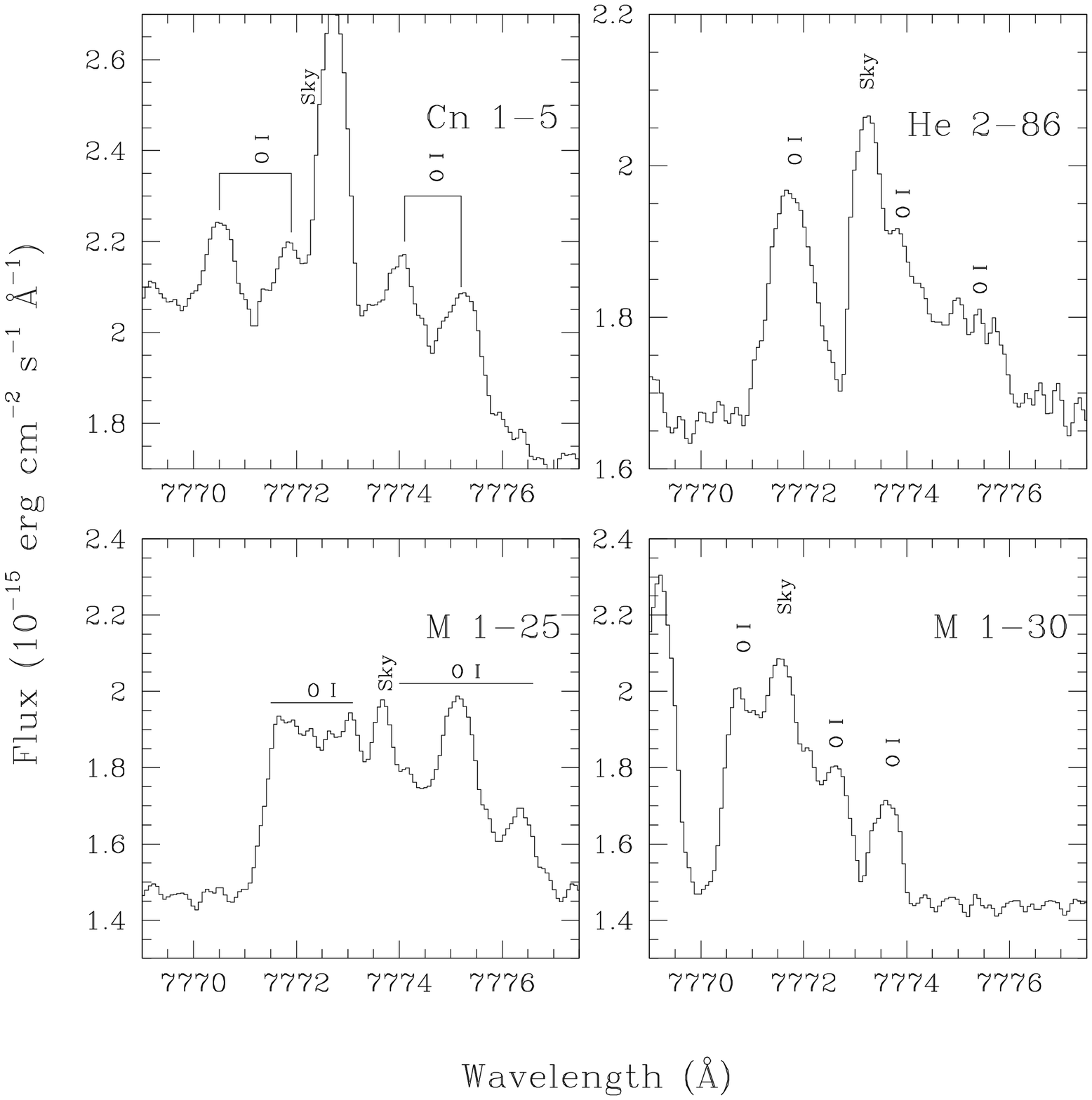}
\includegraphics[width=\columnwidth]{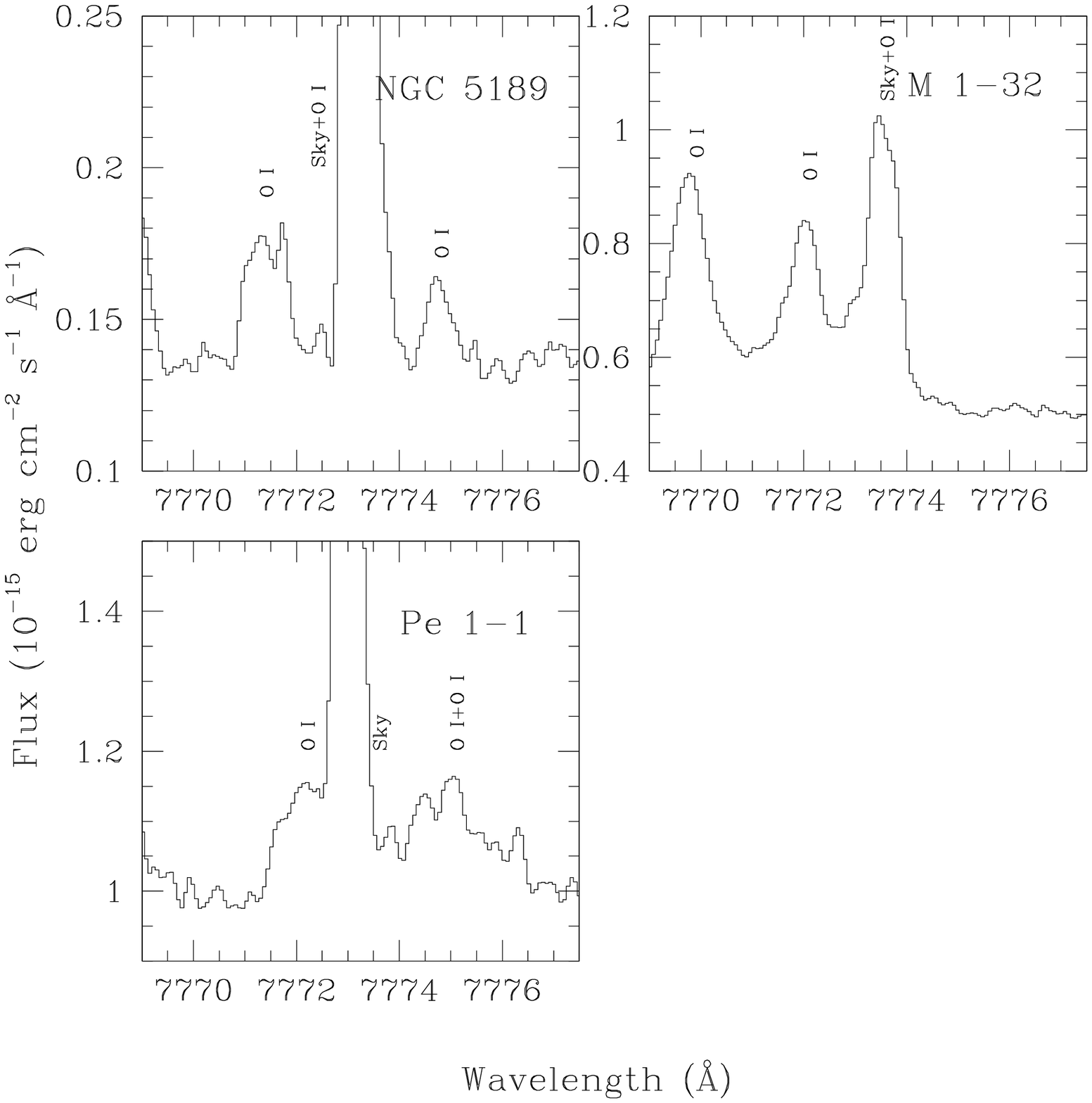}
\caption{Section of the echelle spectra showing the recombination emission lines of multiplet 1 of {\oi}. The relative position of the telluric emission line changes from one PN to the next due to the different heliocentric velocities of the PNe.}
\label{oi_lines}
\end{center}
\end{figure}

We detected and measured one or several lines of multiplet 1 of {\oi} in several objects. Unfortunately, these lines are in a spectral range with strong telluric OH emission lines that frequently blur the {\oi} lines. In Fig.~\ref{oi_lines} we show the spectral section where {\oi} RLs lie in the seven objects they have been detected in. We were easily able to deblend the three lines in M\,1-30; in Pe\,1-1 we were able to deblend {\oi} $\lambda$7771.94 from a telluric line and measured a blend of the other two components of the multiplet; in M\,1-32 and NGC\,5189 we were able to properly measure two lines; in Cn\,1-5, He\,2-86 we were able only to properly measure one line; and in M\,1-25 we were able to measure only a blend of the whole multiplet with the telluric line. For Pe\,1-1 we applied the theoretical ratio {\oi} $\lambda$7774.17/$\lambda$7775.39 to deblend both lines and derive individual abundances. For M\,1-25, it was impossible to deblend the {\oi} lines from the telluric line at $\sim$7774 \AA , therefore we decided to integrate the entire flux of the three {\oi} lines plus the telluric line between two given limits and then used this value to compute an upper limit to the abundance. In Table~\ref{recomb_oi} we present the abundances obtained from each single line and from the sum of all available lines of the multiplet. The adopted O$^+$/H$^+$ ratio is the average of the abundances obtained using the recombination coefficients from two atomic data sources \citep{pequignotetal91, escalantevictor92}. The results obtained are shown in Table~\ref{recomb_oi}.

\setcounter{table}{4}
\begin{table*}
\begin{tiny}
\caption{Ionic abundance ratios from the {\oi} recombination lines$^{\rm a}$}
\label{recomb_oi}
\begin{tabular}{cccccccccccccc}
\noalign{\hrule} \noalign{\vskip3pt}
Mult.& $\lambda_0$ & \mc{12}{c}{O$^{+}$/H$^+$ ($\times$10$^{-5}$)} 	\\
\noalign{\vskip3pt} \noalign{\hrule} \noalign{\vskip3pt}
&  			&          Cn\,1-5 &        Hb\,4 	& He\,2-86 	& M\,1-25       &  M\,1-30    & M\,1-32          &  M\,1-61   & M\,3-15   &  NGC\,5189    &  NGC\,6369  &    PC\,14   &  Pe\,1-1   \\
\noalign{\smallskip} \noalign{\hrule} \noalign{\smallskip}
1& 7771.94		& 34/27		   & ---  		& 18/14		& 41/32		& 96/74	      & 80/61		 & ---        & ---	  & 26/20         & --- 	& ---	      & 22:/17:	   \\
& 7774.17		& ---		   & ---	  	& ---		& 41/32		& 110/85      & 75/58		 & ---        & ---	  & --- 	  & --- 	& ---	      & 17:/13:	   \\
& 7775.39		& ---		   & ---	  	& ---		& 41/32		& 95/73	      & ---		 & ---        & ---	  & 26:/20:	  & --- 	& ---	      & 22:/17:	   \\
& Sum			& 34/27	   	   & ---	 	& 18/14		& 41/32		& 101/77      & 78/60	         & ---        & ---	  & 26/20	  & --- 	& ---	      & 20:/16:	   \\ 
\noalign{\smallskip} \noalign{\hrule} \noalign{\smallskip}
& Adopted		&{\bf 31$\pm$11}   &{\bf ---}   	&{\bf 16$\pm$4} &{\bf $<$37:} &{\bf 89$\pm$20}&{\bf 67$\pm$22} &{\bf ---}   &{\bf ---}  &{\bf 23$\pm$8}  &{\bf ---}	&{\bf ---}    &{\bf 18:}  \\ 
\noalign{\smallskip} \noalign{\hrule} \noalign{\smallskip}
\end{tabular}
\begin{description}
\item[$^{\rm a}$] Only lines with intensity uncertainties lower than 40 \% were considered (see text).
\item[$^{\rm b}$] Recombination coefficients from \citet{escalantevictor92} and \citet{pequignotetal91}.
\end{description}
\end{tiny}
\end{table*}

Many {\oii} permitted lines were detected in the spectra of our PNe. We only considered those belonging to multiplets 1, 2, 10, 20, and from $3d-4f$ 
transitions, which are mainly excited by recombination \citep[][and references therein]{estebanetal04}; however, some authors did not discard contributions by resonance fluorescense to these lines on relatively low-ionized objects, such as the Orion nebula and the PN IC\,418 \citep[][see below]{escalantemorisset05, escalanteetal12}. In Fig.~\ref{oii_lines} we show the region of the multiplet 1 {\oii} ORLs for each PN of the 
sample. The intensity of the brightest lines of this multiplet is about 10$^{-3} \times$ I({\hb}), which indicates 
the quality of the spectra. \citet{tsamisetal03} and \citet{ruizetal03} pointed out that the upper levels of the transitions of multiplet 1 of {\oii} are not in 
local thermodynamic equilibrium (LTE) for densities $n_e$$<$10$^4$ cm$^{-3}$, which is the case for several of our WRPNe, and the abundances derived from each individual line may differ by 
factors as large as 4. We applied the non-LTE corrections estimated by \citet{apeimbertetal05} to our data, and the abundances obtained from the individual 
lines agree well between them; we also derived the abundance using the sum of all lines of the multiplet following the recipe given by \citet{estebanetal98}. This abundance, which is not affected by non-LTE effects, agrees with that derived from individual lines. 
In Table~\ref{recomb_oii} we present the values obtained for the individual lines, as well as that derived from the sum of all the lines of a given multiplet. We only considered lines with errors lower than 40\%; for $3d-4f$ transitions, when available, we averaged lines with errors $<$40\%, otherwise, we averaged all available lines and quoted the final value with an uncertainty higher than 40\%. {\oii} $\lambda\lambda$4097.26, 4294.92 and 4613.68 were not taken into account because they are severely blended with other lines.  
   
\begin{figure}[!htb] 
\begin{center}
\includegraphics[width=\columnwidth]{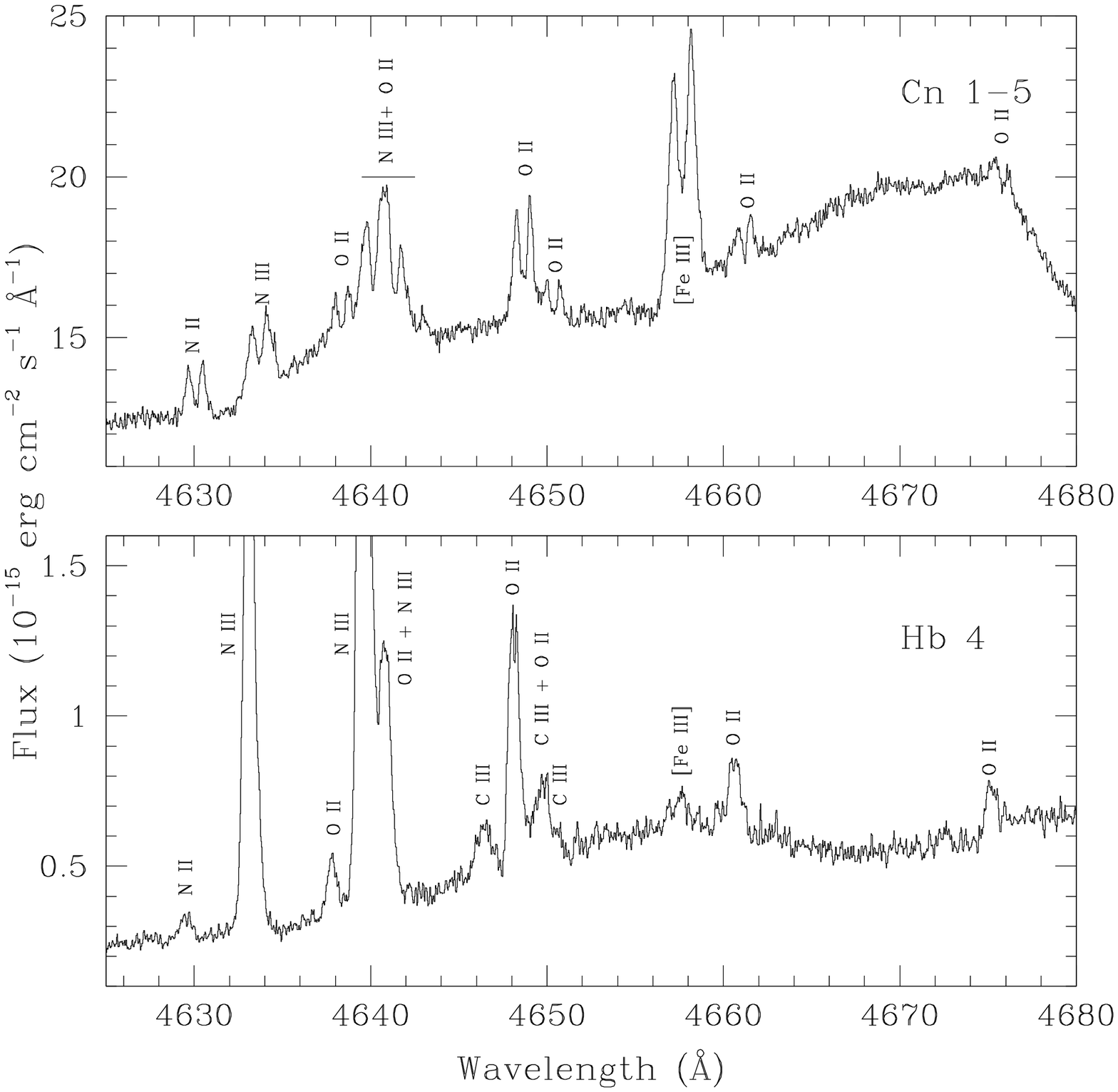}
\includegraphics[width=\columnwidth]{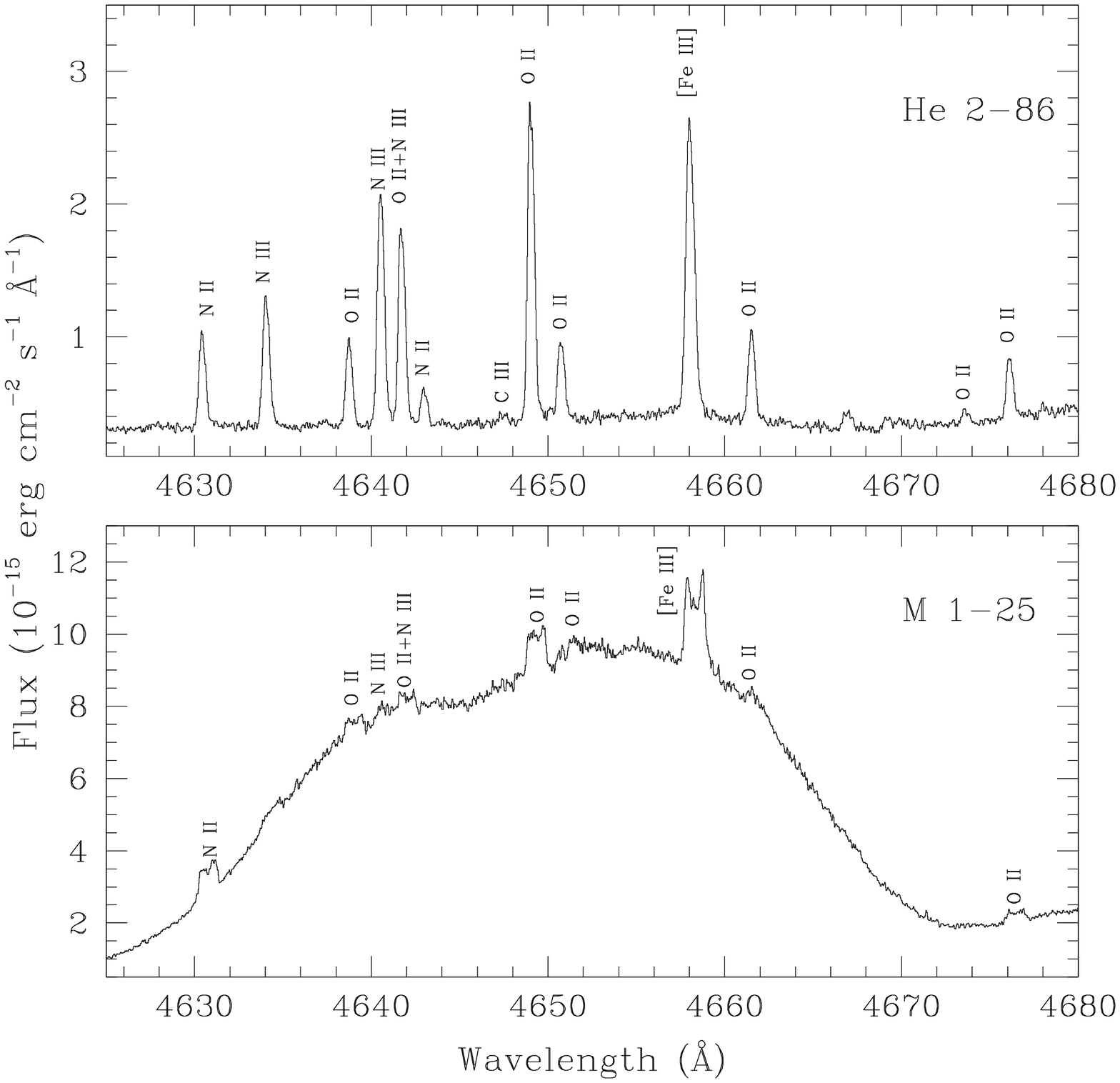}
\caption{Section of the echelle spectra showing the recombination emission lines of multiplet 1 of {\oii} and multiplet 1 of {\ciii}. {\nii} and {\niii} emission lines in these plot are not excited by pure recombination (see text).}
\label{oii_lines}
\end{center}
\end{figure}
\addtocounter{figure}{-1}
\begin{figure}
\begin{center}
\includegraphics[width=\columnwidth]{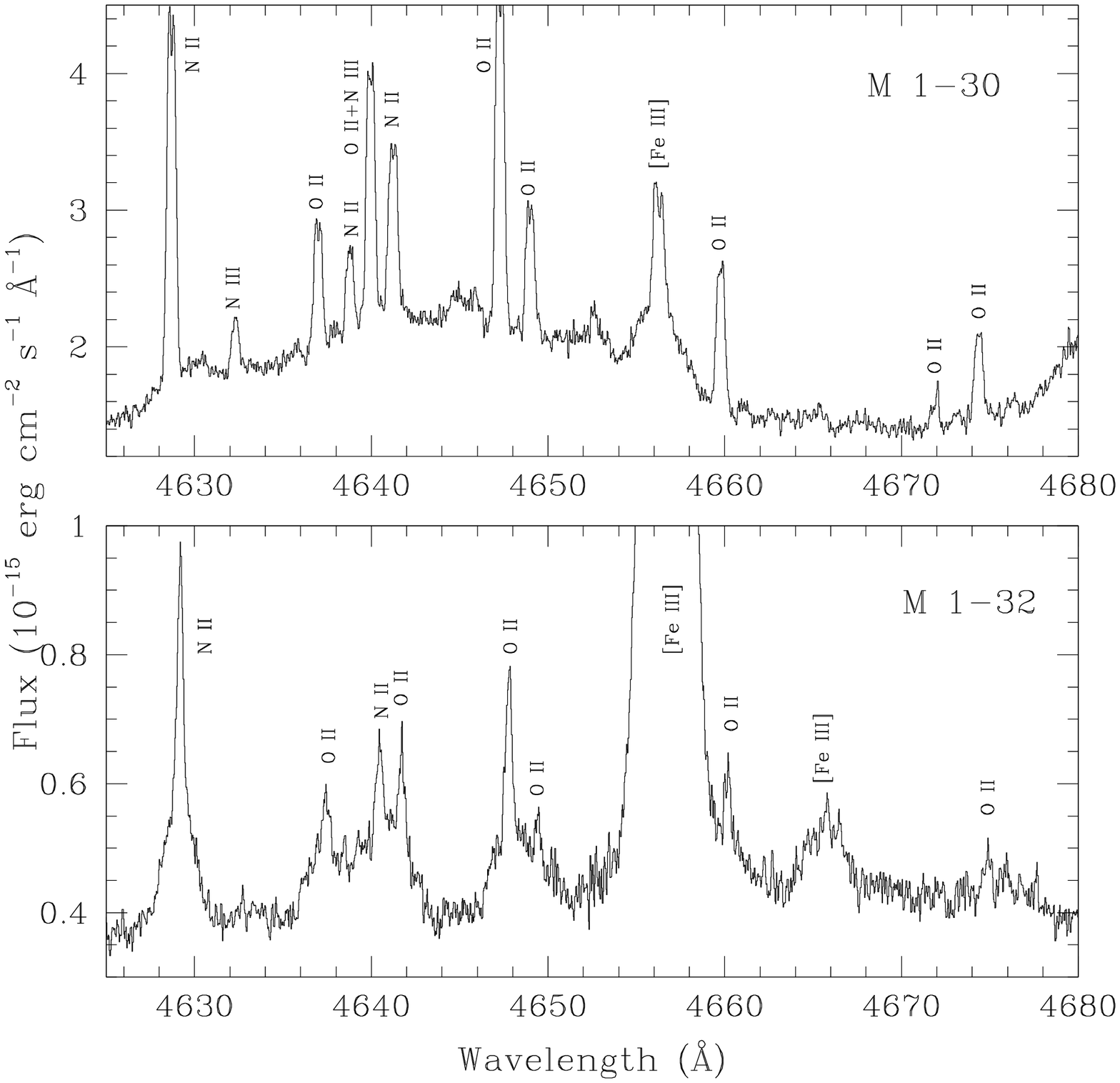}
\includegraphics[width=\columnwidth]{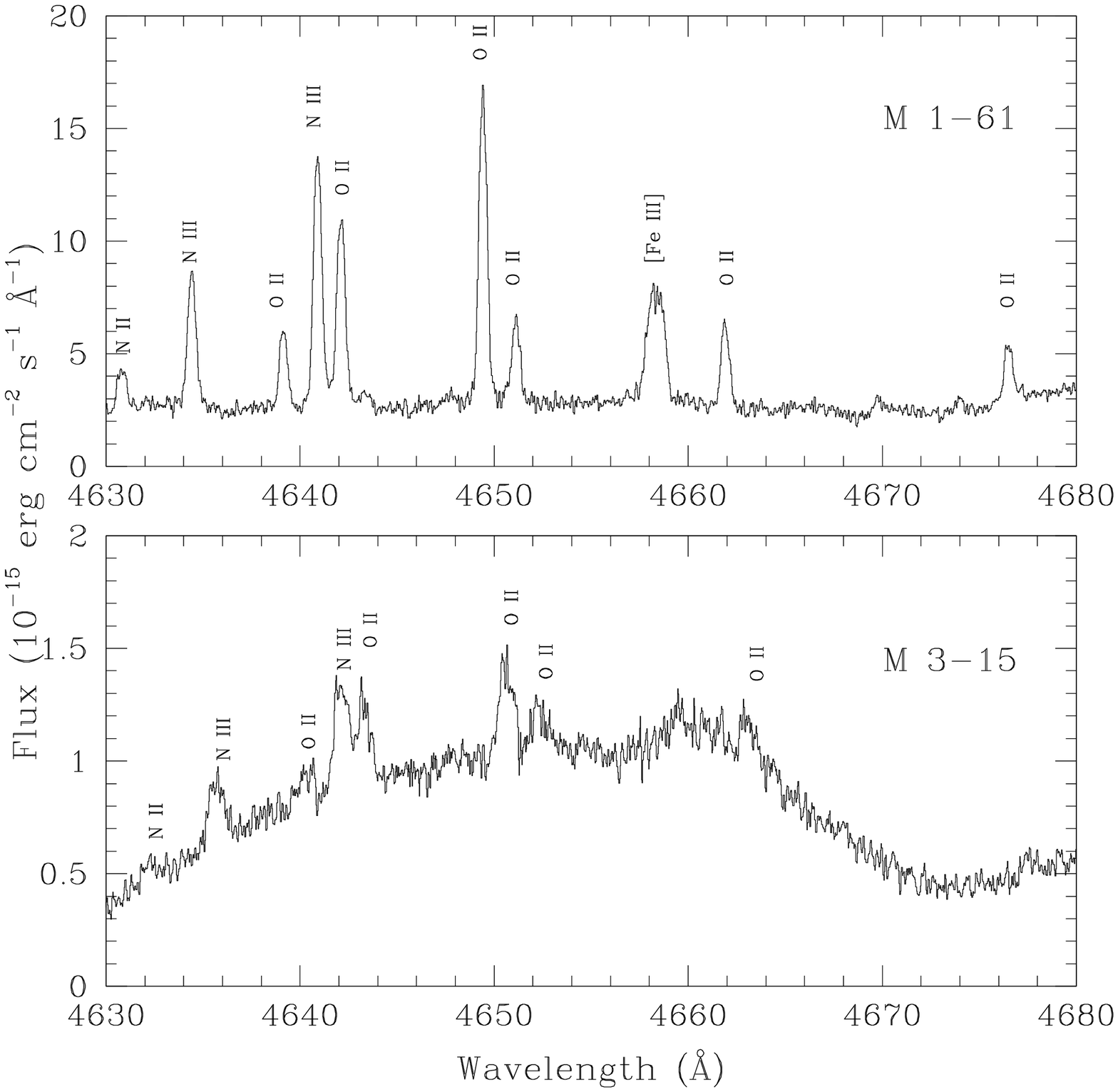}
\caption{Continued.}
\label{oii_lines}
\end{center}
\end{figure}
\addtocounter{figure}{-1}
\begin{figure}
\begin{center}
\includegraphics[width=\columnwidth]{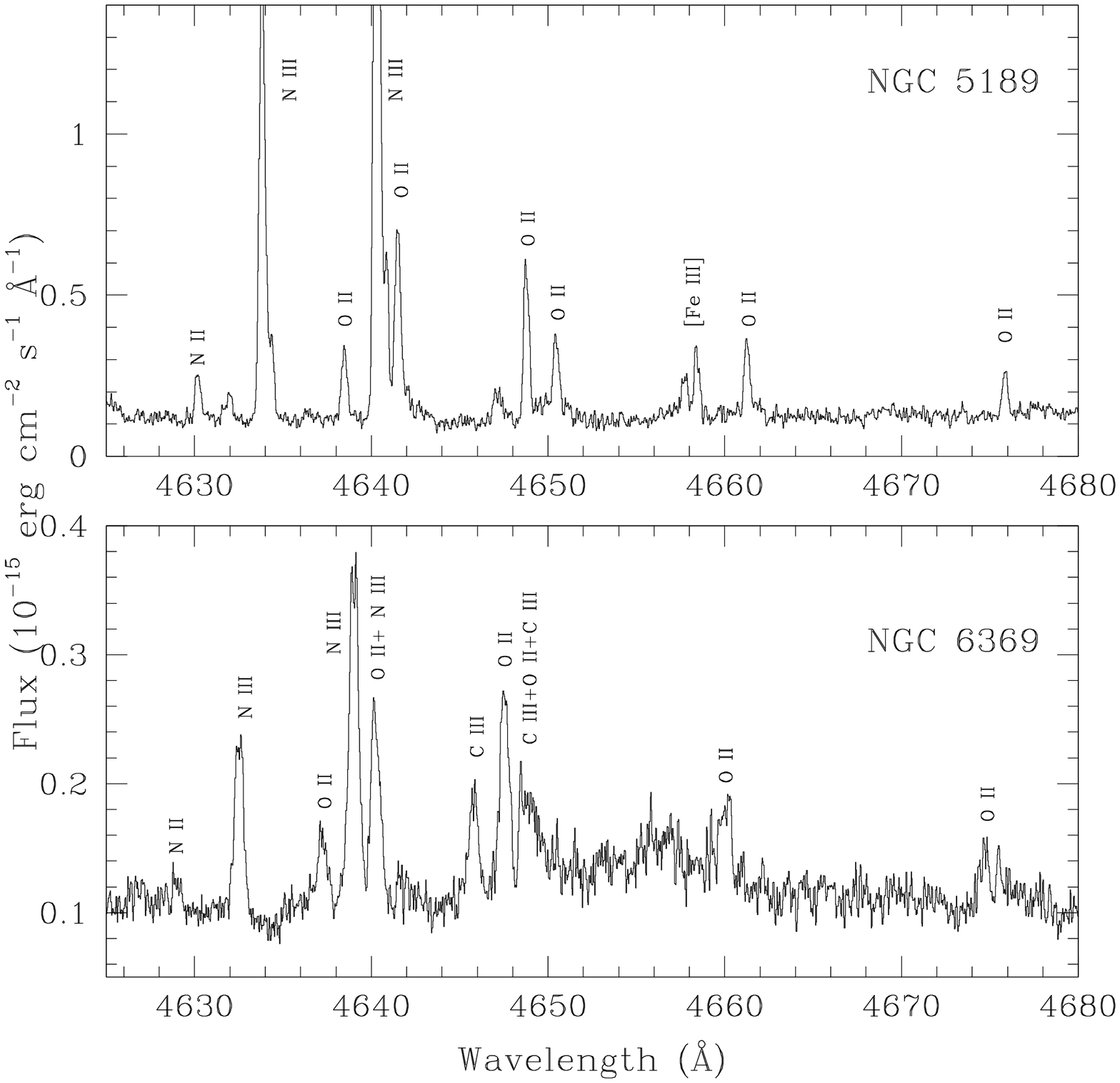}
\includegraphics[width=\columnwidth]{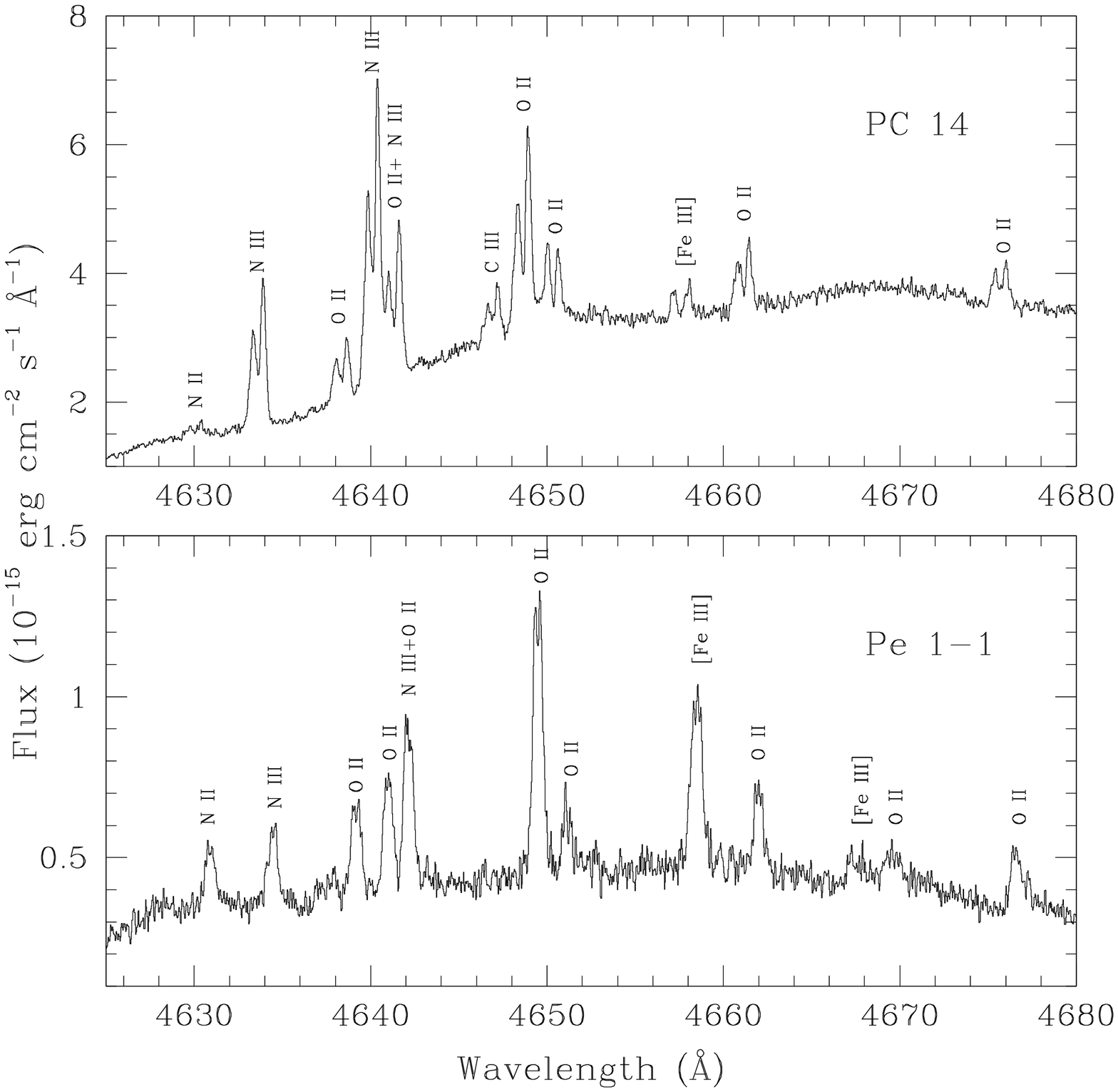}
\caption{Continued.}
\label{oii_lines}
\end{center}
\end{figure}

\begin{figure}[!htb] 
\begin{center}
\includegraphics[width=\columnwidth]{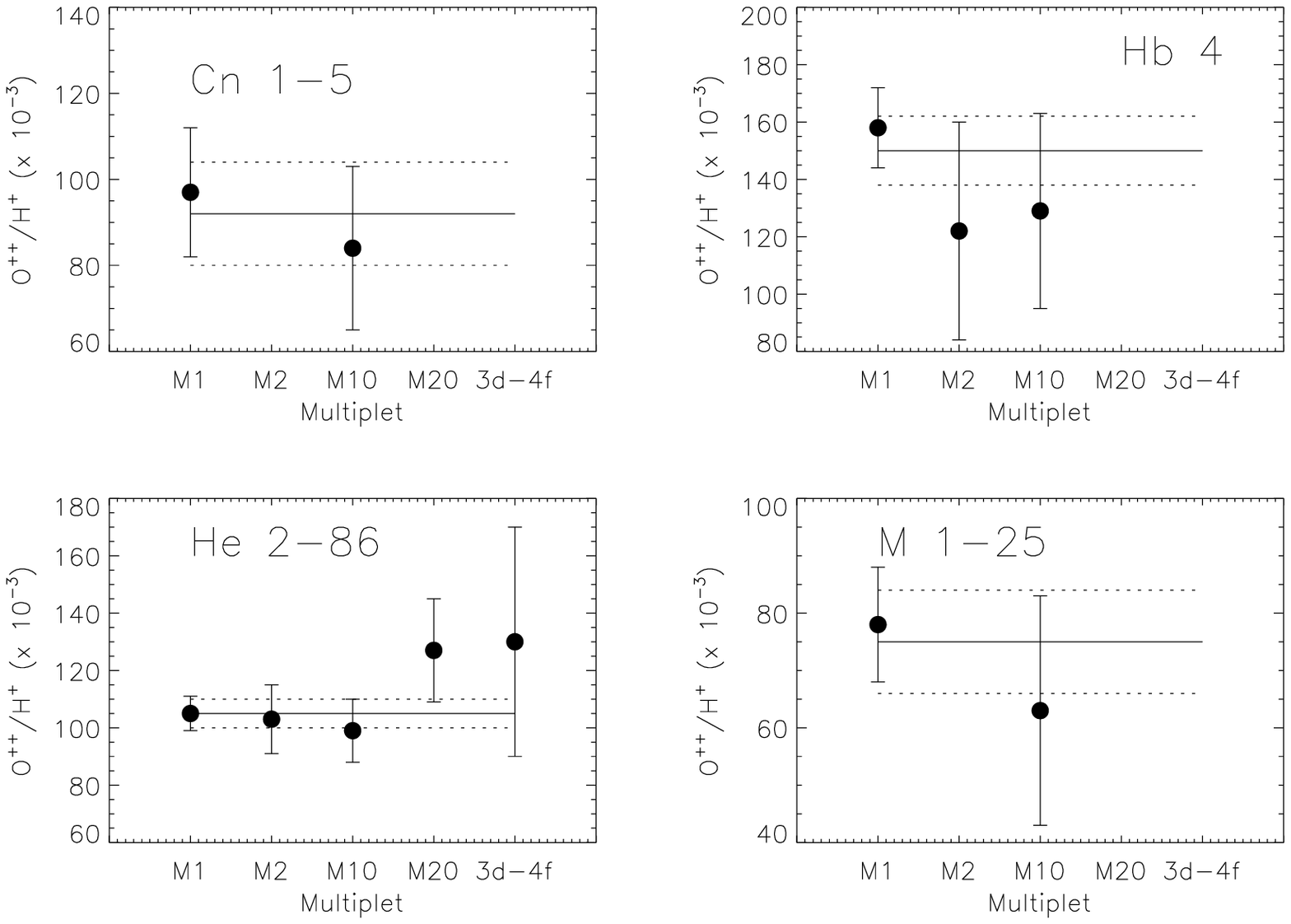}
\includegraphics[width=\columnwidth]{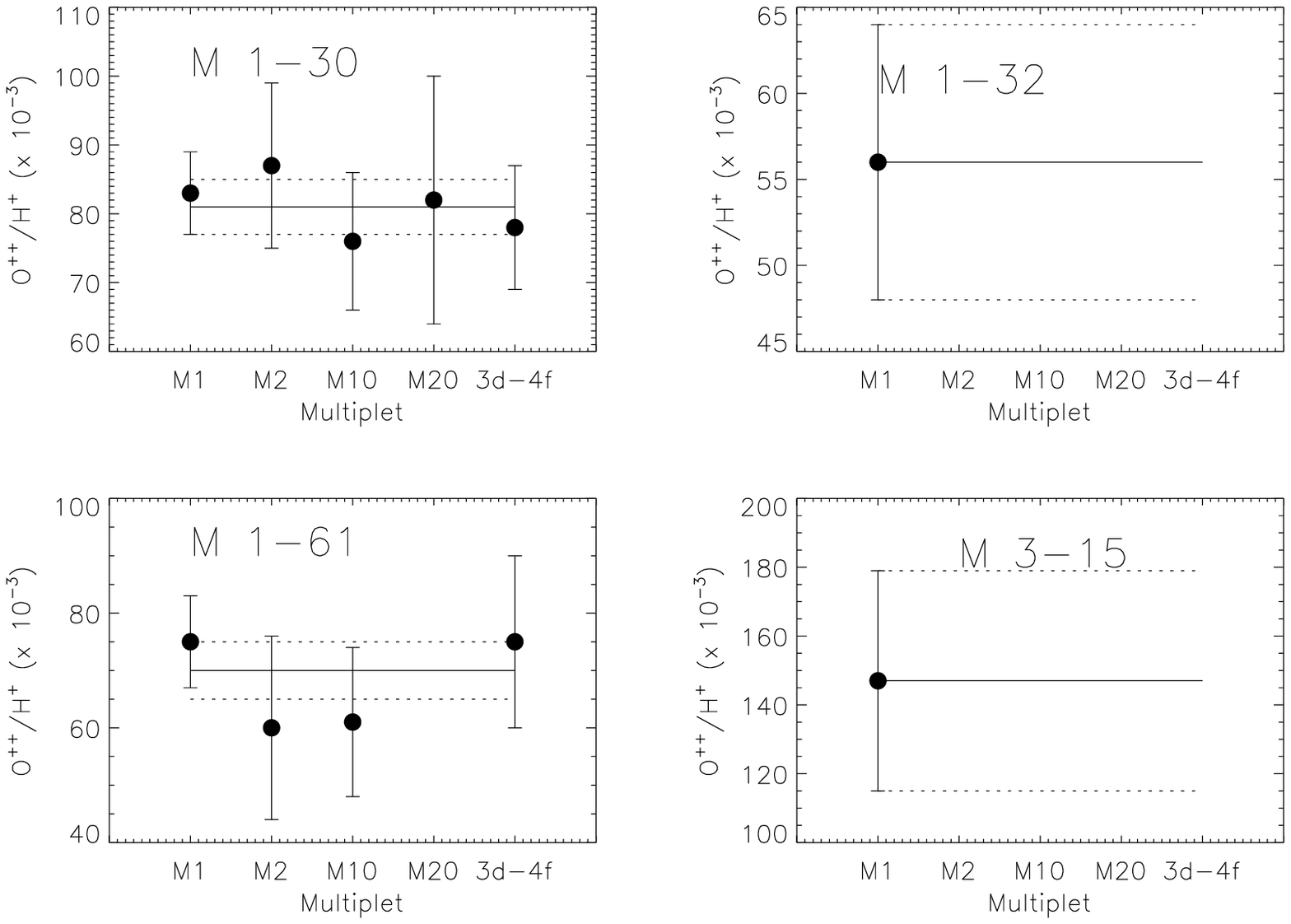}
\includegraphics[width=\columnwidth]{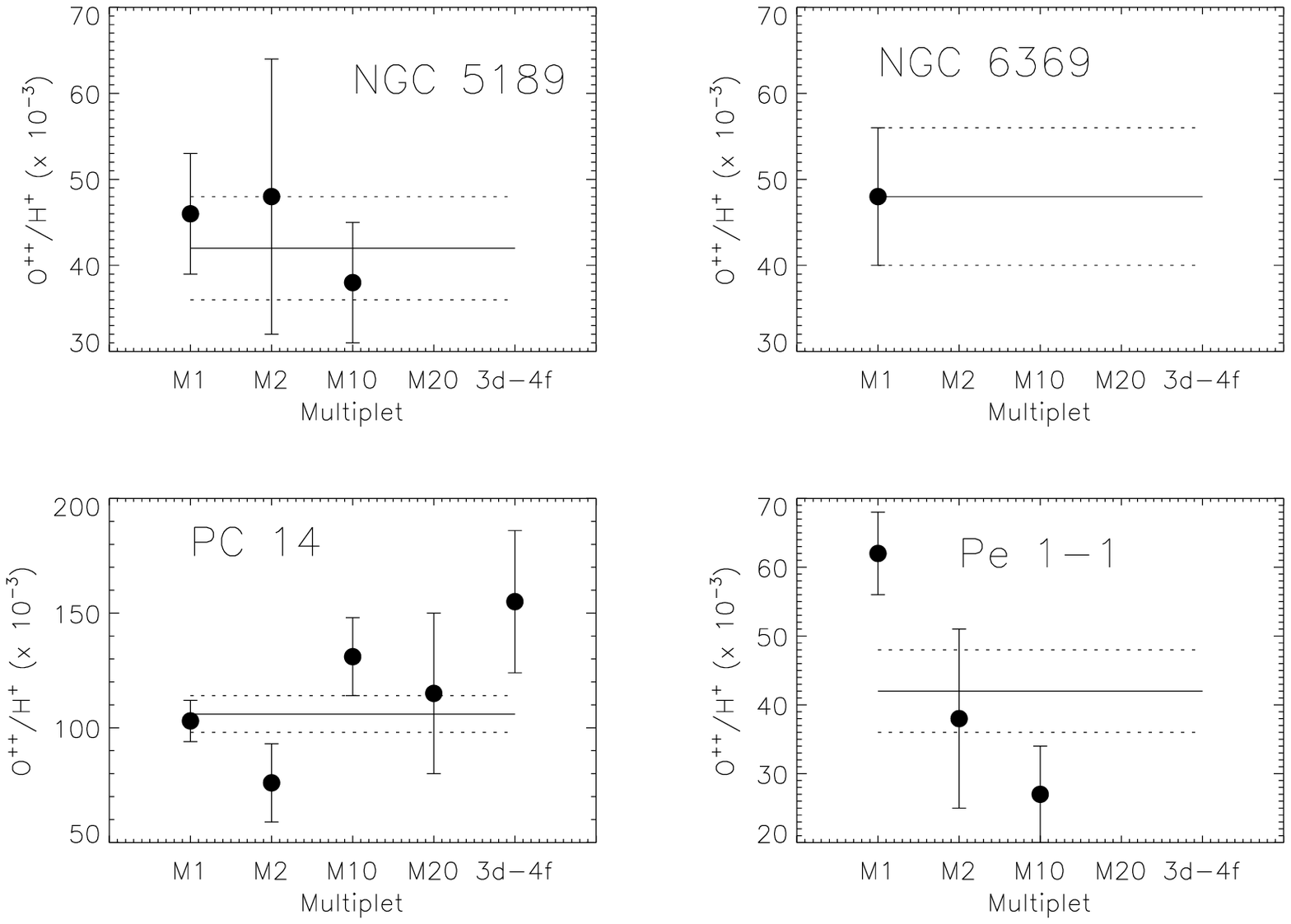}
\caption{O$^{++}$/H$^+$ ratio obtained from different ORL multiplets for PNe of the sample. In general, the values obtained from the different multiplets are consistent. An exception is Pe\,1-1 (see text). Solid lines represent the averaged value and dashed lines the 1$\sigma$ limits to the abundance.}
\label{OII_mult}
\end{center}
\end{figure}

In Fig.~\ref{OII_mult} we show the O$^{++}$ abundances derived for multiplets 1, 2, 10, and 20 ({\it sum} value) and for $3d-4f$ transitions (averaged value) for each object of our sample. In general, all values agree within the uncertainties for the best signal-to-noise spectra, where more {\oii} lines were detected and measured, that is, He\,2-86 and M\,1-30. For PC\,14, abundances from multiplets 1 and 20 agree, but are somewhat different from those of multiplets 2 (lower) and multiplet 10 and $3d-4f$ transitions (higher). However, the behaviour of multiplet 2 can be explained because the brightest line of the multiplet, that is, {\oii} $\lambda$4349.43, gives abundances systematically lower than the rest of the lines of the multiplet, when measured properly. This can be caused by departures of the LTE in the populations of the levels where the {\oii} multiplet 2 lines originate, in a similar way to what occurs for multiplet 1 \citep[see][]{tsamisetal03, ruizetal03, apeimbertpeimbert05}; unfortunately, the quality of the observations in the literature for the {\oii} multiplet 2 lines is lower than that of the {\oii} multiplet 1 ones, and an empirical recipe to correct for non-LTE effects could not be achieved. For the rest of objects there is an overall good agreement between the different multiplets within the uncertainties, except in the particular case of Pe\,1-1, where multiplets 1 and 10 give abundances that are more than 1$\sigma$ different from the average abundance. For this PNe we adopted the value given by multiplet 1 as representative of the O$^{++}$ abundance from ORLs, giving the abundance with the lowest relative uncertainties. Additional arguments favouring this value are presented in Sect.~\ref{tsquared}.

Very recently, \citet{escalanteetal12} have shown that resonance fluorescence is probably responsible for about 10-20 \% of the intensity of lines of multiplets 1, 2, and 20 of {\oii} in the low-ionization PN IC\,418, but this only affects a few (fewer than 5\%) lines of multiplet 10 and does not affect the lines of $3d-4f$ transitions at all. This result is striking because it calls into question the reliability of {\oii} lines as abundance indicators, mainly in {\hii} regions and in low-ionization PNe. In our sample, this effect is mitigated because of the high ionization degrees of most of the PNe.

We detected several lines of multiplets 2 and 5 of {\oiii} in the spectra of the most excited PNe of our sample. We did not consider these lines to compute abundances because they are probably excited by other mechanisms than recombination \citep[see][]{grandi76, garciarojasetal09}.

\setcounter{table}{5}
\begin{table*}
\begin{tiny}
\caption{Ionic abundance ratios from the {\oii} recombination lines$^{\rm a}$}
\label{recomb_oii}
\begin{tabular}{cccccccccccccc}
\noalign{\hrule} \noalign{\vskip3pt}
Mult.& $\lambda_0$ & \mc{12}{c}{O$^{++}$/H$^+$ ($\times$10$^{-5}$)} 	\\
\noalign{\vskip3pt} \noalign{\hrule} \noalign{\vskip3pt}
&  				&          Cn\,1-5 &        Hb\,4 &     He\,2-86 &      M\,1-25 &    M\,1-30  &      M\,1-32     &        M\,1-61   & M\,3-15      &  NGC\,5189   &    NGC\,6369 &    PC\,14   &       Pe\,1-1    \\\\
\noalign{\smallskip} \noalign{\hrule} \noalign{\smallskip}
1$^{\rm b}$& 4638.85		& 43:		   & 79		  & 102		 & 83		& 78	      & 125:		  & 67  	     & 139:	    & 32	   & 44 	  & 88  	& 78		   \\
& 4641.81			& ---		   & 185	  & 102		 & 71		& 83	      & 88:		  & 74  	     & 144	    & 52	   & 57 	  & 97  	& 56		   \\
& 4649.14			& 96		   & 189	  & 109		 & 70		& 88	      & 57		  & 77  	     & 136	    & 96	   & 56 	  & 124 	& 61		   \\
& 4650.84			& 34:		   & 184	  & 110		 & 77		& 80	      & 53		  & 77  	     & 183	    & 30	   & 33 	  & 87  	& 38:		   \\
& 4661.64			& 77		   & 110	  & 100		 & 113		& 77	      & 38:		  & 70  	     & 121:	    & 34	   & 42 	  & 88  	& 60		   \\
& 4673.73			& ---		   & ---	  & 105		 & ---		& 84:	      & ---		  & --- 	     & ---	    & ---	   & ---	  & --- 	& ---		   \\
& 4676.24			& 140		   & 124	  & 100		 & 80		& 89	      & 105:		  & 78  	     & ---	    & 59	   & 41:	  & 130 	& 70		   \\
& 4696.14			& ---		   & ---	  & 140		 & ---		& ---	      & ---		  & --- 	     & ---	    & ---	   & ---	  & --- 	& ---		   \\
& Sum				&{\bf 97$\pm$15}  &{\bf 158$\pm$14}&{\bf 105$\pm$6}&{\bf 78$\pm$10}& {\bf 83$\pm$6}&{\bf 56$\pm$8}&{\bf 75$\pm$8}    &{\bf 147$\pm$32}&{\bf 46$\pm$7} &{\bf 48$\pm$8}&{\bf 103$\pm$9}&{\bf 62$\pm$6} \\
2& 4317.14			& 71:  		   & 105:	  & 122		 & 100:		& 113  	      & 373:$^{\rm g}$     & 84: 	     & ---	    & 55:	   & ---	  & 88: 	& 43		   \\
& 4319.63			& 50:  		   & ---	  & 115		 & 60:		& 67  	      & 372:$^{\rm g}$	  & 63: 	     & ---	    & ---	   & ---	  & 72: 	& 32		   \\
& 4325.76			& ---		   & ---	  & 160		 & 320:		& 211:	      & ---		  & 202:	     & ---	    & ---	   & ---	  & 339:	& 338:  	   \\
& 4336.83			& ---  		   & ---  	  & 161		 & 107:		& 118  	      & ---		  & 103:	     & ---	    & ---	   & ---	  & 137:	& 73:		   \\	   
& 4345.56$^{\rm e}$		& 117:  	   & 58:	  & 136		 & 128:		& 161: 	      & 95:		  & 104:	     & ---	    & 34:	   & 28:	  & 91: 	& 79:		   \\
& 4349.43			& 52:  		   & 89		  & 68		 & 48:		& 58  	      & 57:		  & 49  	     & ---	    & 32	   & 16:	  & 68  	& 41:		   \\
& 4366.89			& 93:  		   & 206	  & 100		 & 92:		& 146  	      & ---		  & 88  	     & ---	    & 89	   & ---	  & 98  	& 95:		   \\
& Sum				& {\bf 63:}	   &{\bf 122$\pm$38}&{\bf 103$\pm$12} & {\bf 84:}& {\bf 87$\pm$12}&{\bf 67:}      &{\bf 60$\pm$16}   & {\bf ---}    &{\bf 48$\pm$16} &{\bf 19:}   &{\bf 76$\pm$17}&{\bf 38$\pm$13}  \\
10$^{\rm d}$& 4069.62		& 83      	   & 160	  & 110		 & 63		& 80          & --- 	      	  & 70  	     & 142:	    & 162$^{\rm c}$& ---	  & 165 	& 29		   \\
& 4069.89			& *            	   & *  	  & *		 & *		& *           & *	      	  & *		     & *	    & * 	   & *  	  & *		& *		   \\	     
& 4072.15			& 85		   & 115	  & 94		 & 63		& 71	      & ---		  & 57  	     & ---	    & 59	   & ---	  & 116 	& 26		   \\
& 4075.86			& 85		   & 115	  & 94		 & 63		& 71	      & ---		  & 57  	     & 110:	    & 22	   & 91:	  & 116 	& 26		   \\
& 4078.84			& 134:		   & 134	  & 134		 & ---		& 120	      & ---		  & 161:	     & ---	    & ---	   & ---	  & 151:	& ---		   \\
& 4085.11			& ---		   & ---	  & 83		 & ---		& 83	      & ---		  & 117:	     & ---	    & 88:	   & ---	  & 156:	& ---		   \\
& 4092.93			& ---	     	   & ---	  & 102		 & ---		& ---	      & ---	      	  & --- 	     & ---	    & ---	   & ---	  & 96: 	& ---		   \\
& Sum				& {\bf 84$\pm$19}  &{\bf 129$\pm$34}&{\bf 99$\pm$11}&{\bf 63$\pm$20}& {\bf 76$\pm$10}&{\bf ---}   &{\bf 61$\pm$13}   &{\bf 124:}    &{\bf 38$\pm$7} &{\bf 91:}    &{\bf 131$\pm$17}&{\bf 27$\pm$7}  \\ 
20$^{\rm d}$& 4097.22$^{\rm c}$	& ---		   & ---	  & ---		 & ---		& ---	      & ---		  & --- 	     & ---	    & ---	   & ---	  & --- 	& ---		   \\
& 4104.79			& ---	     	   & ---	  & 115		 & 82:		& 100	      & ---	      	  & 103:	     & ---	    & ---	   & ---	  & 54: 	& ---		   \\
& 4110.79$^{\rm f}$		& 165:	     	   & ---	  & 206:	 & 193:		& 318:	      & ---	      	  & 128:	     & ---	    & ---	   & ---	  & --- 	& ---		   \\
& 4119.22			& ---   	   & 148:	  & 132		 & 69:		& 65   	      & ---		  & 66: 	     & ---	    & 52:	   & ---	  & 115 	& ---		   \\
& Sum				& {\bf 165:} 	   & {\bf 148:}	  &{\bf 127$\pm$18}& {\bf 91:}	& {\bf 82$\pm$18}& {\bf ---}      & {\bf 87:}	     & {\bf ---}    & {\bf 52:}    & {\bf ---}    &{\bf 115$\pm$35}& {\bf ---}     \\ 
3d--4f$^{\rm d}$& 4083.90	& ---		   & ---          & 143          & ---          & ---         & ---               & --- 	     & ---	    & ---	   & ---	  & 146:	& ---		   \\
& 4087.15			& ---		   & ---          & 147          & ---          & 116         & ---               & --- 	     & ---	    & 113:	   & ---	  & 173:	& ---		   \\
& 4089.29			& 74: 	    	   & ---	  & 126		 & 56:		& 65	      & ---	      	  & 72  	     & ---	    & 40:	   & ---	  & 126:	& 63:		   \\	   
& 4095.64			& ---		   & ---          & 212          & ---          & ---         & ---               & --- 	     & ---	    & ---	   & ---	  & 167:	& ---		   \\
& 4097.26$^{\rm c}$		& ---		   & ---          & ---          & ---          & ---         & ---               & --- 	     & ---	    & ---	   & ---	  & --- 	& 218:  	   \\
& 4275.55			& 120:		   & 95:          & 171          & ---          & 82          & 184:              & 77  	     & ---	    & ---	   & ---	  & 77: 	& ---		   \\
& 4276.75			& 118:		   & 94:          & 95           & ---          & ---         & 182:              & 79  	     & ---	    & ---	   & ---	  & 93: 	& ---		   \\
& 4277.90			& ---		   & ---          & ---          & ---          & ---         & ---               & --- 	     & ---	    & ---	   & ---	  & 163:	& ---		   \\
& 4282.96			& ---		   & ---          & ---          & ---          & 59:         & ---               & --- 	     & ---	    & ---	   & ---	  & 109:	& ---		   \\
& 4285.69			& ---		   & ---          & 115          & ---          & ---         & ---               & 75: 	     & ---	    & ---	   & ---	  & 112:	& ---		   \\
& 4288.82			& ---		   & ---	  & 608$^{\rm f}$& ---		& ---	      &	---		  & --- 	     & ---	    & ---	   & ---	  & --- 	& ---		   \\
& 4291.25			& ---		   & ---	  & 95:	 	 & ---		& 69:	      &	---		  & --- 	     & ---	    & ---	   & ---	  & 97: 	& ---		   \\
& 4292.21			& ---		   & ---          & 156:         & ---          & 205:        & ---               & --- 	     & ---	    & ---	   & ---	  & --- 	& ---		   \\
& 4294.78$^{\rm c}$		& ---		   & ---          & 112          & ---          & ---         & ---               & --- 	     & ---	    & ---	   & ---	  & --- 	& ---		   \\
& 4303.61			& 108:	    	   & 79:	  & 122		 & 75:		& 87	      & ---	      	  & 62: 	     & ---	    & 82:	   & ---	  & 156 	& ---		   \\
& 4303.82			& 106:	    	   & 152:	  & 124		 & 70:		& 84	      & ---	      	  & 56: 	     & ---	    & 86:	   & ---	  & 150 	& ---		   \\
& 4491.23			& 230:		   & ---          & 114          & 131:         & 181:        & ---               & --- 	     & ---	    & 231:	   & ---	  & 259:	& 283:  	   \\
& 4602.13			& ---		   & ---          & 224          & ---          & ---         & ---               & --- 	     & ---	    & ---	   & ---	  & 130:	& ---		   \\
& 4609.44			& ---      	   & 172: 	  & 109		 & 58:		& 87   	      & ---	      	  & 56: 	     & ---	    & ---	   & ---	  & 159 	& ---		   \\
& 4613.68$^{\rm c}$		& ---		   & ---          & ---          & ---          & ---         & ---               & --- 	     & ---	    & ---	   & ---	  & --- 	& ---		   \\
& Average			& {\bf 126:}	   & {\bf 118:}	  &{\bf 130$\pm$40}& {\bf 78:}	& {\bf 78$\pm$9}& {\bf 183:}      &{\bf 75$\pm$15}   & {\bf ---}    & {\bf 110:}   & {\bf ---}    &{\bf 155$\pm$31} &{\bf }		   \\  
\noalign{\smallskip} \noalign{\hrule} \noalign{\smallskip}
& Adopted			&{\bf 92$\pm$12}   &{\bf 150$\pm$12}&{\bf 105$\pm$5}&{\bf 75$\pm$9}&{\bf 81$\pm$4} &{\bf 56$\pm$8}&{\bf 70$\pm$7}    &{\bf 147$\pm$32}&{\bf 42$\pm$6}&{\bf 48$\pm$8}&{\bf 106$\pm$8}  &{\bf 62$\pm$6}	   \\ 
\noalign{\smallskip} \noalign{\hrule} \noalign{\smallskip}
\end{tabular}
\begin{description}
\item[$^{\rm a}$] Only lines with intensity uncertainties lower than 40 \% have been considered (see text).
\item[$^{\rm b}$] Corrected for non-LTE effects (see text).
\item[$^{\rm c}$] Blended with another line or affected by internal reflections or charge transfer in the CCD.
\item[$^{\rm d}$] Recombination coefficients for intermediate coupling \citep{liuetal95}.
\item[$^{\rm e}$] Blended with the {\oii} 4345.55 line.
\item[$^{\rm f}$] Blended with an unknown line.
\item[$^{\rm g}$] Lines are at the very edge of one order and the overlapping between orders was unsuitable. 
\end{description}
\end{tiny}
\end{table*}

\subsubsection{Permitted lines of carbon}

Several permitted lines of {\cii} were measured in the spectra of our PNe. 
All detected multiplets are $3d-4f$ transitions and are, in principle, excited by pure 
recombination \citep[see][]{grandi76}. In two objects, M\,1-32 and NGC\,5189, the intensity of multiplet 17.04 {\cii} $\lambda$6461.95 line was not reliable because it was affected by charge transfers in the CCD. The C$^{++}$/H$^+$ ratios obtained are shown in Table~\ref{recomb_cii}.

We detected lines of multiplets 1, 16, and 18 of {\ciii} in several objects of our sample. The excitation mechanism of these multiplets is pure recombination \citep[see][and references therein]{garciarojasetal09}, therefore they are suitable for ionic abundance determinations. Results are shown in Table~\ref{recomb_ciii}.  

\setcounter{table}{6}
\begin{table*}
\begin{tiny}
\caption{Ionic abundance ratios from the {\cii} recombination lines$^{\rm a}$}
\label{recomb_cii}
\begin{tabular}{cccccccccccccc}
\noalign{\hrule} \noalign{\vskip3pt}
Mult.& $\lambda_0$ & \mc{12}{c}{C$^{++}$/H$^+$ ($\times$10$^{-5}$)} 	\\
\noalign{\vskip3pt} \noalign{\hrule} \noalign{\vskip3pt}
&  			&          Cn\,1-5 &        Hb\,4 	&     He\,2-86 	&      M\,1-25 &    M\,1-30   &      M\,1-32     &        M\,1-61   & M\,3-15      &  NGC\,5189   &    NGC\,6369 &    PC\,14   &       Pe\,1-1    \\
\noalign{\smallskip} \noalign{\hrule} \noalign{\smallskip}
6& 4267.15		& 130		   & 72	  		& 68		& 53		& 86	      & 192		 & 43		    & 50		 & 29		 & 77		 & 83	       & 107		  \\
17.06& 5342.38		& 174		   & 49		  	& 73		& 62		& 92	      & 276		 & 50		    & 82		 & ---  	 & ---  	 & 66	       & 104		  \\
16.04& 6151.43		& 130		   & 56		  	& 63		& 53		& 78	      & 157		 & 47		    & 85		 & 50		 & 101  	 & 72	       & 112		  \\
17.04& 6461.95		& 92		   & 74		  	& 59		& 55		& 90	      & 84$^{\rm b}$	 & 43		    & 69		 & 56$^{\rm b}$  & 87		 & 77	       & 98		  \\
\noalign{\smallskip} \noalign{\hrule} \noalign{\smallskip}
& Adopted		&{\bf 121$\pm$7}   &{\bf 68$\pm$6}   	&{\bf 65$\pm$3} &{\bf 56$\pm$3}	&{\bf 86$\pm$3}&{\bf 194$\pm$14} &{\bf 44$\pm$3}    &{\bf 69$\pm$7}	 &{\bf 30$\pm$3} &{\bf 81$\pm$6} &{\bf 81$\pm$5}&{\bf 104$\pm$6}  \\ 
\noalign{\smallskip} \noalign{\hrule} \noalign{\smallskip}
\end{tabular}
\begin{description}
\item[$^{\rm a}$] Only lines with intensity uncertainties lower than 40 \% were considered (see text).
\item[$^{\rm b}$] Affected by charge transfer in the CCD.
\end{description}
\end{tiny}
\end{table*}

\setcounter{table}{7}
\begin{table*}
\begin{tiny}
\caption{Ionic abundance ratios from the {\ciii} recombination lines$^{\rm a}$}
\label{recomb_ciii}
\begin{tabular}{cccccccccccccc}
\noalign{\hrule} \noalign{\vskip3pt}
Mult.& $\lambda_0$ & \mc{12}{c}{C$^{+3}$/H$^+$ ($\times$10$^{-5}$)} 	\\
\noalign{\vskip3pt} \noalign{\hrule} \noalign{\vskip3pt}
&  			&          Cn\,1-5 &        Hb\,4 	&     He\,2-86 	&      M\,1-25 &    M\,1-30   &      M\,1-32     &        M\,1-61   & M\,3-15      &  NGC\,5189   &    NGC\,6369 &    PC\,14   &       Pe\,1-1    \\
\noalign{\smallskip} \noalign{\hrule} \noalign{\smallskip}
1& 4647.42		& ---		   & 34	  		& 4			& ---		& ---	      & ---		 & 3:	  	    & ---		 & 6:	  	 & 17		 & 19	       & ---		 \\
& 4650.25		& ---		   & 34		  	& 4			& ---		& ---	      & ---		 & ---  	    & ---		 & ---  	 & 17		 & ---         & ---		 \\
& 4651.47		& ---		   & 31:	  	& ---		& ---		& ---	      & ---		 & ---  	    & ---		 & ---  	 & 17		 & ---         & ---		 \\
& Sum			& ---		   & 34$\pm$7	& 4$\pm$1	& ---		& ---	      & ---	     & 3:  	    	& ---		 & 6:	  	 & 17$\pm$4	 & 19$\pm$4    & ---		\\ 
16& 4067.94		& ---		   & ---	  	& ---		& ---		& ---	      & ---		 & ---  	    & ---		 & ---  	 & 32		 & ---         & ---		  \\
& 4068.91$^{\rm b}$	& ---		   & ---	  	& ---		& ---		& ---	      & ---		 & ---  	    & ---		 & 41:  	 & 32		 & ---         & ---		  \\
& 4070.26		& ---		   & ---	  	& ---		& ---		& ---	      & ---		 & ---  	    & ---		 & 24		 & 32		 & ---         & ---		  \\
& Sum			& ---		   & ---	  	& ---		& ---		& ---         & --- 		 & ---  	    & ---		 & 24$\pm$4	 & 32$\pm$5	 & ---         & ---		  \\  
18& 4186.90		& ---		   & 13:	  	& ---		& ---		& ---	      & ---		 & ---  	    & ---		 & 4:  	 & ---  	 & 13$\pm$5    & ---		  \\ 
\noalign{\smallskip} \noalign{\hrule} \noalign{\smallskip}
& Adopted		&{\bf ---} 	   &{\bf 34$\pm$7}   	&{\bf 4$\pm$1}    &{\bf ---}	&{\bf ---}    &{\bf ---}    	 &{\bf 3:}	    &{\bf ---}  	 &{\bf 24$\pm$4} &{\bf 23$\pm$3} &{\bf 17$\pm$3}&{\bf ---}	  \\ 
\noalign{\smallskip} \noalign{\hrule} \noalign{\smallskip}
\end{tabular}
\begin{description}
\item[$^{\rm a}$] Only lines with intensity uncertainties lower than 40 \% were considered (see text).
\item[$^{\rm b}$] Deblended from {\fsii} $\lambda$4068.60 by assuming the theoretical ratio {\ciii} $\lambda$4067.94/$\lambda$4068.91.
\end{description}
\end{tiny}
\end{table*}

\subsubsection{Permitted lines of neon}

We detected several permitted lines of {\neii} in our objects, belonging to multiplets 1, 39, and 55. Transitions from multiplet 1 and 55 are probably results of recombination because they correspond to quartets and their ground level has a doublet configuration \citep{estebanetal04}. However, the multiplet 1 {\neii} $\lambda$3694.22 line gives abundances much higher than those derived from the multiplet 55 lines; because the multiplet 55 lines correspond to $3d-4f$ transitions, whose upper levels can hardly be populated by fluorescense, and the multiplet 1 lines correspond to $3s-3p$ transitions, we rely more on the multiplet 55 values.  Multiplet 39 {\neii} $\lambda$3829.77 was detected only in M\,1-30; this line corresponds to an intercombination transition (3p$^2$P$^0$--3d$^4$D) and the abundance derived from it is much higher than that derived from multiplet 55, hence, it is probably excited by other mechanisms than recombination. To derive Ne$^{++}$ abundances, we used effective recombination coefficients from recent calculations by  Kisielius \& Storey (unpublished), assuming LS-coupling. 
We adopted the {\it sum} value derived from multiplet 55 as representative of the Ne$^{++}$ abundance. This is the first time that Ne$^{++}$/H$^+$ has been derived from recombination lines for all these PNe. 

\setcounter{table}{8}
\begin{table*}
\begin{tiny}
\caption{Ionic abundance ratios from the {\neii} recombination lines}
\label{recomb_neii}
\begin{tabular}{cccccccccccccc}
\noalign{\hrule} \noalign{\vskip3pt}
Mult.& $\lambda_0$ & \mc{12}{c}{Ne$^{++}$/H$^+$ ($\times$10$^{-5}$)} 	\\
\noalign{\vskip3pt} \noalign{\hrule} \noalign{\vskip3pt}
&  			&          Cn\,1-5 &        Hb\,4 	& He\,2-86 	& M\,1-25       &  M\,1-30    & M\,1-32          &  M\,1-61   & M\,3-15   &  NGC\,5189    &  NGC\,6369  &    PC\,14   &  Pe\,1-1   \\
\noalign{\smallskip} \noalign{\hrule} \noalign{\smallskip}
1& 3694.22		& 50 		   & ---  		& ---		& 53 		& ---	      & ---		 & 14:        & ---	  & 19            & --- 	& 54	      & ---	   \\
39& 3829.77		& ---		   & ---	  	& ---		& ---		& 46          & ---		 & ---        & ---	  & --- 	  & --- 	& ---	      & ---	   \\
55& 4391.94		& 17:		   & 33:   	  	& 16		& 25:		& 10	      & ---		 & 6:         & 54:	  & 12:  	  & --- 	& 16:	      & 7:	   \\
& 4409.30		& 22:		   & 17: 	  	& 27		& 34:		& 11:	      & ---		 & 13:        & 39:	  & 9:   	  & --- 	& 19:	      & ---	   \\
& Sum			& 19: 	   	   & 27:	 	& 20$\pm$9 & 28:   	& 10          & ---	         & 9:         & 49:	  & 11: 	  & --- 	& 17:	      & 7:	   \\ 
\noalign{\smallskip} \noalign{\hrule} \noalign{\smallskip}
& Adopted$^{\rm a}$	&{\bf 19:}         &{\bf 27:}   	&{\bf 20$\pm$9} &{\bf 28:}      &{\bf 10:}    &{\bf --- }        &{\bf 9:}   &{\bf 49:}  &{\bf 11:}       &{\bf ---}	&{\bf 17:}    &{\bf 7:}  \\ 
\noalign{\smallskip} \noalign{\hrule} \noalign{\smallskip}
\end{tabular}
\begin{description}
\item[$^{\rm a}$] Lines from multiplet 1 and 39 have not been considered (see text).
\end{description}
\end{tiny}
\end{table*}

\subsubsection{Permitted lines of nitrogen}

Many {\nii} permitted lines, belonging to different multiplets, were measured in our spectra.  
\citet{grandi76}, \citet{escalantemorisset05}, and, very recently, \citet{escalanteetal12} have discussed the formation mechanism of several permitted lines of {\nii} 
in the Orion nebula and the PN IC\,418, and concluded that recombination cannot account for the observed intensities of most of them, because resonance 
fluorescence by line and starlight excitation are the dominant mechanisms. Moreover, \citet{liuetal01} suggested that 
continuum fluorescence by starlight could be the excitation mechanism of several permitted lines of {\nii} in some PNe. This can be especially important for multiplet 3, for which the resonance fluorescence contributes $\sim$85\% and $\sim$75\% in the Orion nebula and IC\,418, respectively \citep{escalantemorisset05, escalanteetal12}∫. 
On the other hand, the upper term of the {\nii} $\lambda$4236.91, $\lambda$4237.05 and $\lambda$4241.78 lines of multiplet 48 
is 4$f$ $^3$$F$, and of the {\nii} $\lambda$4041.31 line of multiplet 39 is 4$f$ $G$[9/2] and cannot be populated by permitted resonance transitions, therefore this line is probably excited mainly by recombination. For the multiplet 48 lines, we co-added the intensities of these lines, 
along with the characteristic wavelength of the whole multiplet, 4239.40 \AA , to compute the abundance. The N$^{++}$/H$^+$ ratios are shown in Table~\ref{recomb_nii}.

\begin{figure}[!htb] 
\begin{center}
\includegraphics[width=\columnwidth]{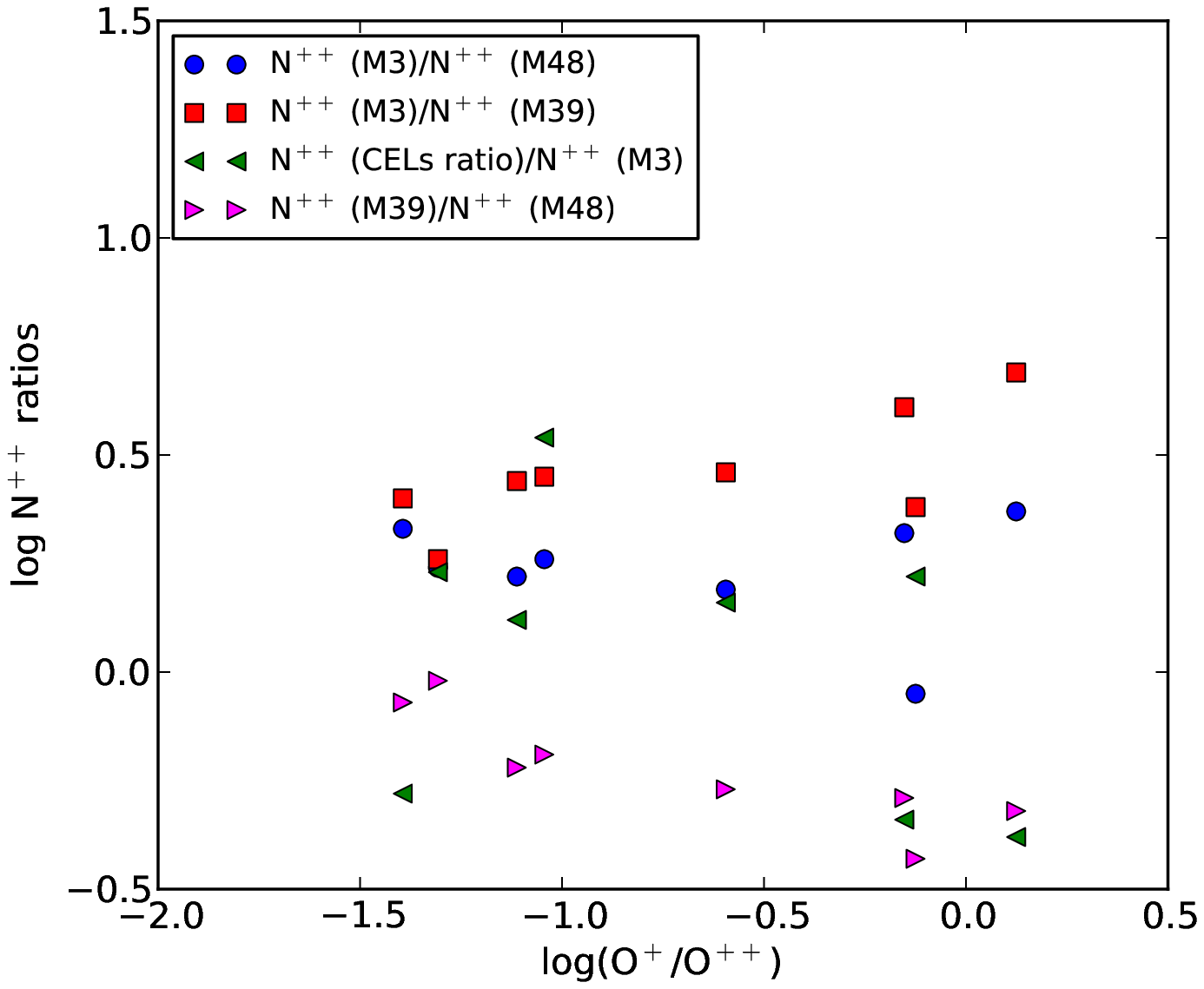}
\caption{N$^{++}$ ratios between different multiplets plotted against the ionization degree. The value N$^{++}$ (CELs ratio) corresponds to N$^{++}$ computed from the relation (N$^{++}$/H$^+$)$_{\rm ORLs}$=(N$^+$/O$^+$)$_{\rm CELs}$ $\times$ (O$^{++}$/H$^+$)$_{\rm ORLs}$. Evidently, the dispersion of the values obtained from different multiplets of N$^{++}$ is very large, but some trends seem to be present and are discussed in the text. }
\label{NII_ratios}
\end{center}
\end{figure}

The effect of fluorescense needs to be treated with caution because it depends on several factors: the fluorescense contribution is strongest in low-ionization PNe and, as the fluorescence contribution to a given line is emitted from the recombined ion, its level is related to the presence of this ion. In our case, the relative strength of the N$^+$ ions depends on the object, and we can only neglect the fluorescence in the emission of the {\nii} lines for objects with a high ionization degree; additionally, the strength of the effect also depends on the form of the spectral energy distribution (SED); in figure~1 of \citet{escalanteetal12} we can see that a detailed high-resolution model of the ionizing continuum and of the velocity field is needed to compute the exact contribution  of fluorescence excitation. The stellar temperature and the form of the SED are reflected in the ionization degree of the PN. Given the complexity of fluorescence, the biases of our sample and that it deserves a detailed treatment for each object, it is beyond the scope of this paper to analyse this in detail. However, taking into account the quality of our data, some discussion is of interest. In Fig.~\ref{NII_ratios} we present N$^{++}$ abundance ratios obtained from different multiplets and methods: i) from multiplet 3, ii) from multiplet 39, iii) from multiplet 48, and iv) assuming that the ratios N$^{++}$/N$^+$ and O$^{++}$/O$^+$ are similar and are also similar computed from CELs or ORLs (i.~e. N$^{++}$/H$^+$)$_{\rm ORLs}$=(N$^+$/O$^+$)$_{\rm CELs}$ $\times$ (O$^{++}$/H$^+$)$_{\rm ORLs}$. The first result is evident: there is a large scatter among the different values at a given ionization degree. Multiplet 3, M3, yields the highest values and multiplet 39, M39, the lowest ones. The tendencies of the different ratios also give us some information: apparently N$^{++}$ (M3)/N$^{++}$ (M39) increases with the ionization degree, while N$^{++}$ (M39)/N$^{++}$ (M48) decreases; this behaviour could be explained taking into account the relative strength of fluorescence in IC\,418, which increases from M39 (nothing), M48 ($\sim$20\%) to M3 ($\sim$75\%) and, as we have pointed out above, it is stronger at low-ionization degrees. However,  the ratio N$^{++}$ (M3)/N$^{++}$ (M39) should show a similar increasing tendency, but the dispersion of the values does not let us confirm this beyond doubt. Finally, the values obtained from the assumption that (N$^{++}$/H$^+$)$_{\rm ORLs}$=(N$^+$/O$^+$)$_{\rm CELs}$ $\times$ (O$^{++}$/H$^+$)$_{\rm ORLs}$ compared with abundances derived from multiplet 3 do not match any of the values reported here from permitted lines. Therefore, the N$^{++}$ abundances obtained from permitted lines should be treated carefully, especially when correcting for the recombination contribution to the auroral {\fnii} $\lambda$5755 line (Paper I).  

We did not consider the lines of multiplets 1 and 2 of {\niii} detected in the spectra of several objects because they appear to 
be excited by the Bowen mechanism and are not reliable for abundance determinations \citep{grandi76}.

\setcounter{table}{9}
\begin{table*}
\begin{tiny}
\caption{Ionic abundance ratios from the {\nii} recombination lines.}
\label{recomb_nii}
\begin{tabular}{cccccccccccccc}
\noalign{\hrule} \noalign{\vskip3pt}
Mult.& $\lambda_0$ & \mc{12}{c}{N$^{++}$/H$^+$ ($\times$10$^{-5}$)} 	\\
\noalign{\vskip3pt} \noalign{\hrule} \noalign{\vskip3pt}
		&  			&          Cn\,1-5 &        Hb\,4 	&     He\,2-86 	&      M\,1-25 	&    M\,1-30   	&      M\,1-32  &  M\,1-61 	& M\,3-15      	&  NGC\,5189	&    NGC\,6369 &    PC\,14   	&       Pe\,1-1    \\
\noalign{\smallskip} \noalign{\hrule} \noalign{\smallskip}
3$^{\rm a}$	& 5666.64		& 61		   & 30	  		& 60		& 65		& 107	      	& 73		& 21		& 31		& 21		& ---		& 15	      	& 20		 \\
  		& 5676.02		& 50		   & 28  		& 50		& 51		& 98	      	& 63		& 18		& ---		& 15		& ---		& ---	      	& ---		 \\
 		& 5679.56		& 52		   & 32	  		& 60		& 54		& 89	      	& 55		& 20		& 26		& 15		& 8.9		& 14	      	& 17		 \\
  		& 5686.21		& 65		   & 34  		& 50		& 58		& 102	      	& 96		& ---		& ---		& ---		& ---		& ---	      	& ---		 \\
  		& 5710.76		& 52		   & 31  		& 58		& 66		& 102	      	& 83		& 17		& ---		& ---		& ---		& ---	      	& ---		 \\
		& Sum			& 55$\pm$5	   & 31$\pm$4		& 58$\pm$4	& 58$\pm$4	& 97$\pm$6	& 66$\pm$9     	& 20$\pm$2	& 29$\pm$4	& 17$\pm$3	& 8.9$\pm$3.2  	& 15$\pm$3	& 18$\pm$2   	\\ 
39$^{\rm b}$	& 4041.31		& 19:		   & 11:	  	& 21$\pm$4	& 14:		& 20$\pm$4     	& ---		& 6.5:		& ---		& 7.1: 	 	& ---  	 	& 8.3:         	& ---		  \\
48$^{\rm c}$	& 4239.40$^{\rm c}$	& 41/30:	   & 20/14:	  	& 40/29$\pm$6	& 32/23	:	& 49/35$\pm$9   & ---		& ---		& ---		& 22/16: 	& ---  	 	& 10/7.4:	& ---		  \\
\noalign{\smallskip} \noalign{\hrule} \noalign{\smallskip}
		& Adopted		&{\bf 27:}   	   &{\bf 14:}   	&{\bf 28$\pm$4} 	&{\bf 21:}	&{\bf 31$\pm$4}	&{\bf $\le$66} &{\bf 6.5:}	&{\bf $\le$29}	& {\bf 14:} 	& {\bf $\le$8.9} & {\bf 8.5:}	& {\bf $\le$18}  	\\ 
\noalign{\smallskip} \noalign{\hrule} \noalign{\smallskip}
\end{tabular}
\begin{description}
\item[$^{\rm a}$] Recombination coefficients reported by \citet{kisieliusstorey02}
\item[$^{\rm b}$] Recombination coefficients reported by \citet{pequignotetal91}. 
\item[$^{\rm c}$] We considered the sum of all lines of the multiplet. Recombination coefficients reported by \citet{pequignotetal91}/\citet{escalantevictor92} 
\end{description}
\end{tiny}
\end{table*}

\subsubsection{Permitted lines of magnesium}

We detected the recombination {\mgii} $3d-4f$ $\lambda$4481.21 in four objects of our sample. However, in three of them the line is near the detection limit, therefore abundances derived from this line need to be treated with some caution. 
Following \citet{yliuetal04b}, we took advantage of the similarity of the {\cii} and {\mgii} atomic structure and assumed that the {\mgii} $3d-4f$ $\lambda$4481.21 line has an effective recombination coefficient equal to that of the {\cii} $3d-4f$ $\lambda$4267 line. Therefore, from the {\mgii} $\lambda$4481.21/{\hb} intensity ratio and the {\cii} line effective recombination coefficient we can derive Mg$^{++}$/H$^+$, which is shown for each PN in Table~\ref{recomb_mgii}.

\setcounter{table}{10}
\begin{table*}
\begin{tiny}
\caption{Ionic abundance ratios from the {\mgii} recombination lines$^{\rm a}$}
\label{recomb_mgii}
\begin{tabular}{cccccccccccccc}
\noalign{\hrule} \noalign{\vskip3pt}
Mult.& $\lambda_0$ & \mc{12}{c}{Mg$^{++}$/H$^+$ ($\times$10$^{-5}$)} 	\\
\noalign{\vskip3pt} \noalign{\hrule} \noalign{\vskip3pt}
&  			&          Cn\,1-5 &        Hb\,4 	&     He\,2-86 	&      M\,1-25 &    M\,1-30   &      M\,1-32     &        M\,1-61   & M\,3-15      &  NGC\,5189   &    NGC\,6369 &    PC\,14   &       Pe\,1-1    \\
\noalign{\smallskip} \noalign{\hrule} \noalign{\smallskip}
4& 4481.21		& ---		   & 5.02:	  	& 4.45		& 3.14:		& 2.66:	      & ---		 & ---		    & ---		 & ---		 & ---		 & ---	       & ---		  \\
\noalign{\smallskip} \noalign{\hrule} \noalign{\smallskip}
& Adopted		&{\bf ---}   	&{\bf 5.02:}   	&{\bf 4.45$\pm$0.85} &{\bf 3.14:}	&{\bf 2.66:}	&{\bf ---} &{\bf ---}    &{\bf ---}	 &{\bf ---} &{\bf ---} &{\bf ---}&{\bf ---}  \\ 
\noalign{\smallskip} \noalign{\hrule} \noalign{\smallskip}
\end{tabular}
\begin{description}
\item[$^{\rm a}$] Recombination coefficient equal to that of the {\cii} $3d-4f$ $\lambda$4267 line (see text).
\end{description}
\end{tiny}
\end{table*}

\section{Total abundances
\label{totalabundances}}

To correct for the unseen ionization stages and then derive
the total gaseous abundances of chemical elements in our
PNe, we should have adopted a set of ionization correction factors (ICF). For oxygen and nitrogen, we adopted
the classical ICF scheme reported by \citet{kingsburghbarlow94}, but for the other elements, special cases need to be discussed. 
In Table~\ref{total_ab} we show the total abundances obtained for our sample. 

\subsection{Helium}

When {\heii} lines were detected, the total helium is the sum of He$^+$ and He$^{++}$, otherwise we used the ICF(He) reported by \citet{peimbertetal92}. Their ICF is somewhat uncertain, but does not significantly affect the total He abundance, given the low amount of neutral He expected in our objects.

\subsection{Carbon}

For carbon, when {\cii} and {\ciii} lines were detected and {\heii} lines as well (Hb\,4, NGC\,5189, NGC\,6369 and PC\,14),  we applied the ICF given by  \citet{leisydennefeld96}. When {\cii} and {\ciii} lines were detected, but no {\heii} lines (He\,2-86 and M\,1-61) or when {\cii} lines only are detected, we applied the recipe given by \citet{kingsburghbarlow94}.  

\subsection{Neon}

When {\fneiii}, {\fneiv} and {\fnev} lines were observed, we assumed that the total Ne is the sum of all the observed ionic species. When  {\fneiii} lines only were observed, we applied the ICF scheme reported by \citet{kingsburghbarlow94}.

\subsection{Sulfur}

For sulfur, we used the recipe given by \citet{stasinska78} from photoionization models, which is

\begin{equation}
\frac{S}{H} =
            \left[ 1-\left( \frac{O^{+}}{O}\right)^{\alpha} \right]
            ^{\left( \frac{-1}{\alpha}\right)}  
            \times \left( \frac{S^{+} + S^{++}}{H^{+}}\right), 
\end{equation}
where $\alpha$=3. \citet{kwitterhenry01} developed a grid of detailed photoionization models including state-of-the-art atomic data as well as charge transfer and dielectronic recombination processes and found a more sophisticated ICF expression that, for the range of ionization degrees of our sample, gives total abundances  similar to those derived using \citet{stasinska78} expression. We finally adopted the expression by \citet{stasinska78} because it is widely used and can be more useful to achieve meaningful comparisons with the literature.    

\subsection{Chlorine}

The total abundance of Cl was considered as the sum of all ionic species: Cl$^+$, Cl$^{++}$ and Cl$^{+3}$ in objects with log(O$^+$/O$^{++}$) $>$-0.1 ; in these objects chlorine is not expected to be more highly ionized. However, for the most ionized objects of our sample we computed the total abundance of Cl by using the ICF from \citet{kwitterhenry01} to take into account the presence of Cl$^{+4}$.  Hence, the total chlorine was computed from
\begin{equation}
\rm  \frac{Cl}{H} = \Big( \frac{He^+ + He^{++}}{He^{+}} \Big) \times %
     \Big( \frac{Cl^{+}+Cl^{++}+Cl^{+3}}{H^{+}} \Big).
\end{equation}
This scheme was also applied when no Cl$^+$ was detected in the PN. As a general result, this ICF gives total abundances that are similar to that derived from the simple sum of all ionic species, except for the cases of the high ionized PNe Hb\,4 and NGC\,5189 for which the abundances increase by 0.09 and 0.16 dex, respectively. For the lowest-ionized objects (M\,1-30 and M\,1-32), where only  Cl$^+$ and Cl$^{++}$ are seen, we assumed that the total abundance of Cl is the sum of  these two species. For PB\,8, where the Cl$^{++}$ abundance only could be computed we assumed the ICF scheme proposed by \citet{girardetal07}, which is
\begin{equation}
\rm  \frac{Cl}{H} = \Big( \frac{He}{He^{+}} \Big)^2 \times %
     \frac{Cl^{++}}{H^{+}}.
\end{equation}

\subsection{Argon}

For Ar, we applied the recipe given by \citet{kingsburghbarlow94} when {\fariii} lines only were detected, or when {\fariii}, {\fariv} and {\farv} were detected. These authors do not give any recipe when {\fariii} and {\fariv} only are detected, which is the case for most of our objects. In such cases, we followed \citet{wangliu07}, who used the expression of \citet{kingsburghbarlow94} when {\fariii}, {\fariv} and {\farv} are detected, because they considered that the fraction of {\farv} is negligible which is, indeed, our case.
 
\subsection{Magnesium}

The total abundances of Mg were estimated from recombination lines. Following \citet{yliuetal04b}, we considered that most of the Mg is in the form of Mg$^{++}$ for the excitation classes of the PNe where the faint {\mgii} ORL was detected. Therefore, we assumed Mg/H=Mg$^{++}$/H$^+$. 

\subsection{Iron}

We measured lines of one, two, or three Fe ionization stages (Fe$^+$, Fe$^{++}$ and Fe$^{+3}$)  in our PNe. For He\,2-86, we can derive the total Fe abundance 
from the sum of the three ionization stages. For the other objects --and also for He\,2-86 for comparison-- we used an ICF to obtain the total Fe/H ratio. \citet{rodriguezrubin05} suggested a correction scheme for iron based on two ICFs, one from photoionization models (their equation 2) and one from an observational fit that takes into account all the uncertainties in the atomic data involved in the calculations and that allows us to constrain real iron abundances. Here, we only used the ICF from observational data given by

\begin{equation}
\label{eq2}
\rm  \frac{Fe}{H} = 1.1 \times \Big( \frac{O^+}{O^{++}} \Big)^{0.58} \times %
     \frac{Fe^{++}}{O^{+}} \times \frac{O}{H} , 
\end{equation}
which is valid in the range $-$1.35 $<$ log(O$^+$/O$^{++}$) $<$ $-$0.1. For log(O$^+$/O$^{++}$) $>$ $-$0.1 and, as Fe$^{++}$ and O$^+$ become the dominant ions, the ICF used is
\begin{equation}
\label{eq3}
\rm  \frac{Fe}{O} = \frac{Fe^+ + Fe^{++}}{O^{+}}.
\end{equation}

\begin{figure}[!htb] 
\begin{center}
\includegraphics[width=\columnwidth]{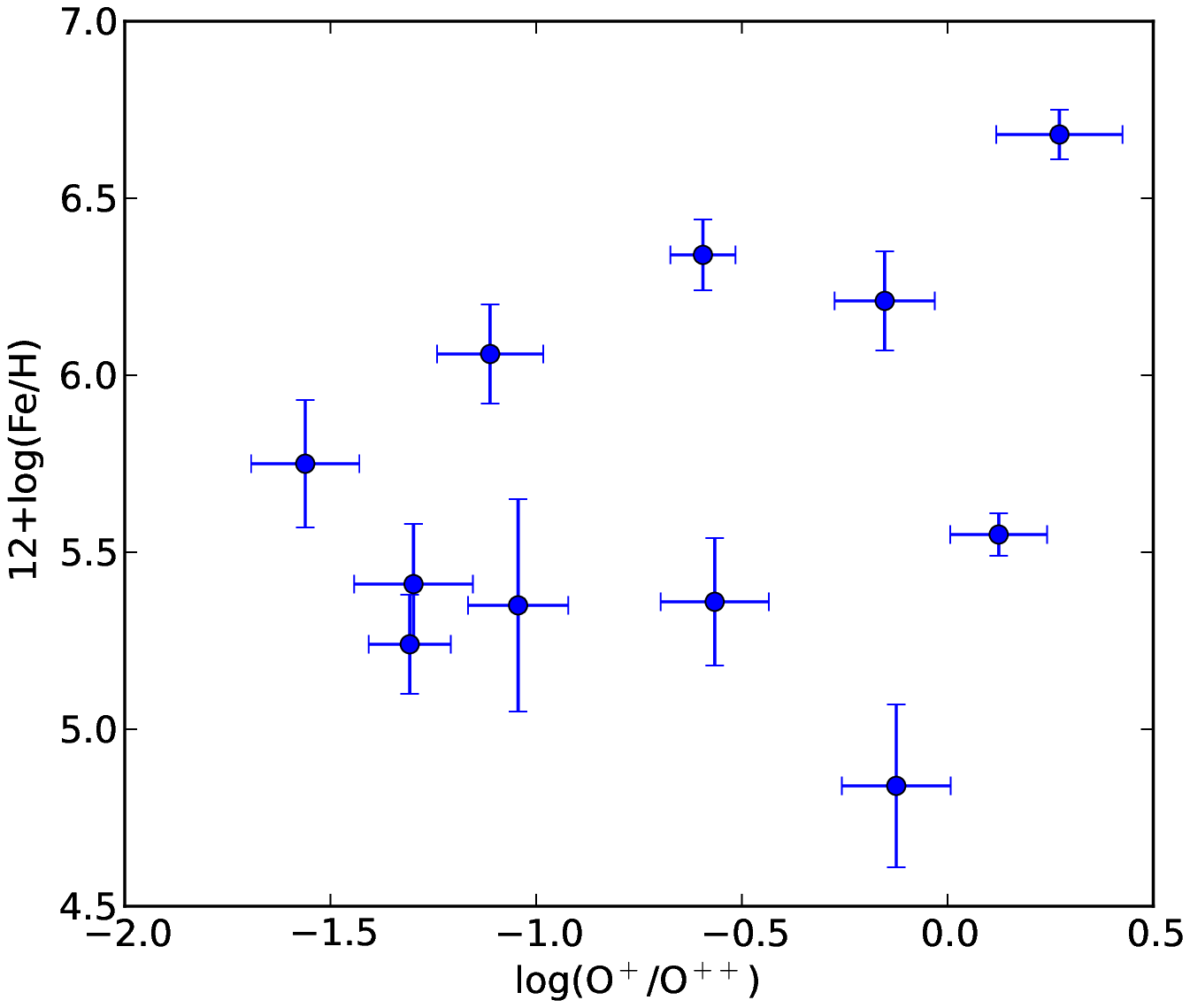}
\caption{ Values of Fe/H as a function of the ionization degree. }
\label{feo_ioniz}
\end{center}
\end{figure}

In Fig.~\ref{feo_ioniz} we present the Fe/H ratio plotted against the ionization degree, which shows no particular tendency. From this plot, we can see that the iron abundances, 12+log(Fe/H), derived for our sample range from 4.84$\pm$0.23 to 6.68$\pm$0.07, which are well below  the solar value \citep[(Fe/H)$_{\odot}$ = 7.46$\pm$0.08, ][]{lodders10} as a consequence of iron depletion into dust grains. We obtained that more than 85\% of the iron atoms are deposited into dust grains, in agreement with previous findings in PNe \citep{delgadoingladaetal09}. If we had used the ICF from photoionization models from \citet{rodriguezrubin05}, the derived iron abundances would be up to 0.69 dex higher, but our conclusions about iron depletions would be similar, with more than 80\% of the iron atoms in the nebulae condensed onto dust grains.

\subsection{Nickel}

There is no ICF available in the literature to correct for the presence of Ni$^{+3}$ to calculate the total Ni abundance for the four PNe where we detected 
Ni emission lines. 
However, we followed the procedure developed by \citet{mesadelgadoetal09} to compute the total Ni abundance in the Orion nebula, who applied a first-order ICF 
scheme based on the similarity between the ionization potentials of Ni$^{+3}$ (35.17 eV) and O$^{++}$ (35.12 eV): 
\begin{equation}
   {\rm \frac{Ni^{+3}}{Ni} = \frac{O^{++}}{O}}.
\end{equation}
Therefore,  they used the following expression to compute the total Ni abundance 
\begin{equation}
  {\rm  \frac{Ni}{H} = \frac{O}{O^{+}} \times
                         \Big(\frac{Ni^{+}}{H^{+}} +\frac{Ni^{++}}{H^{+}} \Big)}.
\end{equation}                   
Because we did not compute Ni$^+$ abundances, we estimated the total Ni abundance by applying this expression only to the Ni$^{++}$ abundance. This is a good approximation when log(O$^+$/O$^{++}$) $<$-0.1, which is the case for Cn\,1-5 and He\,2-86, but for M\,1-30 and M\,1-32 this expression provides only a lower limit to the real  Ni abundance.

\setcounter{table}{11}
\begin{table*}
\begin{tiny}
\begin{center}
\caption{Total abundances.}
\label{total_ab}
\begin{tabular}{lccccccc}
\noalign{\smallskip} \noalign{\smallskip} \noalign{\hrule} \noalign{\smallskip}
 & \multicolumn{7}{c}{12 + log(X/H)} \\
\noalign{\smallskip} \noalign{\hrule} \noalign{\smallskip}
             Element &         Cn1-5 &           Hb4 &        He2-86 &         M1-25 &         M1-30 &         M1-32 &         M1-61\\
\noalign{\smallskip} \noalign{\hrule} \noalign{\smallskip}
                 He  &11.21$\pm$0.01 &11.06$\pm$0.02 &11.09$\pm$0.01 &11.09$\pm$0.01 &11.12$\pm$0.01 &11.10$\pm$0.01 &11.05$\pm$0.01\\
        C$^{\rm a}$  & 9.18$\pm$0.05 & 9.07$\pm$0.06 & 8.87$\pm$0.06 & 8.96$\pm$0.07 & 9.30$\pm$0.07 & 9.75$\pm$0.09 & 8.68$\pm$0.06\\
                  N  & 8.72$\pm$0.05 & 8.55$\pm$0.09 & 8.62$\pm$0.09 & 8.40$\pm$0.06 & 8.59$\pm$0.05 & 8.44$\pm$0.06 & 8.28$\pm$0.11\\
                  O  & 8.79$\pm$0.03 & 8.70$\pm$0.04 & 8.79$\pm$0.04 & 8.87$\pm$0.05 & 8.90$\pm$0.06 & 8.74$\pm$0.10 & 8.67$\pm$0.04\\
        O$^{\rm b}$  & 9.07$\pm$0.04 & 9.27$\pm$0.04 & 9.08$\pm$0.02 & 9.05$\pm$0.05 & 9.23$\pm$0.05 & 9.11$\pm$0.08 & 8.89$\pm$0.04\\
       Ne$^{\rm c}$   & 8.37$\pm$0.04 & 8.17$\pm$0.06/7.73$\pm$0.12 & 8.27$\pm$0.05/8.46$\pm$0.12 & 7.71$\pm$0.11 & 8.04$\pm$0.08 & 7.69$\pm$0.12 & 8.08$\pm$0.07/8.34$\pm$0.19\\
                  S  & 7.20$\pm$0.04 & 7.01$\pm$0.09 & 7.23$\pm$0.08 & 7.22$\pm$0.06 & 7.25$\pm$0.07 & 7.17$\pm$0.09 & 7.11$\pm$0.09\\
                 Cl  & 5.53$\pm$0.03 & 5.37$\pm$0.04/5.31$\pm$0.04 & 5.43$\pm$0.04 & 5.50$\pm$0.04 & 5.63$\pm$0.04 & 5.47$\pm$0.06 & 5.17$\pm$0.05\\
        Ar$^{\rm c}$  & 6.73$\pm$0.04 & 6.55$\pm$0.05/6.45$\pm$0.06 & 6.61$\pm$0.12/6.63$\pm$0.11 & 6.92$\pm$0.05 & 7.11$\pm$0.08 & 6.99$\pm$0.11 & 6.45$\pm$0.05/6.49$\pm$0.05\\
	   Mg$^{\rm d}$  &    \nodata    &    7.70:      & 7.65$\pm$0.08 &    7.50:      &     7.42:     &    \nodata    &    \nodata    \\
       Fe$^{\rm e}$  & 6.05$\pm$0.05 &    \nodata    & 5.96$\pm$0.08 & 5.94$\pm$0.06 & 5.55$\pm$0.06 & 6.68$\pm$0.07 & 4.88$\pm$0.10\\
       Fe$^{\rm f}$  & 6.34$\pm$0.10 & 5.35$\pm$0.30 & 6.06$\pm$0.14 & 6.21$\pm$0.14 & 5.55$\pm$0.06 & 6.68$\pm$0.07 & 5.41$\pm$0.17\\
       Ni$^{\rm g}$  & 5.37$\pm$0.22 &    \nodata    & 5.56$\pm$0.38 &    \nodata    & 4.52$\pm$0.18 & 5.72$\pm$0.19 &    \nodata   \\

\noalign{\smallskip} \noalign{\hrule} \noalign{\smallskip}
             Element &         M3-15 &       NGC5189 &       NGC6369 &          PC14 &         Pe1-1 &           PB8 &       NGC2867\\
\noalign{\smallskip} \noalign{\hrule} \noalign{\smallskip}
                 He  &11.03$\pm$0.02 &11.09$\pm$0.01 &11.02$\pm$0.01 &11.03$\pm$0.01 &11.02$\pm$0.01 &11.09$\pm$0.01 &11.06$\pm$0.01\\
        C$^{\rm a}$  & 8.85$\pm$0.13 & 9.02$\pm$0.10 & 9.00$\pm$0.05 & 9.02$\pm$0.04 & 9.13$\pm$0.06 & 8.85$\pm$0.03 & 9.25$\pm$0.03\\
                  N  & 8.33$\pm$0.10 & 8.60$\pm$0.07 & 8.17$\pm$0.06 & 8.15$\pm$0.07 & 8.06$\pm$0.08 & 8.20$\pm$0.08 & 8.09$\pm$0.06\\
                  O  & 8.81$\pm$0.05 & 8.77$\pm$0.06 & 8.53$\pm$0.04 & 8.77$\pm$0.04 & 8.67$\pm$0.04 & 8.77$\pm$0.05 & 8.58$\pm$0.04\\
        O$^{\rm b}$  & 9.18$\pm$0.08 & 8.92$\pm$0.07 & 8.69$\pm$0.06 & 9.06$\pm$0.04 & 8.88$\pm$0.05 & 9.11$\pm$0.02 & 8.78$\pm$0.03\\
       Ne$^{\rm c}$  & 8.03$\pm$0.13 & 8.27$\pm$0.13/7.57$\pm$0.13 & 7.90$\pm$0.04/7.54$\pm$0.16 & 8.20$\pm$0.04/7.70$\pm$0.21 & 8.03$\pm$0.06 & 8.13$\pm$0.06 & 8.02$\pm$0.10\\
                  S  & 7.21$\pm$0.10 & 7.11$\pm$0.05 & 6.92$\pm$0.05 & 7.09$\pm$0.06 & 6.81$\pm$0.06 & 7.30$\pm$0.09 & 6.68$\pm$0.05\\
                 Cl  & 5.30$\pm$0.05 & 5.57$\pm$0.03/5.54$\pm$0.03 & 5.05$\pm$0.04/5.02$\pm$0.04 & 5.27$\pm$0.03/5.22$\pm$0.04 & 5.14$\pm$0.05 & 5.29$\pm$0.08 & 5.10$\pm$0.04\\
       Ar$^{\rm c}$ & 6.49$\pm$0.06 & 6.71$\pm$0.06/6.65$\pm$0.06 & 6.26$\pm$0.04/6.22$\pm$0.04 & 6.39$\pm$0.04/6.33$\pm$0.04 & 6.43$\pm$0.05 & 6.61$\pm$0.05 & 6.17$\pm$0.04\\
	   Mg$^{\rm d}$  &   \nodata     &    \nodata    &    \nodata    &   \nodata     &   \nodata     &    \nodata    &    \nodata    \\
       Fe$^{\rm e}$  &   \nodata     &    \nodata    &    \nodata    & 4.69$\pm$0.10 & 5.17$\pm$0.11 &    \nodata    &    \nodata   \\
       Fe$^{\rm f}$  & 5.75$\pm$0.18 & 4.84$\pm$0.23 &    \nodata    & 5.24$\pm$0.14 & 5.36$\pm$0.18 &    \nodata    &    \nodata   \\
       Ni$^{\rm g}$  &   \nodata     &    \nodata    &    \nodata    &    \nodata    &    \nodata    &    \nodata    &    \nodata   \\
\noalign{\smallskip} \noalign{\hrule} \noalign{\smallskip}
\end{tabular}
\end{center}
\begin{description}
\item[$^{\rm a}$] From recombination lines (see text).
\item[$^{\rm b}$] From recombination lines (see text).
\item[$^{\rm c}$] Considering low to medium ionization or high-ionization physical conditions when different (see text).
\item[$^{\rm d}$] From recombination lines (see text).
\item[$^{\rm e}$] Fe/H=Fe$^+$/H$^+$+Fe$^{++}$/H$^+$, except for He\,2-86, where Fe/H=Fe$^+$/H$^+$+Fe$^{++}$/H$^+$+Fe$^{+3}$/H$^+$.
\item[$^{\rm f}$] ICF from an observational fit reported by \citet{rodriguezrubin05}. Valid for -0.1 $<$ log(O$^{+}$/O$^{++}$) $<$ -1.35.
\item[$^{\rm g}$] Empirical ICF, following \citet{mesadelgadoetal09} 
\end{description}
\end{tiny}
\end{table*}

\section{Discussion of the elemental abundances.
\label{discuss_ab}}
 
It is a well-known fact predicted by stellar evolution models that PN central stars synthesize He in their nuclei, which is pulled-up to the surface in some of the dredge-up events, thus contaminating the nebular material. The enrichment also occurs for nitrogen, although in this case, N is not a primary element but it is enriched via the CNO-cycle, which is much more effective in most massive progenitors. Another element that is modified in the stellar surface as a consequence of nucleosynthesis processes and dredge-up events is carbon, while at solar metallicity, oxygen, neon, and other elements are not predicted to be significantly modified \citep[see, e.g.,][and references therein]{marigoetal03, karakasetal06}. Thus elemental abundance ratios such as N/O and C/O  reflect the occurrence of different nucleosynthesis processes in AGB stars. 

In this section the total abundances obtained  from ORLs or CELs are discussed. We assumed that the abundance ratios (C/O, N/O, Ne/O, Ar/O,  S/O, etc.) are almost independent of the procedure used to calculate abundances, provided that the same types of lines (ORLs or CELs) are used for each ratio, therefore these  ratios can be compared with results from other PNe samples and with predictions of stellar evolution models. In the following, the abundance ratios are expressed in the usual logarithmic form.

A word of caution: because our PN sample is highly biased in the sense that our objects were selected among WRPNe that showed a considerable fraction of O in the form of O$^{++}$ to find measurable {\oii} and {\cii} ORLs, the comparison with other more general samples needs to be made with care.

Aditionally, for two objects (namely M\,1-61 and NGC\,6369) the N$^+$ and O$^{++}$ temperatures show discrepancies that cannot be reproduced by simple photoionization models (Morisset, in preparation). The problem may be solved by using models that combine different regions that differ in their densities, in which case it is obvious that the direct method used for determining the temperature (especially that of N$^+$) may be inadequate.
These two objects are of the low-ionization class, thus the effect of a misestimated {\te}(N$^+$) does not affect the total O/H abundance, but it does have consequences on N/H and S/H, which both use O$^+$ in their ICFs. To evaluate the effect, we determined abundances for the two objects using {\te}(O$^{++}$) for the low-ionization zone. The N/H (S/H) abundances were then reduced by 0.23 (0.13) and 0.13 (0.10) dex for M\,1-61 and NGC\,6369, respectively. These variations do not affect the general results presented here.

\subsection{N and He abundances
\label{NandHe}}

\citet{peimbert90} classified PNe into four groups according to their metallicity and kinematics in the Galaxy. In this classification, type I PNe present He- and N-rich envelopes (He/H$>$0.125 and N/O$>$0.5). This classification was revised by \citet{kingsburghbarlow94}, who classified as type I PNe whose central star suffered hot-bottom burning (HBB) and showed N/O abundance ratio higher than 0.8. In the following, we use Peimbert's classification for our objects to be consistent with previous papers of our group \citep{penaetal13}.
For completeness we mention that Peimbert's type II and III PNe are nebulae that are not particularly H- or N-rich and have heliocentric radial  velocities lower and higher than 60 km s$^{-1}$, respectively. Type IV PNe are  objects from the Galactic halo.

\begin{figure}[!htb] 
\begin{center}
\includegraphics[width=\columnwidth]{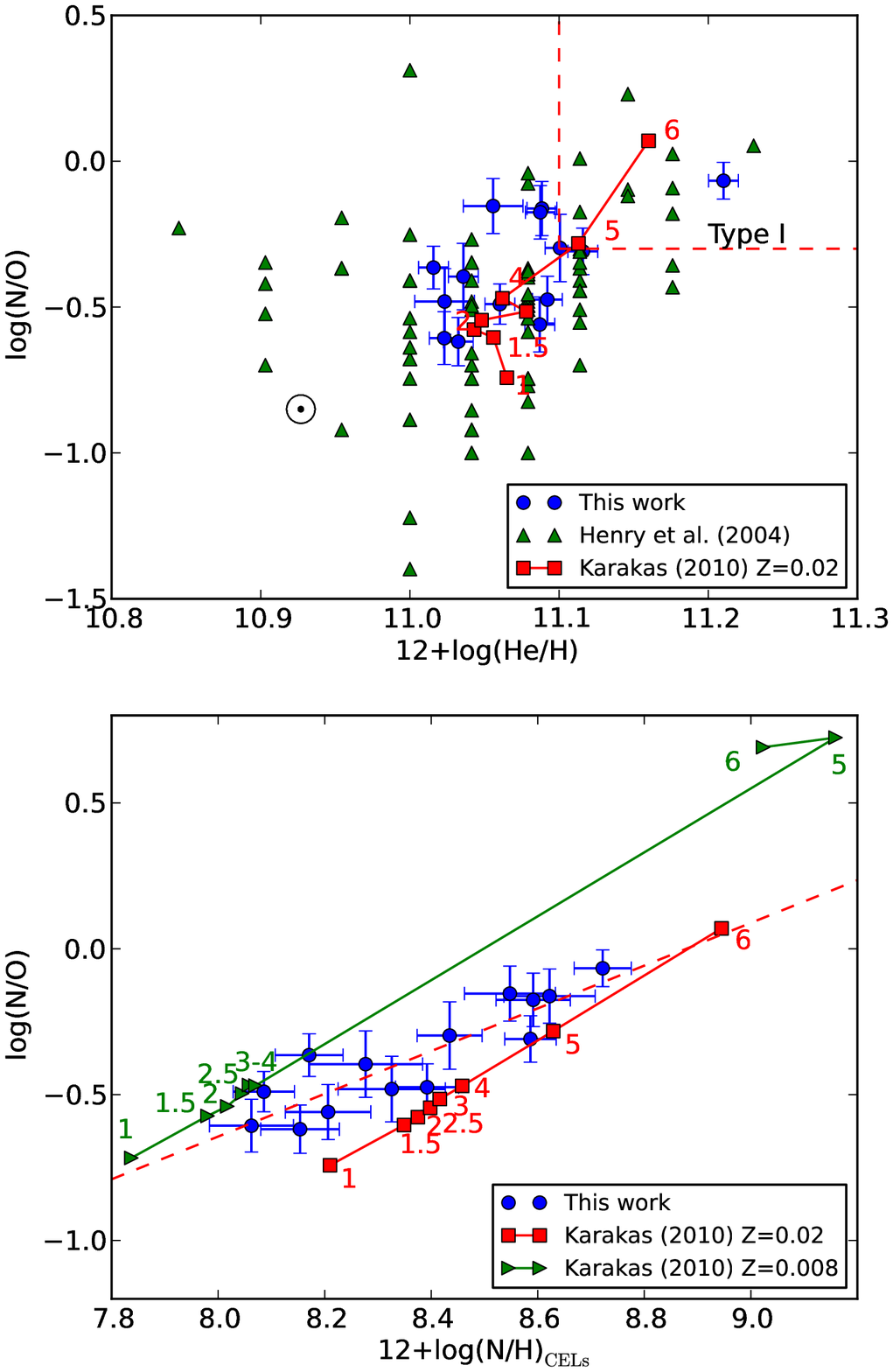}
\caption{Upper panel: N/O abundance ratio plotted against He abundance for our PN sample. The top-right box marks the criteria  given by \citet{peimbert90} to define  Type I PNe. In green we included the data of the Galactic PNe sample by \citet{henryetal04}. Predictions of stellar evolution models by \citet{karakas10} for  Z=0.02 and different initial masses (indicated in units of {\msun}) are also included.  The solar symbol represents the solar values. Lower panel: Relationship between N/O abundance ratio and the total N abundance; the fit to our data is shown as a dashed line (see text). The predictions of stellar evolution models by \citet{karakas10} for stars with different initial masses and for two different metallicities are included.}
\label{no_he_nh}
\end{center}
\end{figure}

The upper panel of Fig.~\ref{no_he_nh} shows the N/O $vs.$ He/H  distribution for our sample (N and O derived from CELs). With green triangles we 
show data given for a sample of 85 Galactic PNe  by Henry et al. (2004) for comparison (although care must be taken when comparing abundances obtained with different methods, different atomic parameters, etc.). Only one of our objects (Cn\,1-5) is a clear type I PN and another two (M\,1-30 and M\,1-32) are within the limits to be considered type I PNe. Three other objects (Hb\,4, He\,2-86 and NGC\,5189) show a higher N/H than M\,1-30 and M\,1-32, but their low He/H  values exclude them from the Peimbert's type I classification. The number of PNe in our sample is small, therefore it is difficult to assess whether our sample shows a higher or similar proportion of type I PNe as the Henry et al. sample, but our three possible type I PNe (according to Peimbert's criterion) represent about 20\%  against 11\% (9 of 85) in the Henry et al. sample. Interestingly, in the Henry et al. sample log (N/O) values cover from -1.4 to 0.4 (a similar range is found for the sample of  \citet{kingsburghbarlow94}, see their figure 4), while our sample shows a much more constrained log (N/O) range, from -0.65 to 0.0, that is, in our sample (certainly biased towards a particular type of PNe, as explained above)  all objects seem richer in N (although not as rich as to be considered type I)  than those of the Henry et al.  sample. In fact, the N/O average value of our sample  (excluding the three WRPNe marked as Type I) is $<$N/O$>$ = 0.42$\pm$0.18, which is significantly higher than the average value of  0.31$\pm$0.14 found for non-Type I PNe  by \citet{henryetal04}. This indicates that our non-Type I WRPNe are on average richer in N than PNe in more general samples, which may be due to a slightly  higher initial mass of the progenitors (see discussion below).
This is consistent with the recent result by \citet{penaetal13}, showing that Galactic WRPNe are located in a disk thinner than that of average PNe, which means that they are younger (and probably more massive) objects. 

For a comparison, in the upper panel of Fig.~\ref{no_he_nh} we have included  the predictions of stellar evolution models by \citet{karakas10} for a metallicity of Z=0.02. It can be seen that the progenitors of Type I PNe  had initial masses larger than 5 M$_{\odot}$, but also that all the PNe with log(N/O)\,$>-0.47$ (which are about half of our sample) are compatible with initial masses of the progenitors larger than 4 M$_{\odot}$.

\begin{figure}[!htb] 
\begin{center}
\includegraphics[width=\columnwidth]{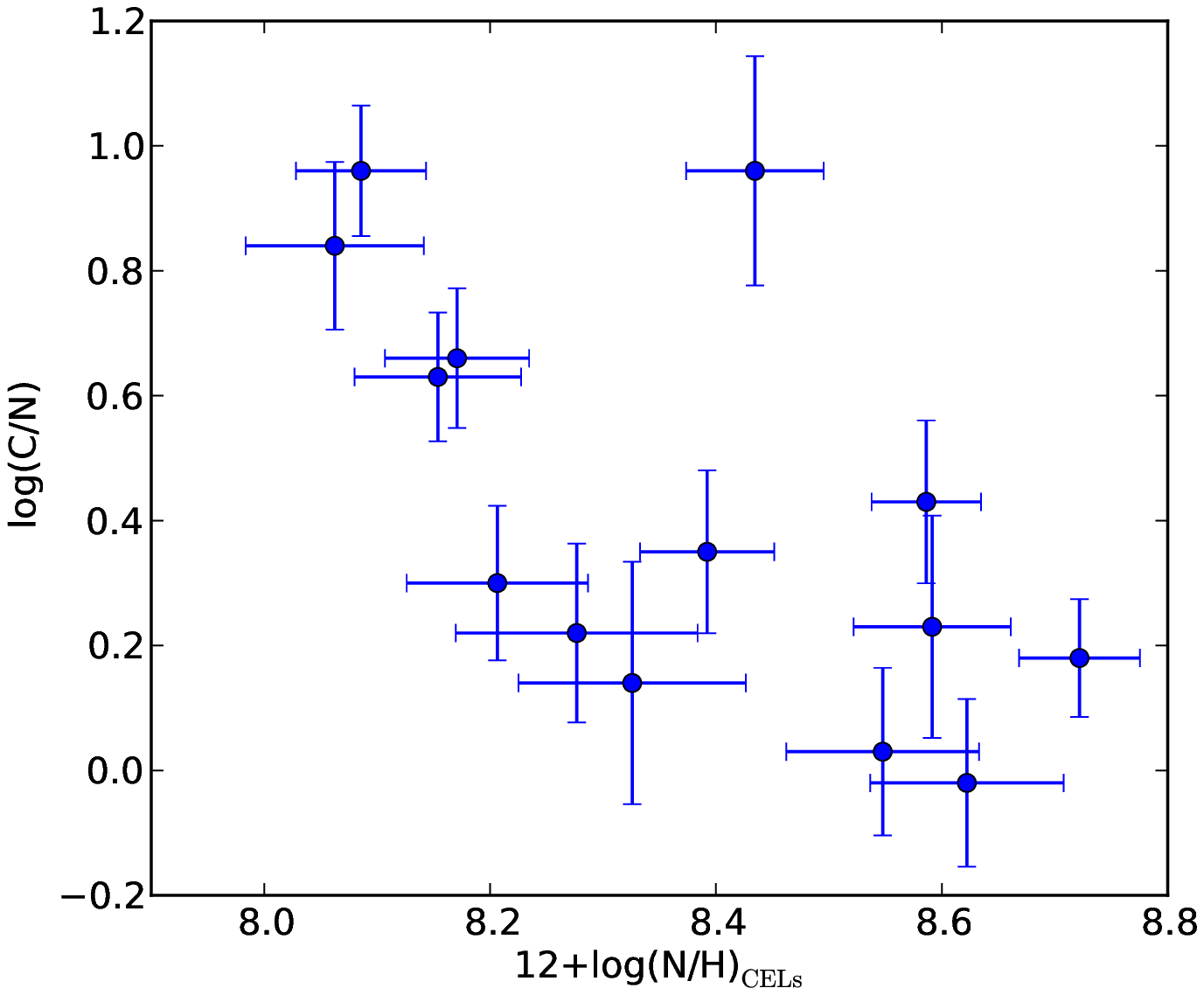}
\caption{C/N abundance ratio plotted against the total N abundance for our sample of PNe. For the C/N ratio, C was computed from ORLs and N from CELs and assuming a {\tsq} $>$ 0 (see text). The total N abundances shown in the x-axis are computed from CELs and {\tsq}=0.}
\label{nc_nh}
\end{center}
\end{figure}

A relation between N/O $vs.$ N/H is presented in the bottom panel of Fig.~\ref{no_he_nh}. As in other samples of PNe, a correlation is found between these quantities \citep[see e.g.,][]{henry90}. A linear fit to our data gives log(N/O) = 0.73$\times$[12+log(N/H)]$-$6.50, with a Spearman correlation 
factor r=0.86. This correlation indicates that N-enrichment in PNe occurs independently of the O abundance, therefore it probably is mainly a consequence of the CN-cycle (N increasing at the expense of C) and not of the ON-cycle. To verify this, in Fig.~\ref{nc_nh} we show the behaviour of the C/N ratio $vs.$ N abundance. In the C/N ratio, C was calculated from ORLs and N from CELs and assuming {\tsq}$>$ 0 (see Sect.~\ref{tsquared}), which would correspond to abundances derived from ORLs.  A clear dependence exists, showing that C/N decreases when N abundance increases up to a value  12+log (N/H) $\sim$ 8.4. In this range, evidently N is increasing at the expense of C through the CN-cycle (initial stellar masses lower than 3 M$_\odot$). For N higher than this value, the tendency is less clear, N continues to increase but the C/N ratio stops to decrease, indicating that C is being enriched through the third dredge-up process. A very interesting object (M\,1-32) shows a huge C enrichment, which deserves a more detailed analysis.

In Fig.~\ref{no_he_nh} (bottom) we have included the prediction of stellar evolution  models for stars with initial masses between 1 and 6 M$_\odot$ and metallicities Z=0.02 and Z=0.008 obtained from Karakas (2010). Evidently, our objects are between both sets of models. If N abundances were derived from ORLs, the points would move to the right by about 0.3 dex (0.5 dex in the extreme cases) and would be more in line with models of Z=0.020.  Another interesting observation is that the linear fit to our points is less steep than models at a given metallicity. At low N/H our objects are closer to the models with Z=0.008 and at 12+log(N/H)$>$8.4 they are closer to models with Z=0.02. This indicates that the objects with low 12+log(N/H)  are probably older and started with a lower metallicity. By comparing the observed data with the models, an initial mass can be assigned to the objects. We find that our WRPNe have central stars with initial masses between about 1.5 and more than 4 solar masses, where Cn\,1-5 (the richest in N) is the most massive. Other PNe with initial masses larger than 4 M$_\odot$ (those with 12+log(N/H)$\geq$8.44) could be He\,2-86, NGC\,5189, Hb\,4, M\,1-30, and M\,1-32.

\subsection{C/O ratios and infrared dust features
\label{coratios}}

\begin{figure}[!htb] 
\begin{center}
\includegraphics[width=\columnwidth]{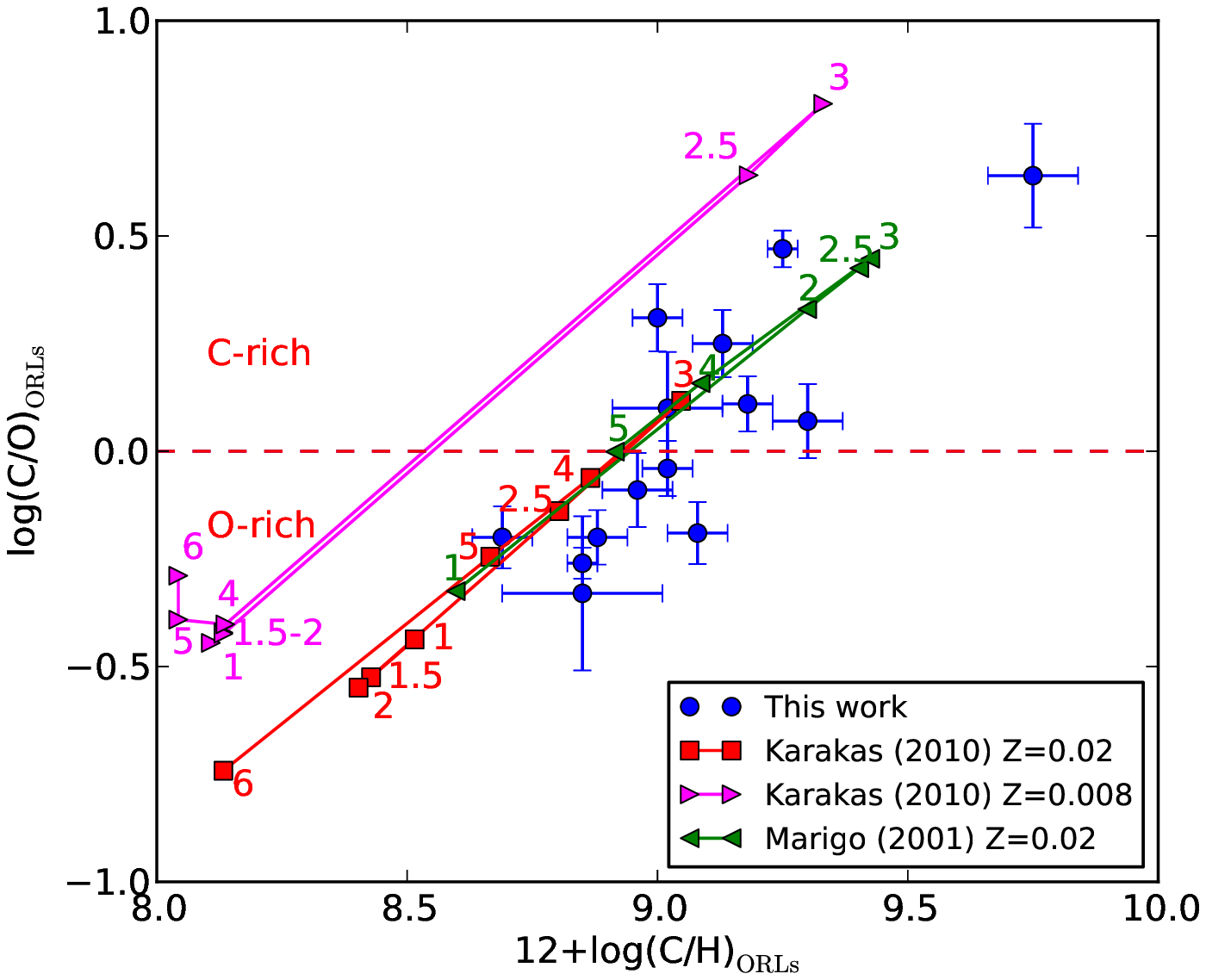}
\caption{C/O  abundance ratio plotted against C abundance for our sample of PNe. The horizontal dashed line represents the separation between C-rich and O-rich objects. The predictions of stellar evolution models by \citet{karakas10} (red models: Z=0.02, magenta models: Z=0.008) and by \citet{marigo01} (green models: Z=0.02)  for stars with different initial masses (indicated in units of {\msun})  are included.}
\label{CO_C}
\end{center}
\end{figure}

As said above, C/O abundance ratios reflect the occurrence of different nucleosynthesis processes in AGB stars. 
Theoretical models predict enhanced carbon abundances in the surface of 
stars with initial masses above $\sim$2 M$_{\odot}$ as a consequence of the third 
dredge-up, whereas they will be decreased in stars with masses above $\sim$4-5 M$_{\odot}$ 
due to the HBB process (see models represented in Fig.~\ref{CO_C}). 
Therefore, we expect that PNe coming from progenitors with M$\lesssim$1.5 M$_{\odot}$ or
 M$\gtrsim$4 M$_{\odot}$ will have C/O$<$1, while those coming from stars with masses 
 within 1.5 M$_{\odot}$ and $\sim$4 M$_{\odot}$ will show C/O$>$1. 

Using the values of O/H and C/H calculated from ORLs from Table~\ref{total_ab}, we obtain 
that half of our sample is C-rich (C/O$>$1), whereas the other half is O-rich (C/O$<$1).  
This is not unusual; in the sample analysed by \citet{kingsburghbarlow94} 40\% of the PNe were C-rich, and 
\citet{rolastasinska94}, using data from the literature, found that 35\% of their sample of 85 PNe were C-rich 
(in both cases C and O were determined via CELs). In our sample the fraction of  C-rich nebulae is slightly larger than in these two samples.

Fig.~\ref{CO_C} shows the C/O values as a function of C/H for our PNe, all of them calculated 
from ORLs.  A clear correlation exists, reflecting the effect of the third dredge-up, which pulls out freshly produced carbon to the stellar surface.
In this figure we overplotted the predictions from theoretical models calculated by \citet{karakas10} with metallicities 
Z=0.020 and 0.008, and \citet{marigo01} with metallicity Z=0.020, for different initial masses. Models show 
the behaviour described above in the sense that C abundance and C/O ratio increase for masses increasing from 1 to about 
3 M$_\odot$ and then decrease for larger masses. 

We discuss our objectscompared with models from \citet{marigo01}, which better reproduce our data.
Apparently, our C-rich objects (C/O$>$1) were produced by stars with initial masses between 1.5 and about 
5 M$_\odot$. Among them we find Cn\,1-5, M\,1-30 and NGC\,5189 (with initial masses near or above 4 M$_\odot$ as derived from their N abundances)  and  NGC\,2867, NGC\,6369, and Pe\,1-1, which probably descend from stars with masses between 1.5 and 2.5 M$_\odot$. 
The case of M\,1-32, which is the PN richest in C in our sample, is difficult to understand because there is no model that reproduces 
its extremely large C abundance. Its progenitor may have had an initial mass around 3 M$_\odot$, for which models predict the highest C enrichment, and for some reason, the third dredge-up was very efficient in this object.

It is difficult to assign an initial stellar mass to the objects with C/O$<$1 because the model curve is bi-valued  in this zone. 
Among these objects we find Hb\,4 and He\,2-86, which in the N-diagram appear as objects coming from stars with an initial  mass 
larger than 4 M$_\odot$, hence they probably suffered HBB, loosing C in the process. They are not classified as Peimbert type I due to 
their low He/H abundance (perhaps this criterion should be revised). The rest (five PNe with C/O$<$1) are expected to have descended from stars with initial 
masses between 1 and 1.5 M$_\odot$.
  
Summarizing, from N and C abundances and their ratios with O, we find consistent results with regard to the initial stellar masses of the progenitors. 
About half of our objects had a progenitor with an initial mass around or larger than four solar masses.
\smallskip

Another aspect of C/O abundance ratio is that its value in the atmospheres of 
AGB stars determines the composition of the grains that are formed. In principle, we expect 
carbon-based grains (such as PAHs or SiC) in PNe with C/O$>$1 and oxygen-based grains (mainly silicates) in PNe with C/O$<$1 \citep[see, e.g.,][]{whittet03}. 
Therefore a correlation between 
C/O and dust features is expected. We compiled all 
available information for our sample of PNe from the literature and also inspected the unpublished {\it Spitzer} 
spectra for NGC\,2867 and PB\,8. As shown in Table~\ref{tab:dust}, the PNe can be divided 
into two categories. The first one contains two C-rich PNe showing 
only C-rich features: NGC\,6369 with PAHs and SiC, and Pe~1-1 with PAHs and the 
broad feature at 30 $\mu$m that is usually associated with MgS \citep{cohenbarlow05, stanghellinietal12}. 
The second group corresponds to PNe with C/O either below or above one and a mixed chemistry 
of dust grains, that is, where crystalline silicates and PAHs are simultaneously found in the spectra. 
The explanation for this may be different in C-rich and O-rich environments. 
The preferred scenario for the PNe with  C/O$>$1, Cn~1-5, M~1-32, and NGC~2867, is that 
the silicates were formed when the star was still O-rich, whereas the PAHs formed 
more recently, after the third dredge-up transformed the O-rich star to a C-rich star 
\citep{watersetal98}. For the O-rich (C/O$<$1) PNe, \citet{guzmanramirezetal11} suggested that PAHs 
may form in the dense torus. Alternatively, PAH emission may come from the surrounding ISM.  
The other three nebulae are M~1-30, PC~14, and NGC~5189, for which we do not 
have information of the type of grains present. 

\setcounter{table}{12}
\begin{table*}
\caption{C/O values and infrared dust features$^{\rm a}$.\label{tab:dust}}
\begin{tabular}{cccc}
\noalign{\smallskip} \noalign{\smallskip} \noalign{\hrule} \noalign{\smallskip}
               & SiC/30 $\mu$m + PAHs & PAHs + crystalline silicates & no information\\
\noalign{\smallskip} \noalign{\hrule} \noalign{\smallskip}
C/O$>$1 & NGC~6369, Pe~1-1 & Cn~1-5, M~1-32, NGC~2867   & M~1-30, NGC~5189\\
C/O$<$1 &             --                   & Hb~4, He~2-86, M~1-25,  & PC~14\\
                &                                  & M~1-61, M~3-15, PB~8    &  \\
\noalign{\smallskip} \noalign{\hrule} \noalign{\smallskip}
\end{tabular}
\begin{description}
\item[$^{\rm a}$]Dust identifications from \citet{cohenbarlow05}, \citet{pereacalderonetal09}, and \citet{stanghellinietal12}.
\end{description}
\end{table*}

\subsection{$\alpha$-elements}

\begin{figure*}[!htb] 
\begin{center}
\includegraphics[width=\textwidth]{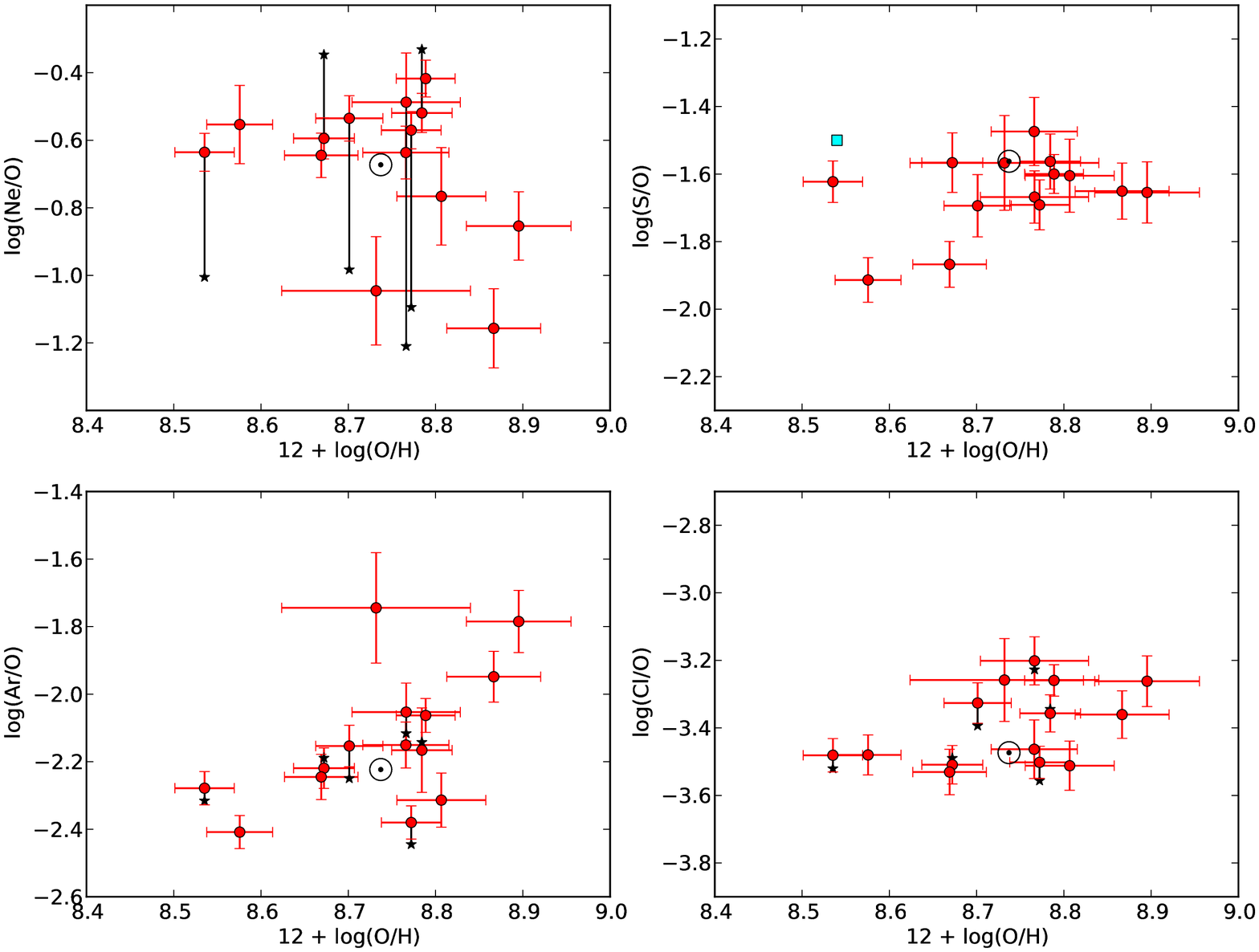}
\caption{$\alpha$-element/O abundance ratios for our sample of PNe. Stars indicate abundances computed using {\te}({\fariv}) for ions with IP$>$ 39 eV, while red circles represent abundances computed using {\te}({\foiii}) (see discussion in text). The solar symbol indicates the solar abundances taken from the compilation by \citet{lodders10}. The blue square in the log(S/O) plot is the value obtained for the Orion nebula from CELs. The y-axis ranges over 1.2 dex.}
\label{abalfa}
\end{center}
\end{figure*}

Oxygen, neon, sulphur, argon, and chlorine are $\alpha$-elements that, in principle, are only synthesized in massive stars, hence they are not affected by nucleosynthesis processes in the PNe progenitor stars \citep{marigo01}. 

In Fig.~\ref{abalfa} we present the abundance ratios of all $\alpha$-elements (all except Mg, which has only been detected in four objects and whose abundance is quite uncertain) over O, plotted againts the total O/H ratio derived from CELs. We overplotted solar abundances from the most recent compilation by \citet{lodders10}.

 In Sect.~\ref{abund_cels} we computed the abundances for the high IP ions (IP $>$ 39 eV) using two sets of physical conditions: the commonly used {\te}({\foiii}) (set A), and the rarely used {\te}({\fariv}) (set B). In Fig.~\ref{abalfa} we represented the abundances calculated using set A as red circles, and we have connected these points with the black stars, which are the points for which we have considered set B. With this comparison, we aim to verify the reliability of using {\te}({\fariv}) as representative of a high-ionization zone in a scheme of three zones.
As we can see in Fig.~\ref{abalfa}, the effect is only strong for Ne, because Ne$^{++}$ is the dominant ionization stage for Ne in the ionization degree range of our sample. Moreover, by assuming set B as representative of the high-ionization zone, the spread of Ne/O ratios is very wide and makes almost half of the sample Ne-poor objects, which cannot be explained by any stellar evolution model. The results obtained with set A are much more reasonable and agree with stellar evolution models, which predict that the Ne/O ratio probably does not vary, because both elements are produced by stars in the same mass range. An explanation for this behaviour would be that the ratio between collisional strengths for {\fariv} $\lambda$$\lambda$4711+40 and $\lambda$$\lambda$7170+262 lines is not reliable. We carried out some tests by increasing the collisional stregths of the $\lambda$$\lambda$4711+40 transitions by a factor of 1.5, which led to a lower {\te}({\fariv}); but  the problem was not solved, as  the abundance of Ne decreased in some objects, but it increased significantly in others. In addition, there are uncertainty sources in the determination of the total Ne abundance due to the used ICF (see below).

\begin{figure}[!htb] 
\begin{center}
\includegraphics[width=\columnwidth]{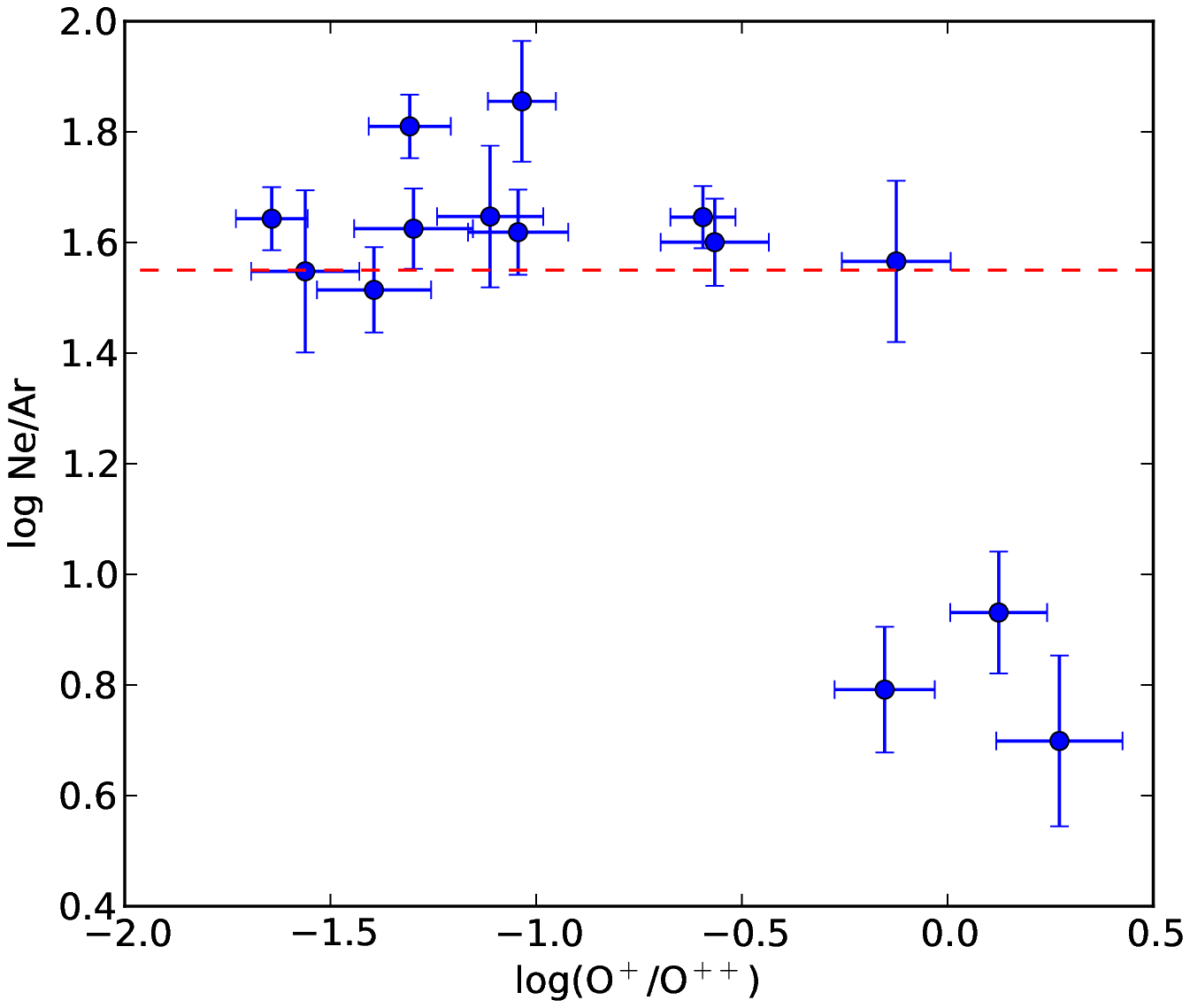}
\caption{Ne/Ar abundance ratio $vs.$ the ionization degree. The dashed red line represents the solar Ne/Ar abundance ratio. Ne/Ar is clearly underestimated for three objects.}
\label{near_ion}
\end{center}
\end{figure}

Despite their relatively large dispersion,  the Ar/O and Cl/O ratios seem to be consistent with what is expected for $\alpha$-elements, in the sense that these elements evolve  in lockstep with O. 
However, interestingly, in Fig.~\ref{abalfa} there are three objects that appear to be Ar-rich and Ne-poor at the same time. These objects are M\,1-25, M\,1-30, and M\,1-32 and are the  objects with the lowest ionization  degree in our sample. In Fig.~\ref{near_ion} we can see this behaviour. The problem with Ne is mainly due to the well-known fact that  the classical ICF scheme for Ne underestimate the total Ne abundance for low ionization nebulae \citep{torrespeimbertpeimbert77,peimbertetal92} because a considerable fraction of Ne$^+$ coexists with O$^{++}$. Additionally, photoionization models have proven that this ICF underestimates the total Ne for PNe with low $T_{eff}$ (Delgado-Inglada et al., in preparation). On the other hand, the ICF(Ar) from \citet{kingsburghbarlow94} seems to overestimate the Ar abundance for the same three objects. This has been proven by recent photoionization models, which  predict that with these ICFs the Ar abundance could be overestimated by between 0.2 and 0.7 dex, depending on the exact ICF scheme and on the assumed central star temperature, being less overestimated if the $T_{eff}$ of the central star is relatively low (about 50000 K, Delgado-Inglada et al. in preparation), which is the case for M\,1-25 and M\,1-32. A drop in the Ar abundance by about 0.2-0.3 dex for these objects would reconcile their abundances with what is expected for an $\alpha$-element (see Fig.~\ref{abalfa}). 

In Fig.~\ref{abalfa} (upper right), the behaviour of S/O vs. O abundance for our sample is shown. We included the solar value \citep[$-$1.57][]{lodders10}  and the Orion nebula value \citep[blue square, log(S/O)=$-$1.5, from CELs, ][]{estebanetal04, garciarojasesteban07}. In general, the PNe sulfur abundances are lower. This anomaly in the sulfur abundance in PNe is a well-known problem that has been discussed frequently in the past years \citep[e. g.][]{henryetal04, henryetal12, shingleskarakas13} and, so far, it is an unresolved question. Several explanations have been proposed, all of them revised by \citet{henryetal12}: a) incorrect ICFs that do not properly take into account the highly ionized S$^{+3}$ state, b) sulfur gas phase depletion due to dust or molecule formation, c) effects of dielectronic recombination in the sulfur ionization balance, and, very recently, d) the possibility that S might be destroyed by nucleosynthetic processes in low- and intermediate-mass stars during stellar evolution, which has been studied in depth by \citet{shingleskarakas13}. Most of these explanations have failed to explain the anomaly, which is still a subject of debate in PN physics. 

For our objects we can analyse the possibility of sulfur atoms to be partially deposited in C-rich dust grains \citep{whittet03}. The lowest sulfur abundances in our sample are found in Pe\,1-1, NGC\,6369, and NGC\,2867. The first two PNe show C-rich features (PAHs, SiC, and/or the broad feature at 30 $\mu$m) in their infrared spectra and therefore their low gaseous sulfur abundance might be partially due to the depletion of sulfur atoms onto dust grains. NGC\,2867 does not show C-rich features in its spectrum, but the C/O value above one might also indicate some depletion of sulfur atoms. 

However, \citet{henryetal12} ruled out this scenario of sulfur depletion in dust to explain the sulfur anomaly in PNe. These authors proposed that the most probable explanation is that ICFs are not properly taking into account the contribution of S ionization stages higher than S$^{++}$. Using photoionization models, \citet{rodriguezdelgadoinglada11} found that the ICF derived by \citet{stasinska78} introduces a bias in the computed sulfur abundances that can be significant even for low-ionization PNe, which would partially explain the situation.

\section{The abundance discrepancy
\label{adfs}}

As discussed in Sect.~\ref{intro}, the discrepancy between ionic abundances of several ions obtained from ORLs and CELs is a well-known problem that has dozens of references in the literature, both in PNe and {\hii} regions \citep[see e.g.][and references therein]{garciarojasesteban07, estebanetal09,mcnabbetal13,fangliu13}. The physical explanation for this discrepancy is still unclear, and several scenarios have been proposed; but, most references are focused on two scenarios: spatial temperature fluctuations in the ionized gas, and a cold, metal-rich ionized gas component embedded in the ambient gas. 

\citet{torrespeimbertetal80} proposed for the first time that the the cause of the discrepancy between abundance computed from CELs and ORLs might be spatial temperature fluctuations (parametrized by {\tsq}). Several mechanisms have been proposed to explain temperature fluctuations in {\hii} regions and/or PNe \citep[see][and references therein]{esteban02, peimbertapeimbert06} but to date, it is still unclear which mechanisms could produce them. 

On the other hand, \citet{liuetal00} proposed that a two-component nebular model with H-deficient (metal-rich) inclusions embedded in the normal metallicity gas, can account for most of the observed patterns. This behaviour has been probed for two born-again PNe: Abell 30 \citep{wessonetal03} and Abell 58 \citep{wessonetal08} and, probably for the PN Hf\,2-2 \citep{liuetal06}, which presents the largest ADF(O$^{++}$) found until now. However, the origin of these metal-rich inclusions in PNe is still a matter of discussion in the astronomical community \citep[see][for an extensive review of this question]{henneystasinska10}. A similar model has been proposed for {\hii} regions by \citet{tsamispequignot05} and a possible origin of these inclusions in these nebulae has been proposed by \citet{stasinskaetal07}. Therefore, the last word about this question has not been written yet. 

Alternatively to these two scenarios, or complementary to them, \citet{nichollsetal12} have proposed that departures of the Maxwell-Boltzmann equilibrium energy distribution of the electrons in the gas in the form of a $\kappa$-distribution of the thermal electrons, might be a potential explanation for the observed electron temperatures and chemical abundance behaviour in {\hii} regions and PNe. Additionally, very recent studies have proposed that high-density objects, such as proplyds, and high-velocity flows, such as Herbig-Haro objects, might play an important role in introducing uncertainties in the determination of physical conditions and chemical abundances using CELs \citep[][]{mesadelgadoetal09b,mesadelgadoetal12,nunezdiazetal12,tsamisetal11}. 

\begin{figure}[!htb] 
\begin{center}
\includegraphics[width=\columnwidth]{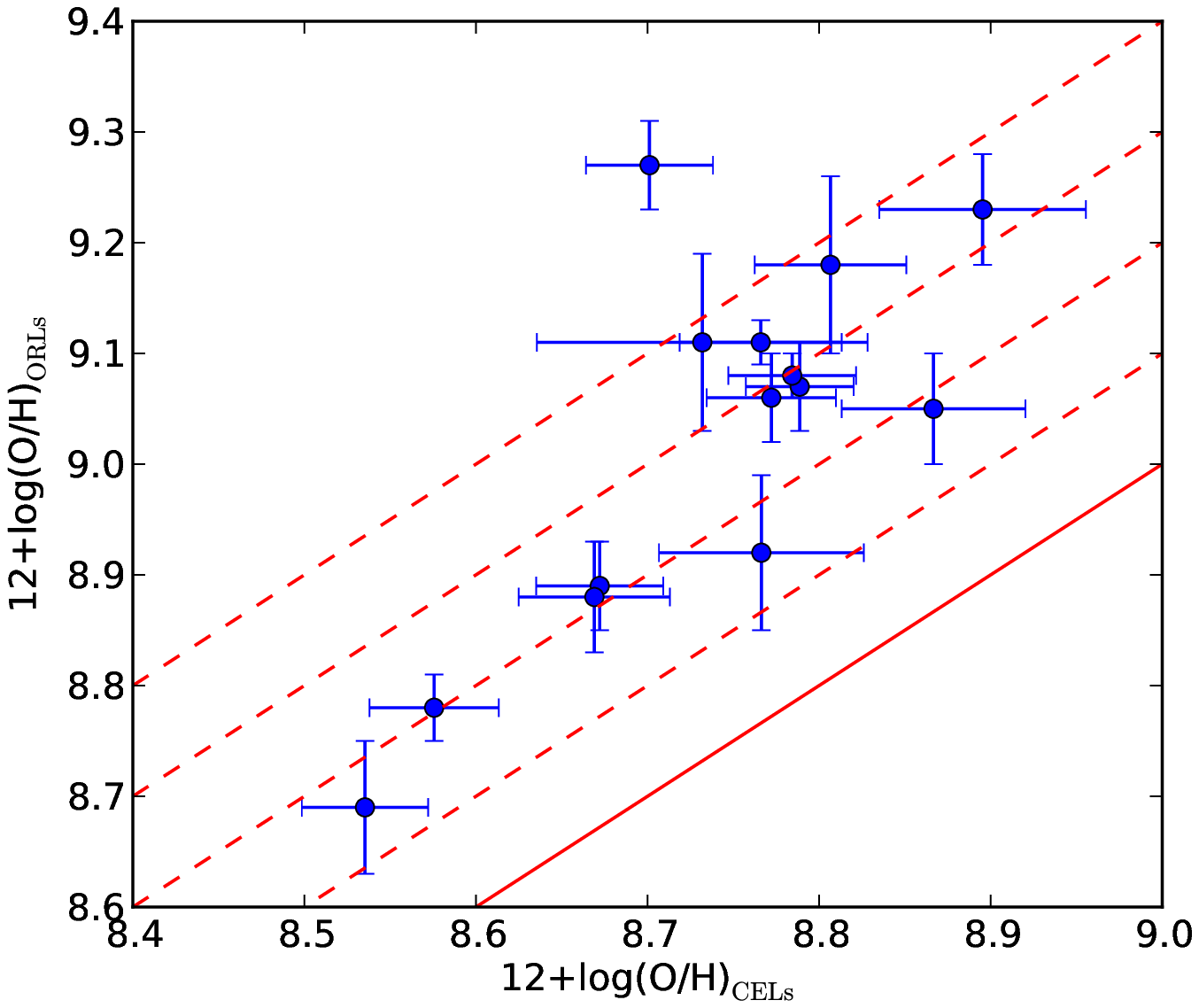}
\caption{O/H derived from ORLs plotted against O/H derived from CELs. The continous line represents the equality. Dashed lines represent ADFs of 0.1, 0.2, 0.3, and 0.4 dex. The O/H derived from ORLs is about a factor of 2-3 higher than that derived from CELs.}
\label{o_orl_cel}
\end{center}
\end{figure}

In Table~\ref{total_ab}, two sets of abundances (from CELs and from ORLs) are presented for O, which so far is the element with the best-determined abundance.  In Fig.~\ref{o_orl_cel} we plot the O/H derived from ORLs $vs.$ O/H derived from CELs. When CELs are used, we obtain 12+ log(O/H) abundances ranging from about 8.5 (NGC\,6369) to about 8.90 (M\,1-30). When ORLs are used, the 12+log(O/H) values range from about 8.7 (NGC\,6369) to more than 9.2 (Hb\,4 and M\,1-30). In general, we found that O/H derived from ORLs is higher than O/H derived from CELs by factors between 1 and 3. In neither case did we find a large ADF (larger than 5), as has been found in  several Galactic PNe \citep[see e.g.][]{mcnabbetal13}. Therefore, it seems that the [WC] nature of the central star does not particularly affect the ADFs in these nebulae. 

At this point we should consider that the solar 12 + log(O/H) value currently accepted is 8.73 \citep{lodders10}\footnote{The most widely used reference for solar abundances is that of \citet{asplundetal09}, but the Lodders compilation includes updated oxygen abundances and a detailed discussion of the evolution of oxygen abundance determinations in the Sun; aditionally, meteoritic and photosferic abundances are given and a value is recommended; given the similarity within the uncertainties between the values published by both authors, we have preferred to cite Lodders's work.} and the same value for the gas+dust Orion nebula (showing the present ISM abundances) based on CELs/ORLs is 8.59/8.73 \citep{estebanetal04}. Progenitors of PNe have ages between 1 and 10 Gyr, and it is expected that their initial O abundances were similar or lower than in the Sun, but not higher. Therefore the O/H abundance ratios derived from ORLs are difficult to reconcile with this view, unless the central stars have modified the O in their surfaces as a consequence of nucleosynthesis. Although some evolution models of LIMS (low to intermediate mass stars) predict such a phenomenon, it occurs at metallicities much lower than in the Milky Way \citep{marigoetal03, karakaslattanzio03}. On the other hand, \citet{lodders03} claimed that solar photospheric abundances are not representative of the proto-Sun, because heavy-element fractionation has altered the photospheric abundances by about $-$0.074 dex in the Sun's lifetime. However, this effect is still too low to explain the large differences ($\sim$0.3 dex in average) found between O/H values derived from ORLs in PNe and those predicted by chemical evolution models. In the following we adopted the abundances given by ORLs (or considering a $t^2$ parameter higher than 0.00) as representative of the nebulae, but we keep in mind the inconsistency reported here.

\begin{figure*}[!htb] 
\begin{center}
\includegraphics[width=\textwidth]{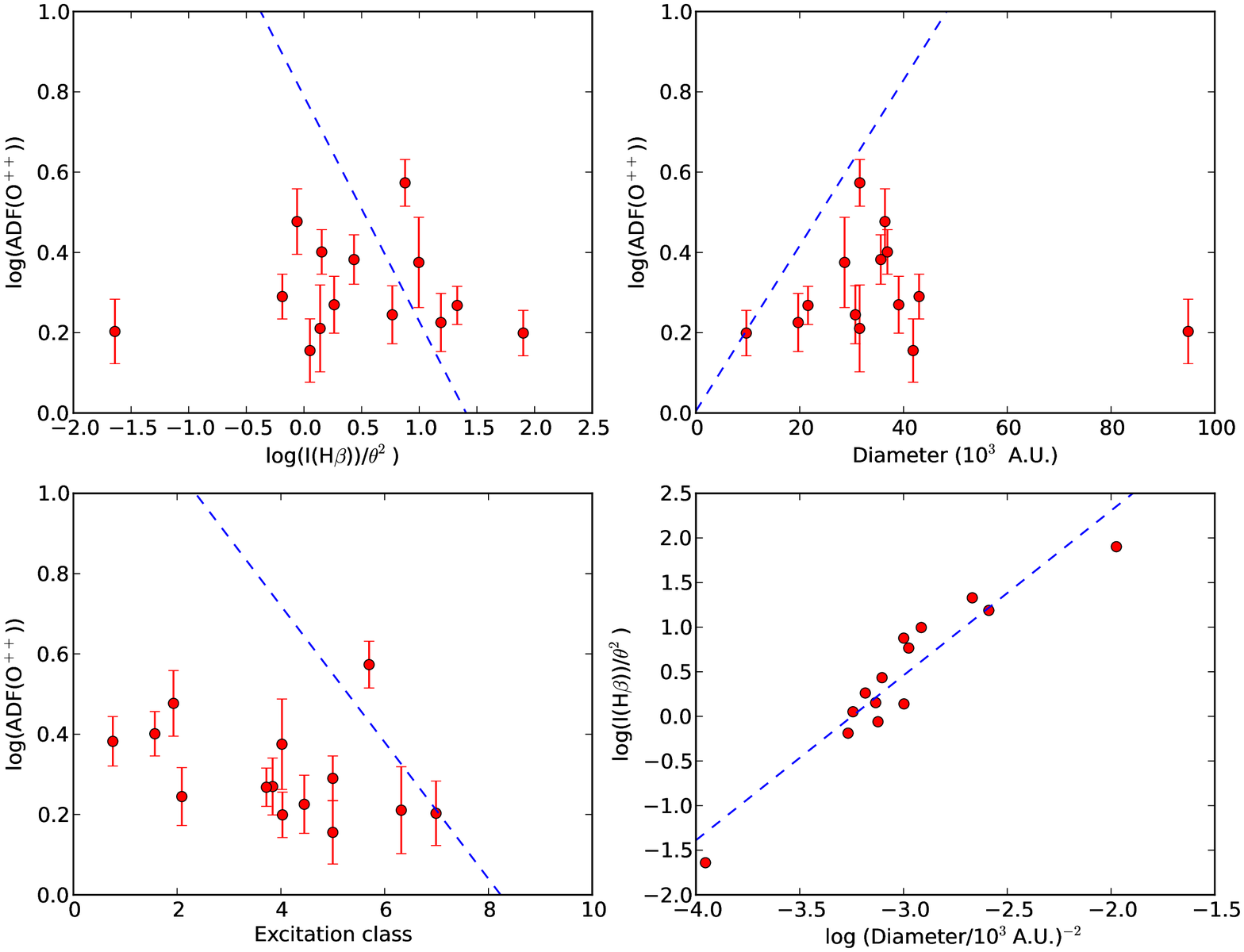}
\caption{ADF(O$^{++}$) plotted against mean nebular surface brightness (upper left panel); absolute nebular diameters (upper right panel) and excitation class as defined by \citet{dopitameatheringham90} (lower left panel). The dashed lines represent the fits obtained by \citet{yliuetal04b} for a sample of 19 PNe. In the lower right panel we show the relation between the nebular surface brightness and the nebular diameter, which are clearly correlated; the line is a fit to the data.}
\label{adf_shb_D}
\end{center}
\end{figure*}

In an extensive study of the abundance discrepancy in several PNe, \citet{yliuetal04b} found correlations between ADF(O$^{++}$) and several nebular properties, such as metallicity, mean surface brightness, diameter, excitation class and density. Some of these correlations were previously found by other authors \citep[e.~g.][]{garnettdinerstein01}. Although our data do not cover a range of ADFs as wide as that of these authors, we consider that a comparison could be interesting. We have obtained the mean surface brightness  and the angular sizes for our PNe from the Strasbourg - ESO catalogue of Galactic planetary nebulae \citep{ackeretal92}; heliocentric distances to compute absolute nebular diameters were taken from \citet{stanghellinihaywood10}, and excitation classes were computed following \citet{dopitameatheringham90}. In Fig.~\ref{adf_shb_D} we explore these relations with our data and found for the range of ADFs of our sample no clear correlation. It seems that the correlations found by \citet{garnettdinerstein01} and \citet{yliuetal04b} are not as evident if we exclude large ADFs from the sample. 

\begin{figure}[!htb] 
\begin{center}
\includegraphics[width=\columnwidth]{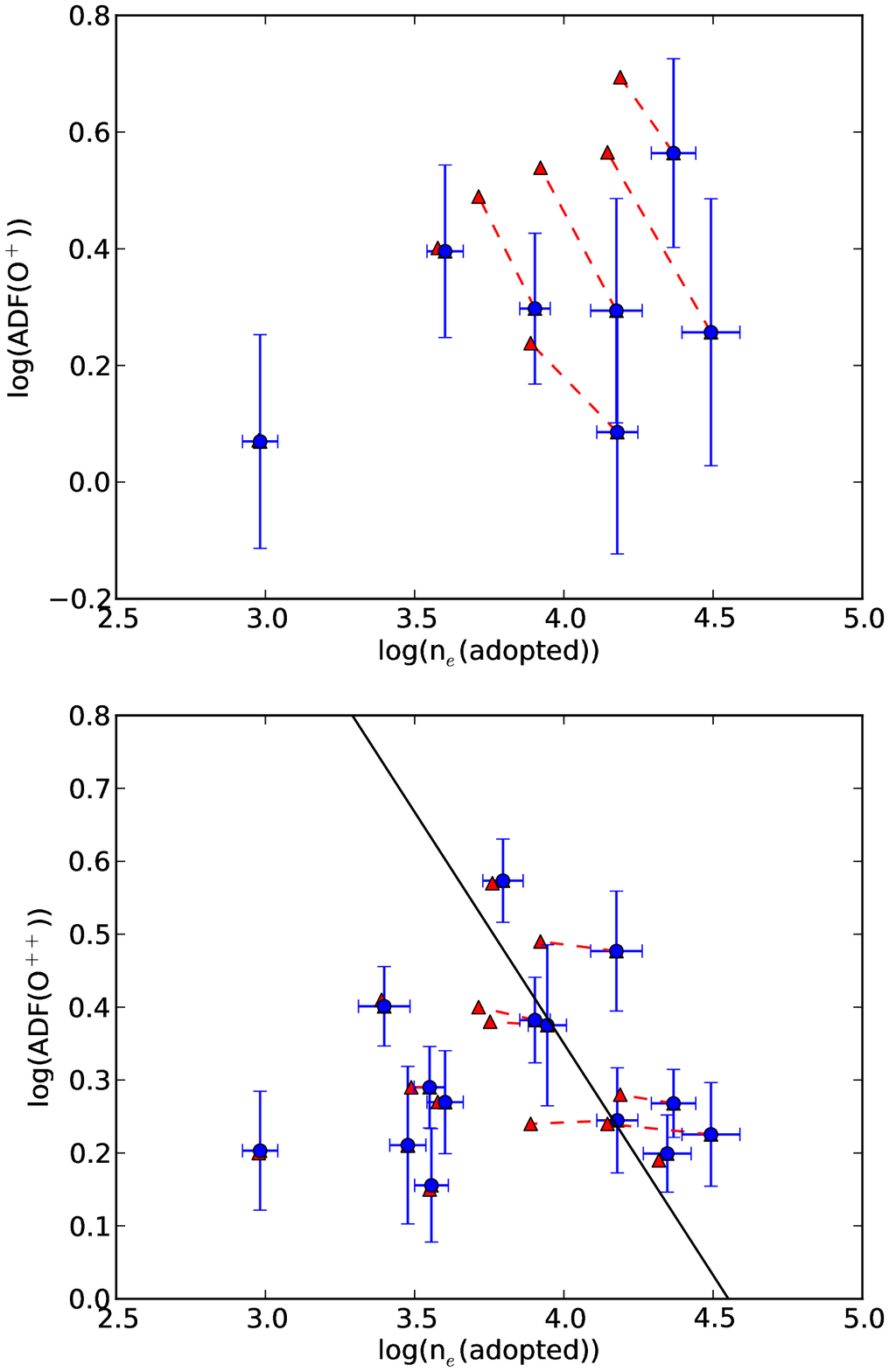}
\caption{ADF(O$^+$) and ADF(O$^{++}$) $vs.$ log({\elecd}). Blue dots represent the values assuming the densities shown in Table~5 of Paper I. Red dots represent the values obtained for {\elecd}({\fsii}) for all the PNe. Note that although ADF(O$^{++}$) remains almost unchanged, the ADF(O$^+$) is very sensitive to the assumed density. The line is the correlation found by \citet{robertsontessigarnett05}.}
\label{adf_ne}
\end{center}
\end{figure}

In Fig.~\ref{adf_ne} we show the ADF(O$^+$) and ADF(O$^{++}$) plotted against the density. In the upper panel, we show that ADF(O$^+$) is extremely sensitive to the assumed density, whereas ADF(O$^{++}$) is not. Hence ADF(O$^+$) should be taken with some caution because this ion comes from the outermost zones of the PNe (especially if the PN is highly excited) and might be only representative for a small volume of the ionized gas.

\begin{figure}[!htb] 
\begin{center}
\includegraphics[width=\columnwidth]{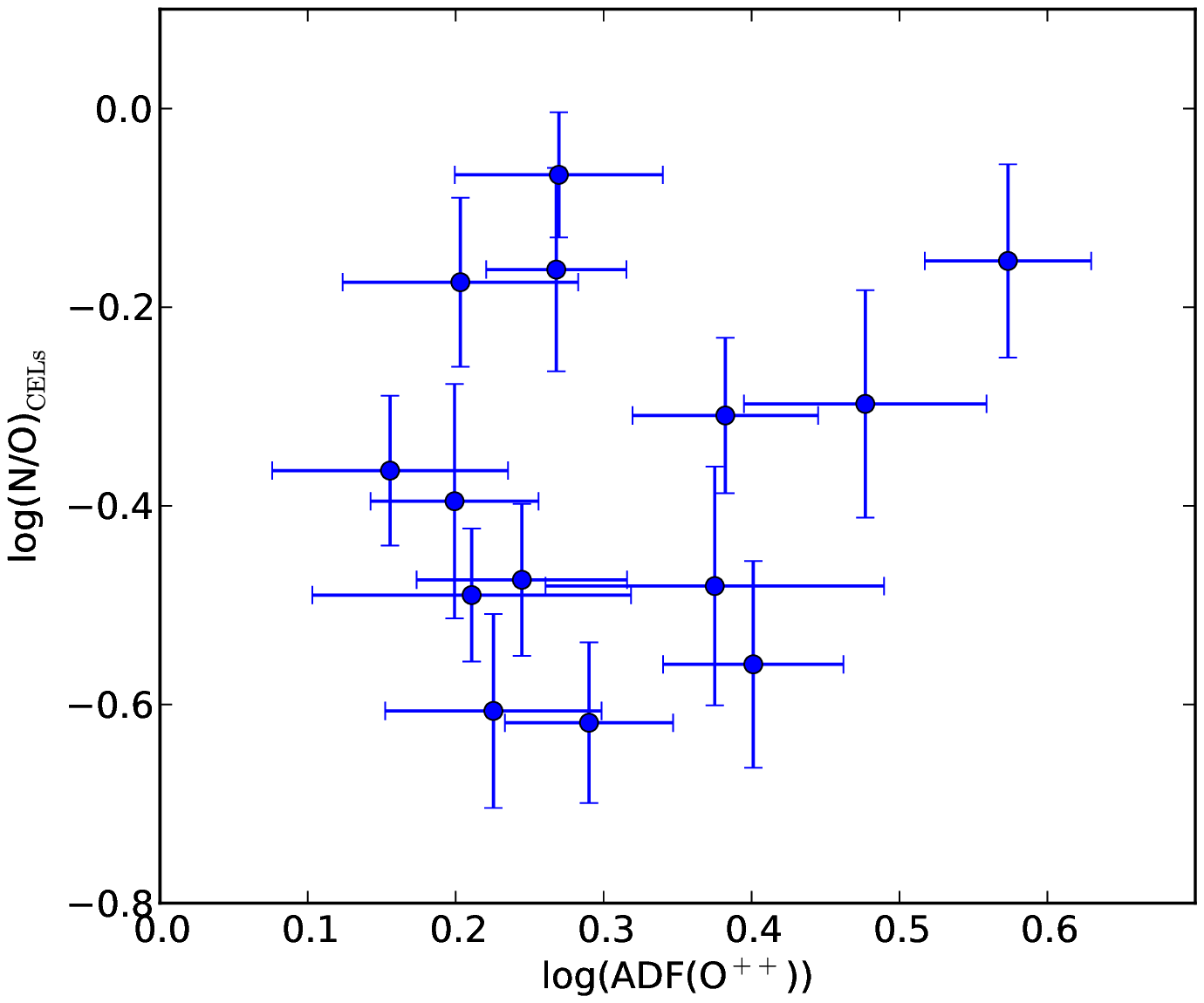}
\caption{N/O $vs.$ ADF(O$^{++}$).}
\label{adf_no}
\end{center}
\end{figure}

Finally, in Fig.~\ref{adf_no} we show the N/O ratio plotted against the ADF(O$^{++}$). We did this to determine whether the ADF depends on the enrichment pattern of the PNe. \citet{yliuetal04b} found a slight correlation between N/O and the ADF(O$^{++}$) although with a large scatter. In our case, the dispersion of the data is large and no apparent correlation is found.

\subsection{Determination of {\tsq}
\label{tsquared}}

Given the different dependence on the electron temperature of the CELs and ORLs emissivities, and assuming the validity of the temperature fluctuations paradigm, we can estimate the mean square temperature fluctuations parameter, {\tsq}, from the comparison between abundances derived from both kind of lines for a given ion \citep[see][and references therein for a detailed description of temperature fluctuations formalism]{peimbertetal04}. 
In particular, we estimated the values of the {\tsq} parameter from the ADFs obtained for O$^+$, O$^{++}$, C$^{++}$ (only for two objects) and Ne$^{++}$. The values obtained from the comparison of C$^{++}$ ionic abundances using CELs and ORLs are quite uncertain because the abundances from CELs were obtained from IUE UV observations, where aperture and extinction effects introduce large uncertainties. The values we computed for our objects are moderate and quite consistent among each other within the uncertainties,  and in principle, they might be caused by spatial temperature fluctuations in the observed volume of gas. 
 
Following \citet{apeimbertetal02}, we also computed the {\tsq} obtained from the application of a maximum-likelihood method (MLM) to search for the physical conditions, including He$^+$/H$^+$ ratios and optical depths, that simultaneously fit all the measured lines of {\hei} (see Sect.~\ref{he_abund}).  

Additionally, from the comparison between electron temperatures obtained from CELs and from the hydrogen recombination continuum discontinuities, we can obtain an indication of {\tsq} \citep[see][and references therein]{torrespeimbertetal80, garciarojasetal09}. Unfortunately, the measurement of Balmer and Paschen discontinuities in our spectra was rather difficult and, therefore, these estimates were unreliable (see Paper I).

\begin{figure}[!htb] 
\begin{center}
\includegraphics[width=\columnwidth]{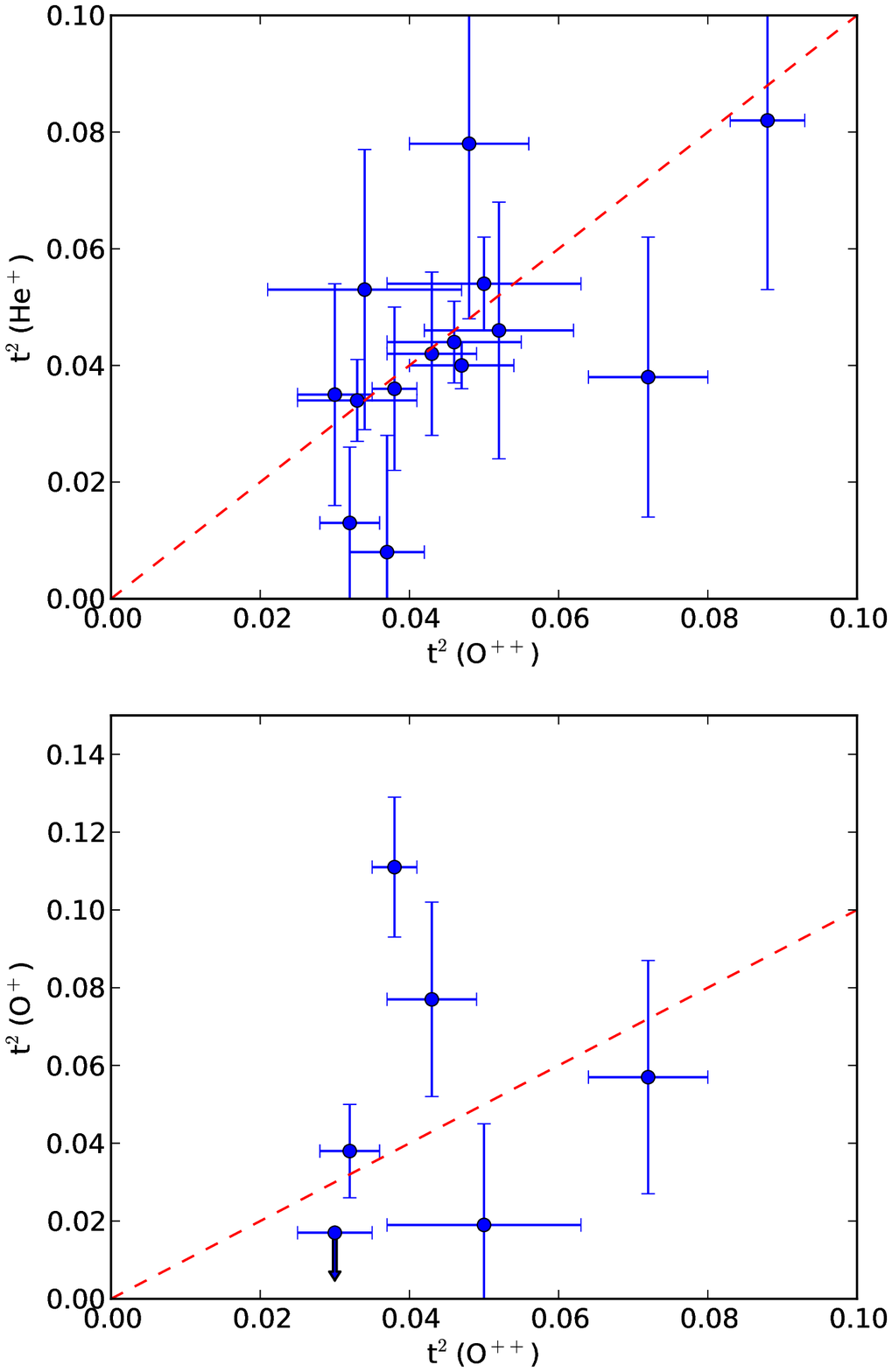}
\caption{ {\tsq}(O$^{++}$) against {\tsq}(He$^{+}$) (upper panel) and against {\tsq}(O$^+$) (lower panel); lines represent equality. Clearly, there is an overall agreement between {\tsq}(He$^+$) and {\tsq}(O$^{++}$). The agreement between the {\tsq}(O$^+$) and {\tsq}(O$^{++}$) values is poorer (see text).}
\label{t2_fig}
\end{center}
\end{figure}

In Fig.~\ref{t2_fig} we show the comparison between {\tsq}(O$^{++}$) and {\tsq}(He$^{+}$) (upper panel) and {\tsq}(O$^{++}$) against {\tsq}(O$^{+}$) (lower panel). Clearly, there is an overall agreement between {\tsq}(He$^+$) and {\tsq}(O$^{++}$), but this agreement is not evident in the second case. It is important to mark that He$^+$ and O$^{++}$ coexist in most of the volume for PNe with excitation classes similar to those of our sample, but O$^+$ and O$^{++}$ represent two distinct zones in each PN. However, given the large uncertainties in the determination of {\tsq}(O$^+$), we can say that most of the {\tsq} values agree within the uncertainties for our sample, with some exceptions that do not affect to the determination of total abundances for {\tsq}$>$0.00 (see next paragraph). 

\begin{figure}[!htb] 
\begin{center}
\includegraphics[width=\columnwidth]{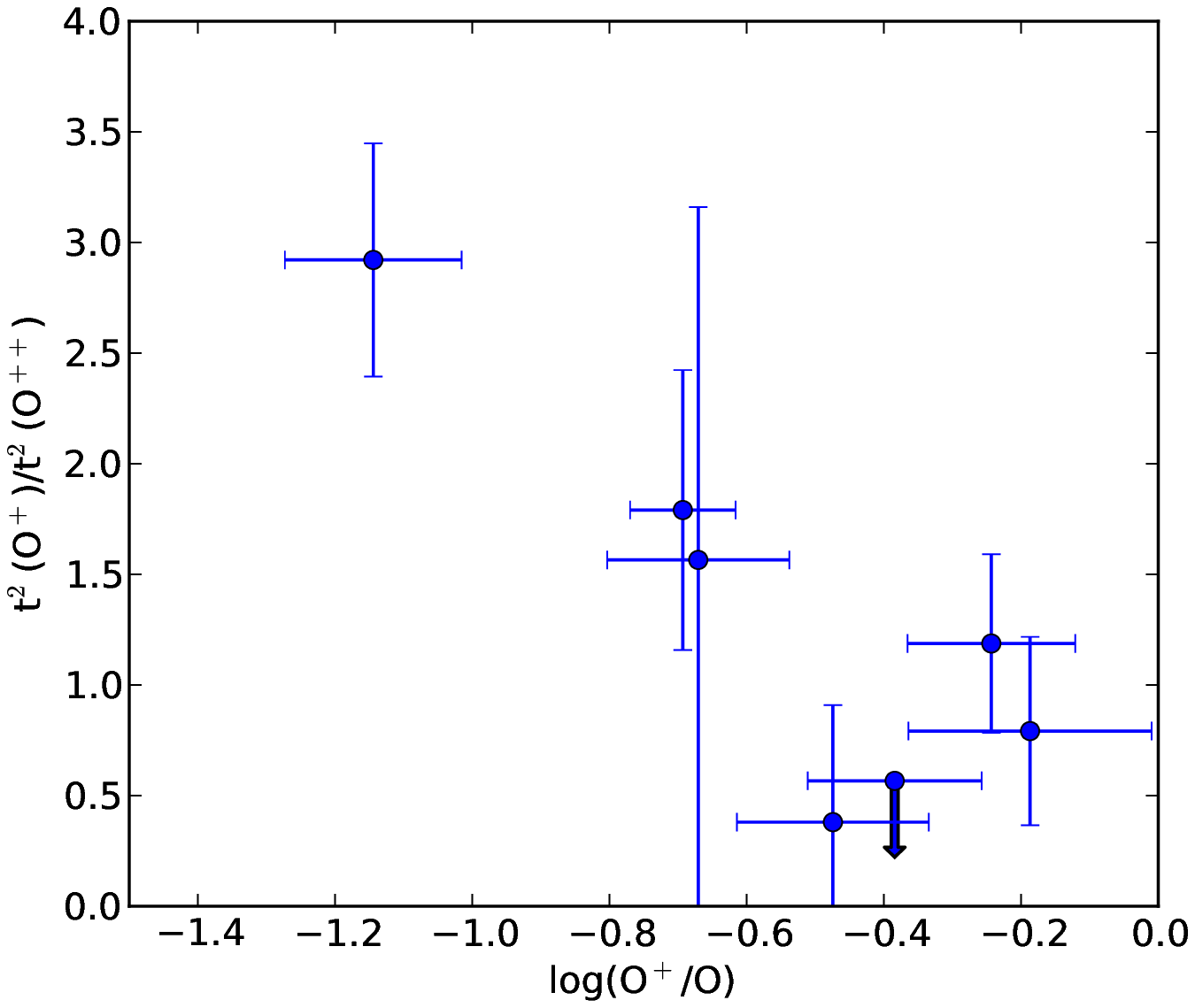}
\caption{{\tsq}(O$^{+}$)/{\tsq}(O$^{++}$) ratio $vs.$ the logarithm of the fraction O$^+$/O. }
\label{coct2_ion}
\end{center}
\end{figure}

In Fig.~\ref{coct2_ion} we plot the {\tsq}(O$^{+}$)/{\tsq}(O$^{++}$) ratio against log(O$^{+}$/O), and found a clear anti-correlation. We ran a generalized Kendall $\tau$-correlation test on these data and found $\tau$=$-$1.2,  and a probability that both quantities are not correlated of $\sim$14\%, hence, they are probably correlated. However, this test should be taken with caution when the data set is small, as in our case. Undoubtedly, more high-quality observations of multiplet 1 {\oi} ORLs are needed to minimize the {\tsq}(O$^+$) uncertainties and confirm (or discard) this tendency. However, if this anti-correlation is real, there must be one (or a set of) physical mechanism to explain why temperature fluctuations in the O$^+$ zone become higher than in the O$^{++}$ zone when the relative O$^+$ volume is smaller. Nevertheless, given the relatively low importance of O$^+$ in these cases, the total O abundance for {\tsq}$>$0.00 is not affected.

To compute the final $t^2$ parameter we averaged the values obtained from the ADFs of O$^+$ and O$^{++}$ and the $t^2$(He$^+$). We weighted the {\tsq} of O$^+$ and O$^{++}$ with the relative abundance of each ion.   We discarded the values derived from the ADFs of Ne$^{++}$ because the {\neii} ORLs are very weak, and those derived from the ADF of C$^{++}$ for the reasons discussed above.  In Table~\ref{t_square} we show all computed {\tsq}.
Following the formulae given by \citet{peimbert67}, \citet{peimbertcostero69}, and \citet{peimbertetal04}, we computed the abundances of the chemical elements in the presence of temperature fluctuations. In table~\ref{total_ab_t2} we show the total abundances of He, C, N, O, Ne, S, Cl, and Ar for $t^2$$>$0.00. 

\setcounter{table}{13}
\begin{table*}
\begin{tiny}
\caption{Determinations of the $t^2$ parameter.}
\label{t_square}
\begin{tabular}{lcccccc}
\noalign{\hrule} \noalign{\vskip3pt}
  				&          Cn\,1-5 		&        Hb\,4 	        & He\,2-86 		& M\,1-25       		&    M\,1-30   			&      M\,1-32     \\
\noalign{\smallskip} \noalign{\hrule} \noalign{\smallskip}
$t^2$(O$^+$)		& 0.077$^{+0.018}_{-0.031}$& \nodata	  	& 0.111$^{+0.015}_{-0.020}$ & $<$0.017: 	& 0.038$^{+0.011}_{-0.013}$ & 0.057$^{+0.024}_{-0.035}$		\\ 
$t^2$(O$^{++}$)		& 0.043$\pm$0.006   & 0.088$\pm$0.005 	& 0.038$\pm$0.003	& 0.030$\pm$0.005	& 0.032$\pm$0.004 		& 0.072$\pm$0.008\\ 
$t^2$(C$^{++}$)		& 0.028:			& \nodata	    	         & \nodata	    		& \nodata	    		& \nodata	    			& \nodata		\\ 
$t^2$(Ne$^{++}$)	& 0.002:			& 0.058:		  	& 0.010: 		& 0.076: 			& 0.026: 				& \nodata		\\ 
$t^2$(He$^+$)		& 0.042$\pm$0.014	& 0.082$\pm$0.029    & 0.036$\pm$0.014	& 0.035$\pm$0.019	& 0.013$^{+0.018}_{-0.013}$& 0.038$\pm$0.024\\ 
			         & 			         &				&				&				&					&		\\ 
$t^2$ (adopted)		& 0.045$\pm$0.006	& 0.088$\pm$0.005	& 0.037$\pm$0.003	&0.031$\pm$0.005	&0.032$\pm$0.004		&0.068$\pm$0.008 \\ 
\noalign{\smallskip} \noalign{\hrule} \noalign{\smallskip}
  			         &  M\,1-61   		& M\,3-15      		&  NGC\,5189   		&    NGC\,6369 		&    PC\,14   		&       Pe\,1-1    \\
\noalign{\smallskip} \noalign{\hrule} \noalign{\smallskip}
$t^2$(O$^+$)		& \nodata	    		& \nodata	  		& 0.019$^{+0.036}_{-0.019}$ & \nodata  	& \nodata 			& 0.072:	\\ 
$t^2$(O$^{++}$)		& 0.033$\pm$0.008    & 0.048$\pm$0.008 	& 0.050$\pm$0.013	& 0.034$\pm$0.013	& 0.047$\pm$0.007 	& 0.046$\pm$0.009\\ 
$t^2$(C$^{++}$)		& \nodata	    		& \nodata	    		&  0.075:		    	& \nodata	    		& \nodata	    		& \nodata		\\ 
$t^2$(Ne$^{++}$)	& $<$0.00:			& 0.068:		  	&  0.021:              	& \nodata 	       		& 0.010: 			& $<$0.00		\\ 
$t^2$(He$^+$)		& 0.034$\pm$0.007	& 0.078$\pm$0.030     & 0.054$\pm$0.008	& 0.053$\pm$0.024	& 0.040$\pm$0.004	& 0.044$\pm$0.007\\ 
			         & 				&				&				&				&				&		\\ 
$t^2$ (adopted)		&  0.034$\pm$0.005	& 0.050$\pm$0.008	& 0.051$\pm$0.008	&0.038$\pm$0.011	&0.042$\pm$0.004	& 0.045$\pm$0.006 \\ 
\noalign{\smallskip} \noalign{\hrule} \noalign{\smallskip}
\end{tabular}
\end{tiny}
\end{table*}

\setcounter{table}{14}
\begin{table*}
\begin{tiny}
\begin{center}
\caption{Total abundances for $t^2 > $0.00}
\label{total_ab_t2}
\begin{tabular}{lccccccc}
\noalign{\smallskip} \noalign{\smallskip} \noalign{\hrule} \noalign{\smallskip}
 & \multicolumn{7}{c}{12 + log(X/H)} \\
\noalign{\smallskip} \noalign{\hrule} \noalign{\smallskip}
             Element &         Cn1-5 &           Hb4 &        He2-86 &         M1-25 &         M1-30 &         M1-32 &         M1-61\\
\noalign{\smallskip} \noalign{\hrule} \noalign{\smallskip}
              $t^2$  & 0.045$\pm$0.006&0.091$\pm$0.005&0.037$\pm$0.003&0.028$\pm$0.005&0.032$\pm$0.004&0.065$\pm$0.008&0.035$\pm$0.005\\
\noalign{\smallskip} \noalign{\hrule} \noalign{\smallskip}
                 He  &11.19$\pm$0.01 &11.06$\pm$0.02 &11.09$\pm$0.01 &11.09$\pm$0.01 &11.15$\pm$0.01 &11.10$\pm$0.02 &11.04$\pm$0.01\\
        C$^{\rm b}$  & 9.18$\pm$0.05 & 9.07$\pm$0.08 & 8.87$\pm$0.07 & 8.93$\pm$0.08 & 9.30$\pm$0.08 & 9.75$\pm$0.10 & 8.67$\pm$0.08\\
                  N  & 9.00$\pm$0.08 & 9.05$\pm$0.12 & 8.90$\pm$0.12 & 8.61$\pm$0.11 & 8.87$\pm$0.11 & 8.79$\pm$0.16 & 8.47$\pm$0.13\\
                  O  & 9.08$\pm$0.04 & 9.27$\pm$0.06 & 9.05$\pm$0.05 & 9.09$\pm$0.05 & 9.20$\pm$0.06 & 9.12$\pm$0.08 & 8.87$\pm$0.05\\
        O$^{\rm c}$  & 9.09$\pm$0.05 & 9.27$\pm$0.04 & 9.08$\pm$0.02 & 9.05$\pm$0.05 & 9.23$\pm$0.05 & 9.10$\pm$0.08 & 8.87$\pm$0.04\\
                 Ne  & 8.69$\pm$0.07 & 8.79$\pm$0.11 & 8.56$\pm$0.09 & 7.95$\pm$0.13 & 8.40$\pm$0.13 & 8.12$\pm$0.13 & 8.01$\pm$0.09\\
                  S  & 7.55$\pm$0.04 & 7.66$\pm$0.07 & 7.55$\pm$0.06 & 7.50$\pm$0.07 & 7.71$\pm$0.08 & 7.62$\pm$0.08 & 7.37$\pm$0.07\\
                 Cl  & 5.80$\pm$0.04 & 5.87$\pm$0.07 & 5.67$\pm$0.05 & 5.73$\pm$0.07 & 5.97$\pm$0.07 & 5.84$\pm$0.07 & 5.35$\pm$0.05\\
                 Ar  & 6.57$\pm$0.11 & 7.06$\pm$0.05 & 6.83$\pm$0.14 & 7.13$\pm$0.11 & 7.36$\pm$0.13 & 7.32$\pm$0.19 & 6.62$\pm$0.06\\
       	        Fe  & 5.75$\pm$0.18 & 5.82$\pm$0.29 &  6.25$\pm$0.15 & 6.38$\pm$0.14 & 6.02$\pm$0.13 &  7.12$\pm$0.16  & 5.53$\pm$0.19 \\
      	        Ni  & 5.63$\pm$0.23 & \nodata     &   5.86$\pm$0.39    &    \nodata    &   4.79$\pm$0.20    &   6.08$\pm$0.20   &    \nodata   \\
\noalign{\smallskip} \noalign{\hrule} \noalign{\smallskip}
             Element &         M3-15 &       NGC5189 &       NGC6369 &          PC14 &         Pe1-1 &           PB8$^{\rm a}$ &       NGC2867 $^{\rm a}$\\
\noalign{\smallskip} \noalign{\hrule} \noalign{\smallskip}
              $t^2$  &0.048$\pm$0.008&0.055$\pm$0.008&0.038$\pm$0.011&0.041$\pm$0.004&0.045$\pm$0.006&	0.033$\pm$0.005&0.052$\pm$0.01\\
\noalign{\smallskip} \noalign{\hrule} \noalign{\smallskip}
                 He  &11.03$\pm$0.02 &11.09$\pm$0.01 &11.02$\pm$0.01 &11.03$\pm$0.01 &11.02$\pm$0.01 &11.09$\pm$0.01 &11.06$\pm$0.01\\
        C$^{\rm b}$  & 8.85$\pm$0.16 & 9.02$\pm$0.11 & 9.00$\pm$0.08 & 9.02$\pm$0.05 & 9.12$\pm$0.09 & 8.85$\pm$0.05 & 9.25$\pm$0.05\\
                  N  & 8.71$\pm$0.11 & 8.79$\pm$0.14 & 8.34$\pm$0.10 & 8.39$\pm$0.09 & 8.29$\pm$0.12 & 8.55$\pm$0.12 & 8.29$\pm$0.10\\
                  O  & 9.20$\pm$0.09 & 8.97$\pm$0.07 & 8.71$\pm$0.07 & 9.03$\pm$0.04 & 8.90$\pm$0.06 & 9.12$\pm$0.07 & 8.78$\pm$0.06\\
        O$^{\rm c}$  & 9.18$\pm$0.08 & 8.92$\pm$0.07 & 8.69$\pm$0.06 & 9.06$\pm$0.04 & 8.90$\pm$0.05 & 9.11$\pm$0.02 & 8.78$\pm$0.03\\
                 Ne  & 8.46$\pm$0.19 & 8.49$\pm$0.07 & 8.09$\pm$0.12 & 8.48$\pm$0.07 & 8.27$\pm$0.10 & 8.52$\pm$0.10 & 8.24$\pm$0.13\\
                  S  & 7.71$\pm$0.11 & 7.33$\pm$0.06 & 7.14$\pm$0.06 & 7.40$\pm$0.05 & 7.07$\pm$0.06 & 7.79$\pm$0.10 & 6.92$\pm$0.06\\
                 Cl  & 5.66$\pm$0.09 & 5.74$\pm$0.04 & 5.20$\pm$0.06 & 5.51$\pm$0.05 & 5.34$\pm$0.07 & 5.69$\pm$0.10 & 5.31$\pm$0.06\\
                 Ar  & 6.83$\pm$0.09 & 6.89$\pm$0.07 & 6.40$\pm$0.06 & 6.62$\pm$0.04 & 6.61$\pm$0.08 & 6.90$\pm$0.08 & 6.35$\pm$0.07\\
       	        Fe  & 5.92$\pm$0.26 & 5.04$\pm$0.23 &    \nodata    & 5.43$\pm$0.14 & 5.79$\pm$0.19 &    \nodata    &    \nodata   \\
      	        Ni  &   \nodata     &    \nodata    &    \nodata    &    \nodata    &    \nodata    &    \nodata    &    \nodata   \\
\noalign{\smallskip} \noalign{\hrule} \noalign{\smallskip}
\end{tabular}
\end{center}
\begin{description}
\item[$^{\rm a}$] From \citet{garciarojasetal09}.
\item[$^{\rm b}$] From recombination lines (see text).
\item[$^{\rm c}$] From recombination lines (see text).
\end{description}
\end{tiny}
\end{table*}

\section{Final discussion.
\label{discuss}}

In this section we discuss additional implications of our results. In particular, we explore possible dependencies of several parameters, such as ADF, C/O, and N/O on the central star spectral type, and we also compare the Galactic gradient of the O abundance obtained from our data using ORLs with that obtained from {\hii} regions. 

\subsection{ADF and the abundance ratios versus the spectral type}

\begin{figure}[!htb] 
\begin{center}
\includegraphics[width=\columnwidth]{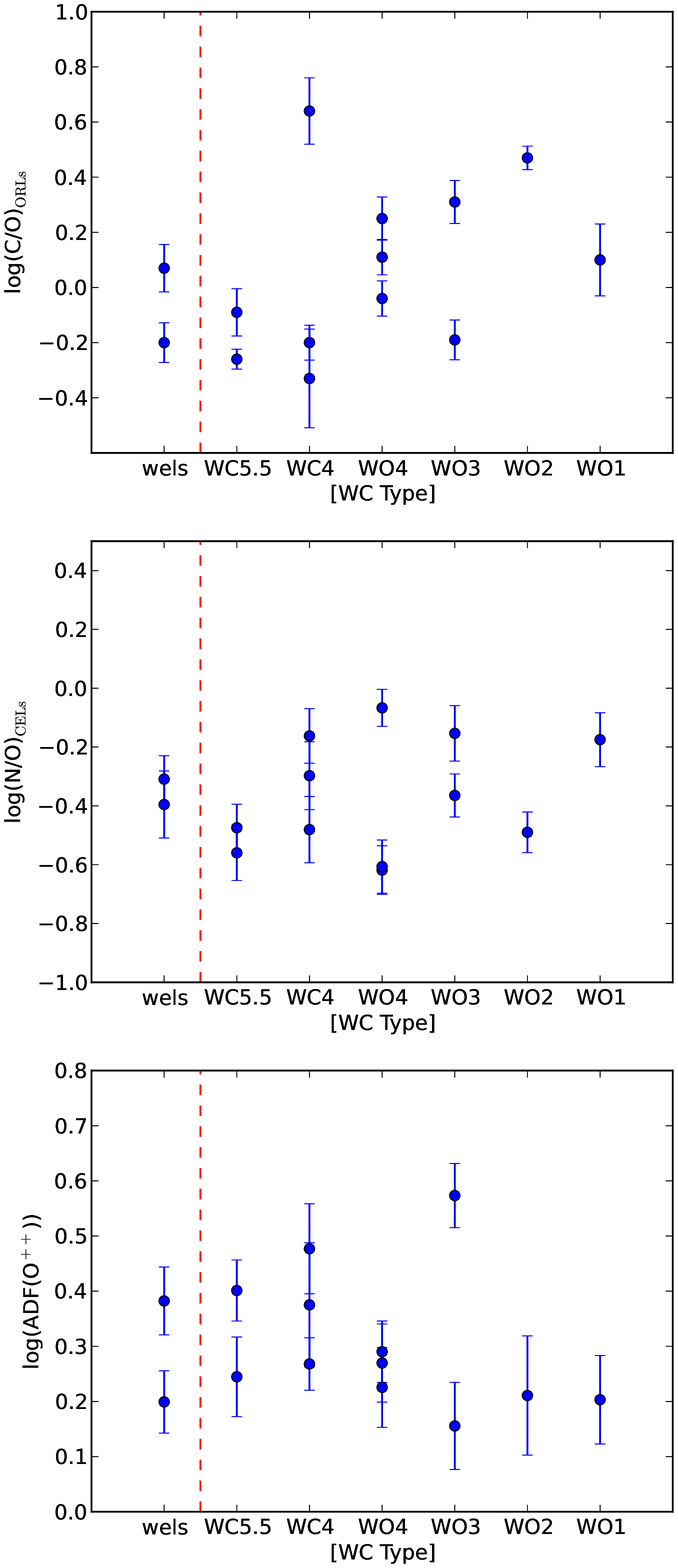}
\caption{C/O, N/O, and ADF $vs.$ central star [WC] type. The abscense of correlations is discussed in the text.}
\label{cono_wc}
\end{center}
\end{figure}

Our objects include central stars of spectral types from [WC]5 to [WO]1, that is,  they are among the hottest
[WC] stars, all of them are early-[WC], with stellar temperatures from 66 kK to more than 100 kK. There are also two {\it wels} among our objects (M\,1-30 and M\,1-61). 
We recall that we worked with a very biased sample and we showed that this sample has several particularities, for instance, the nebulae are more N- and C-rich objects than the average of more extended samples.
We have searched for some particular behaviour of abundances, that might be correlated with the spectral type. In Fig.~\ref{cono_wc} upper and middel panels, the C/O and N/O abundance ratios (which, as we demonstrated in Sect.~\ref{discuss_ab}, are related to the initial stellar mass) are plotted as a function of the spectral type. No correlation is found. C-rich nebulae (C/O $>$ 1) are found around stars with types from [WC]4 to [WO]1 and the same occurs for the O-rich objects (C/O $<$ 1). That is, objects with a particular spectral type can have any C/O ratio.
The N/O ratio is easier to analyse (Fig.~\ref{cono_wc} middle panel), because this ratio has a more direct relation with the initial stellar mass (see Fig.~\ref{no_he_nh}). The objects with log N/O$\lesssim -0.3$ had initial mass $<$ 4 M$_\odot$  and spectral types from [WC]5.5 to [WO]2, and objects with N/O$\gtrsim -0.3$ and initial masses larger than 4 M$_\odot$ also display all  [WC] types.
In conclusion, the central stars that currently show early-[WC] spectra and that lost their H-rich atmosphere had any initial mass from 1.5 to more than 5 M$_\odot$. Therefore the [WC] stage is independent of the initial  stellar mass, and stars with very different initial masses can pass through the same [WC] stage. A similar conclusion was reported by \citet{penaetal01}.

Of particular interest for us is the behaviour of the ADF with the spectral type. From the bottom panel of Fig.~\ref{cono_wc} we can see that no relation is found, hence, the spectral type and, therefore, the temperature and the wind of the central star are probably not related to the ADF physical origin.

\subsection{Abundance gradients in the Galactic disk}

\begin{figure}[!htb] 
\begin{center}
\includegraphics[width=\columnwidth]{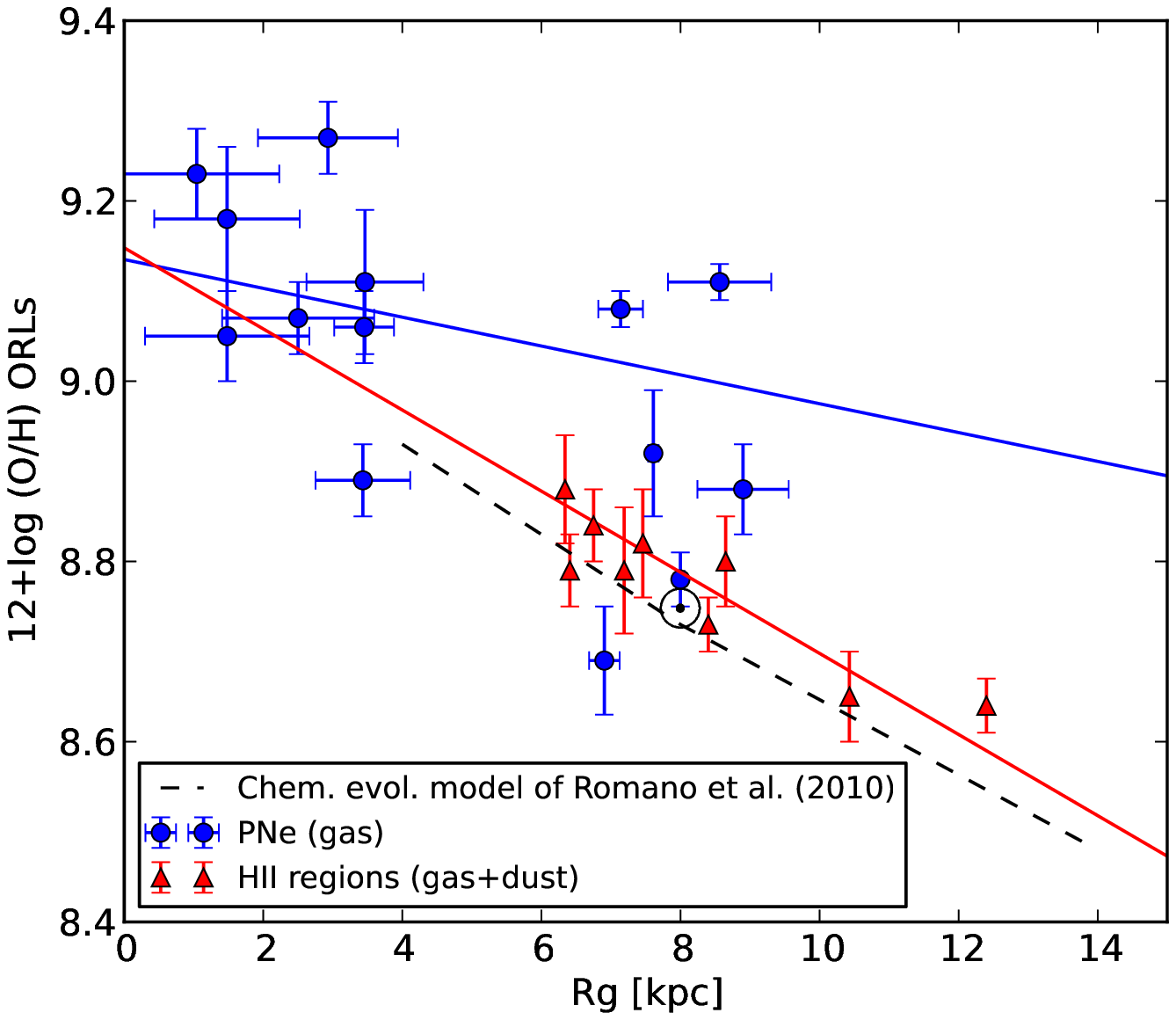}
\caption{Oxygen gradient obtained from ORLs for our sample of PNe (blue dots) and from deep spectrophotometric data of {\hii} regions \citep[red triangles, data taken from][]{estebanetal05, estebanetal13}.  Lines are least-squares fits to the data. The black dashed line represents the results of a chemical evolution model by \citet{romanoetal10}. Clearly, for PNe, the O gradient is flatter and, at the solar Galactocentric distance the O abundance is larger for PNe than for {\hii} regions (see text).}
\label{grad_O}
\end{center}
\end{figure}

Even though our sample is biased towards WRPNe with early-[WC] central stars and $wels$, it is tempting to analyse the behaviour of O abundance of our objects (which according to stellar evolution models is not significantly modified by the central stars at the Galactic metallicity) relative to the position in the Galaxy. Our objects cover a Galactocentric distance from 1 to about 9 kpc. Distances were obtained from \citet{stanghellinihaywood10}.
In Fig.~\ref{grad_O} we present the O/H abundance  ratio (as derived from ORLs) as a function of the Galactocentric distance for our objects. Values for {\hii} regions, also derived from ORLs \citep{estebanetal05, estebanetal13} and corrected for the O depleted in dust \citep[0.08 dex was added, following][]{estebanetal98, mesadelgadoetal09b} are included. The solar position is shown as well. Clear gradients are found in both cases, although the dispersion is much larger for PNe than for {\hii} regions. This large dispersion in O abundances for PNe at the same Galactocentric distance has been found in all spiral galaxies analysed so far by different authors \citep[e.g.,][]{bresolinetal10, kwitteretal12,stasinskaetal13} and it is only poorly understood. It is probably due to a combination of effects: uncertainties in distances (in the Milky Way uncertainties are much larger in PNe than in {\hii} regions), migration of PNe that would be at a Galactocentric distance different from their initial position and, very probably, contamination of nebular O due to stellar nucleosynthesis.
A very interesting fact in Fig.~\ref{grad_O}  is that the O abundance gradient given by PNe is flatter than the one derived from {\hii} regions. Not only that, but it is evident that at the solar distance, the O abundance is larger for PNe than for {\hii} regions (gas+dust) and larger than the solar value (if we were to use abundances from CELs, the same behaviour would be found). Similar results were found by \citet{rodriguezdelgadoinglada11} for some PNe and {\hii} regions in the solar vicinity. In addition,  this also occurs in other spiral galaxies such as NGC\,300 \citep{stasinskaetal13} and M\,31 \citep{zuritabresolin12}, where external PNe show a larger O abundance than {\hii} regions in the same zone. It seems most probable that PNe at these large Galactocentric distances, because they have low metallicity, could be dredging-up O from their nuclei, therefore O would  not be a suitable element for analysing abundance gradients from PNe.

In Fig.~\ref{grad_O} we show the prediction of one of the chemical evolution models presented by \citet{romanoetal10} to compare its behaviour with our gradients (black dashed line). From their Figure 17 (b) we chose their model 5, which best reproduces the observations presented there  
(other models shown in the same figure show a similar behaviour). The main 
characteristics of model 5 are listed in Table 2 of \citet{romanoetal10}. We 
found in the comparison that the gradient predicted in model 5 coincides very 
well with the gradient from {\hii} regions \citep[chemical evolution models from other 
authors produce the same result, see][]{carigietal05}, but it does not 
reproduces the gradient from PNe at all. This corroborates that O in PNe is 
affected by other phenomena than those we described above. 

\section{Conclusions
\label{conclu}}

This paper is the third in a series devoted to studying physical conditions and chemical abundances in PNe with [WC] and $wel$ central stars. 
In \citet{garciarojasetal09} we presented the complete study from deep high-resolution spectrophotometric data obtained at LCO with the 6.5-m Magellan telescope and the spectrograph MIKE for three PNe with [WC] central stars. 
In \citet{garciarojasetal12} (Paper I) we presented the spectrophotometric  data for 12 additional PNe with [WC] and $wel$ central stars obtained with the same configuration. Given the large amount of data, only the physical conditions ({\te} and {\elecd}) were discussed in that paper. In this paper we presented the complete chemical abundance analysis. 

The ionic chemical abundances were derived using the intensity of collisionally excited lines (CELs) for a large number of ions of different
elements. In addition, we determined for the first time in all these objects the O$^{++}$ and C$^{++}$ abundances from optical recombination lines (ORLs). For several objects we also derived for the first time for all of them the ORL abundances of O$^{+}$, C$^{+3}$, and Ne$^{++}$. We carefully analysed the possible phenomena (such as fluorescence) affecting the recombination line intensities. 

From the comparison between CEL and ORL abundances, we derived  ADFs(O$^{++}$) between 1.2 and 4, which are in the range of typical ADFs observed in PNe, far from the high values found in other PNe (such as the born-again PNe). Therefore we concluded that the [WC] nature of the central stars does not significantly affect the ADFs. We concluded that the O$^{++}$/H$^+$ abundance ratio obtained from multiplet 1 is representative of the recombination line abundance, because it is basically the same as that obtained from other multiplets that are excited mainly by recombination (2, 10, 20 and $3d-4f$ transitions), especially for objects with highest-quality spectra.

Total abundances were computed by using a carefully chosen set of ICFs. Total abundances for He, C, N, O, Ne, S, Cl, Ar, Mg, Fe, and Ni were computed. N/O and $\alpha$-element/O abundance ratios were computed from CELs, while C/O ratios were computed from ORLs. We found that our objects are, on average,  richer in N and C than the average of more extended samples. However, there is no correlation between the nebular chemical abundances and the [WC]  stellar type, showing that the [WR] phenomenon can occur in stars of very different stellar masses. By comparing the observed N/O and C/O abundance ratios with those predicted by stellar evolution models, we estimated that about half of our PNe had progenitors with initial masses similar to or larger than 4 {\msun}. This result is consistent with results from kinematical data, which show that WRPNe are located in a disk thinner than that of average PNe and that hence they are younger and probably more massive objects.

Assuming that the ADF and temperature fluctuations are related phenomena, 
we estimated the temperature fluctuations parameter, {\tsq} from two different methods: 1) applying a chi-squared method that minimizes the dispersion of He$^+$/H$^+$ ratios from individual lines, and 2) by comparing the O$^{++}$, O$^+$, C$^{++}$ and/or Ne$^{++}$ ionic abundances derived from ORLs with those derived from CELs. All methods provided {\tsq} values that were consistent within the uncertainties. However, some differences were found for some objects between {\tsq}(O$^+$) and  {\tsq}(O$^{++}$). 
The adopted average value of {\tsq} was used to correct for the ionic abundances derived from CELs.

In our sample, limited to normal values of ADFs, we found no correlations of ADFs with nebular properties such as metallicity, mean surface brightness, diameter, excitation class, or density, in contradiction with correlations found by other authors for samples with very large ADFs. 

The PNe oxygen abundances derived from ORLs are larger than the values from the Sun and Orion nebula. This might be partially explained by the heavy-element fractionation in the Sun since it was formed, but this effect only explains a small percentage of the observed difference. Additionally, chemical evolution models cannot explain the large differences found betweeen PNe and Orion abundances from ORLs. This is an unresolved question that needs to be analysed in more detail.

Our well-determined PN abundances allowed us to analyse the O abundance gradient in the Galaxy (Galactocentric distances  from 1 to 9 Kpc). Comparing this with the gradient determined from {\hii} regions, we found that PNe gradient is highly dispersed (dispersion range of O values at the same Galactocentric distance) and is much flatter. At the solar distance, the O abundances of PNe are larger than the values for {\hii} regions. This fact, which is similar to what was found in other spiral galaxies, is not understood  so far. Phenomena such as migration or O-enrichment of the nebula due to stellar nucleosynthesis may be playing a role in this question.

By comparing our gradients with predictions from Galactic chemical evolution models, we found that these predictions coincide very well with gradients from {\hii} regions, but do not reproduce the behaviour of PNe.


\begin{acknowledgements} 
This work received financial support from the Spanish Ministerio  de Educaci\'on y Ciencia (MEC) under projects AYA2007-63030 and 
AYA2011-22614, from CONACYT-M\'exico under grants \#43121 and CB2010/153985 and from DGAPA-UNAM, M\'exico under grants  IN112708 and IN105511.  AMD acknowledges support from Comit\'e Mixto ESO-Chile and Basal-CATA (PFB-06/2007) grant. MTR received partial support from PB06 CATA (CONICYT). JGR acknowledges all staff, employees and guests of the Instituto de Astronom\'{\i}a at UNAM, where part of this work was done. 
CM received financial support for his sabbatical at the IAC from the Spanish MEC. 
We whish to thank the anonymous referee for his/her comments that helped to improve this paper. The authors whish to thank L. Carigi, C. Esteban, V. Luridiana, D.~A. Garc\'{\i}a-Hern\'andez, A. Peimbert, M. Peimbert, M. Rodr\'{\i}guez, and G. Stasi\'nska for fruitful discussions. 
\end{acknowledgements}





\Online

\end{document}